%% file: Circuit.Photon.summary.FINAL2.tex
\documentclass[preprint,3p,10pt]{elsarticle}
\usepackage[utf8]{inputenc}

\usepackage[english,francais]{babel}

\journal{CR Physique}

\usepackage{xparse}
\usepackage{xspace}
\usepackage{xargs}
\usepackage{ifthen}

\usepackage[x11names,svgnames]{xcolor}

\usepackage[T1]{fontenc}
\usepackage{csquotes}
\usepackage[kerning=true,babel=true]{microtype} 

\usepackage[bitstream-charter]{mathdesign}

\usepackage{amsmath}
\usepackage{stmaryrd}
\usepackage{mathtools}
\usepackage{dsfont}
\makeatletter
\DeclareFontFamily{OMX}{MnSymbolE}{}
\DeclareSymbolFont{MnLargeSymbols}{OMX}{MnSymbolE}{m}{n}
\SetSymbolFont{MnLargeSymbols}{bold}{OMX}{MnSymbolE}{b}{n}
\DeclareFontShape{OMX}{MnSymbolE}{m}{n}{
    <-6>  MnSymbolE5
   <6-7>  MnSymbolE6
   <7-8>  MnSymbolE7
   <8-9>  MnSymbolE8
   <9-10> MnSymbolE9
  <10-12> MnSymbolE10
  <12->   MnSymbolE12
}{}
\DeclareFontShape{OMX}{MnSymbolE}{b}{n}{
    <-6>  MnSymbolE-Bold5
   <6-7>  MnSymbolE-Bold6
   <7-8>  MnSymbolE-Bold7
   <8-9>  MnSymbolE-Bold8
   <9-10> MnSymbolE-Bold9
  <10-12> MnSymbolE-Bold10
  <12->   MnSymbolE-Bold12
}{}

\let\llangle\@undefined
\let\rrangle\@undefined
\DeclareMathDelimiter{\llangle}{\mathopen}%
                     {MnLargeSymbols}{'164}{MnLargeSymbols}{'164}
\DeclareMathDelimiter{\rrangle}{\mathclose}%
                     {MnLargeSymbols}{'171}{MnLargeSymbols}{'171}
\makeatother

\usepackage{braket}
\usepackage{mathtools}
\usepackage{eucal}
\usepackage{upgreek} 

\usepackage[version=3]{mhchem}

\usepackage{graphicx}

\usepackage[unicode]{hyperref}
\hypersetup{
    colorlinks=true,
}

\input{macrosNotations.inc}

\input{specificNotations.inc}

\begin{document}
\hypersetup{
    citecolor=PaleGreen4!80!black,
    linkcolor=DarkRed, 
    urlcolor=DarkSeaGreen4!90!black}

\centerline{}
\begin{frontmatter}


\selectlanguage{english}
\title{Many-Body Quantum Electrodynamics Networks: Non-Equilibrium Condensed Matter Physics with Light}\footnote{French Title: R\' eseaux d'Electrodynamique Quantique: Mati\`ere Condens\' ee et Ph\'enom\`enes Hors-Equilibre de la Lumi\`ere\\ E-mail address: karyn.le-hur@polytechnique.edu}

\selectlanguage{english}
\author{Karyn Le Hur$^{1}$, Lo\" ic Henriet$^{1}$, Alexandru Petrescu$^{2,1}$, Kirill Plekhanov$^{1}$, Guillaume Roux$^{3}$, Marco Schir\'o$^{4}$}

\address{$^{1}$ Centre de Physique Th\'{e}orique, Ecole Polytechnique, CNRS, 91128 Palaiseau Cedex France
\\$^{2}$ Department of Physics, Yale University, New Haven, CT 06520, USA
\\$^{3}$ LPTMS, Univ. Paris-Sud and CNRS, UMR 8626, F-91405 Orsay, France
\\$^{4}$ Institut  de  Physique  Th\' eorique,  Universit\' e  Paris  Saclay,  CNRS,  CEA,  F-91191  Gif-sur-Yvette,  France}

\begin{abstract}

We review recent developments regarding non-equilibrium quantum dynamics and many-body physics with light, in superconducting circuits and Josephson analogues. We start with quantum impurity models addressing dissipative and driven systems. Both theorists and experimentalists are making efforts towards the characterization of these non-equilibrium quantum systems. We show how Josephson junction systems can implement the equivalent of the Kondo effect with microwave photons. The Kondo effect can be characterized by a renormalized light-frequency and a peak in the Rayleigh elastic transmission of a photon. We also address the physics of hybrid systems comprising mesoscopic quantum dot devices coupled to an electromagnetic resonator. Then, we discuss extensions to Quantum Electrodynamics (QED) Networks allowing to engineer the Jaynes-Cummings lattice and Rabi lattice models through the presence of superconducting qubits in the cavities. This opens the door to novel many-body physics with light out of equilibrium, in relation with the Mott-superfluid transition observed with ultra-cold atoms in optical lattices. Then, we summarize recent theoretical predictions for realizing topological phases with light. Synthetic gauge fields and spin-orbit couplings have been successfully implemented with ultra-cold atoms in optical lattices --- using time-dependent Floquet perturbations periodic in time, for example --- as well as in photonic lattice systems. Finally, we discuss the Josephson effect related to Bose-Hubbard models in ladder and two-dimensional geometries. The Bose-Hubbard model is related to the Jaynes-Cummings lattice model in the large detuning limit between light and matter (the superconducting qubits). In the presence of synthetic gauge fields, we show that Meissner currents subsist in an insulating Mott phase. 

\vskip 1\baselineskip


\end{abstract}


\end{frontmatter}


\section{Introduction}

The field of quantum electrodynamics \cite{cohen}  has attracted some attention with novel and modern applications in non-equilibrium quantum optics and quantum information \cite{raimond}. In fact, many experiments have accomplished to couple light and matter in a controlled manner using cavity quantum electrodynamics. Quantum matter here can be either Rydberg atoms (cold atoms) or trapped ions \cite{haroche,Wineland} for example. A step towards the realization of many-body physics has also been made through the realization of model Hamiltonians such as the Dicke Hamiltonian \cite{Esslinger0}, where the associated super-radiant quantum phase transition has been observed in non-equilibrium conditions \cite{Esslinger00}. A  solid-state version of cavity quantum electrodynamics, related to circuit quantum electrodynamics, built with superconducting quantum circuits \cite{SchoelkopfGirvin,Circuits}, is also a very active field both from experimental and theoretical points of view. Theorists have predicted novel emergent quantum phenomena either in relation with strong light-matter coupling \cite{braak} or non-equilibrium quantum physics \cite{Houckreview,Fazio}. 

The goal here is to review developments, both theoretical and experimental, towards realizing many-body physics and quantum simulation in circuit QED starting from small networks to larger ensembles of superconducting elements in the microwave limit. An experimental endeavor has been accomplished towards the realization of larger arrays in circuit QED \cite{experimentHouck,Wallraffnew,Martinis1,Martinis2} and towards controlling trajectories in small systems \cite{Irfan,Nicolas}. This research is complementary to existing efforts on collective effects in ultra-cold atoms \cite{coldatomreview,coldatom2,Optique,AntoineThierry,IOP}, Nitrogen-Vacancy centers \cite{SPEC} and opto-mechanics for example \cite{Ludwig1,Marquardt0}.  We shall also summarize (our) progress related to quantum impurity physics with light, such as Kondo physics \cite{Kondo,Anderson0,Nozieres,Wilson,Affleck,WiegmannTsvelik} of photons \cite{Karyn,Moshe,Saleur1997,Camalet} and non-equilibrium quantum dynamics \cite{loic,PAK1,PAK2,Lesovik,Demler}, as well as theoretical predictions concerning many-body physics in larger network circuits described by the Jaynes-Cummings or Rabi lattice \cite{Greentree,Angelakis1,Angelakis,Hartmann,KochHur,SchmidtBlatter,SchmidtBlatter1,marco,Marconew}. Some progress has been done towards the implementation of an effective chemical potential for photons \cite{Hafezi}.  The effect of dissipation will be discussed for small and larger networks. In Ref. \cite{Karyn00}, we have discussed other theoretical aspects in relation with these quantm impurity models; see also Refs. \cite{Leggett,weiss}.  It is also important to mention recent theoretical developments regarding the strong-coupling light-matter coupling, both at equilibrium \cite{braak,marco,moroz,zhong,gritsev,larson,nataf,Simone} and in non-equilibrium \cite{loic,braak2}. Artificial gauge fields for photons can be realized through nano-circulators, consisting of superconducting Josephson-junction rings penetrated by magnetic fluxes for example \cite{KochT,Andreas,Lehnert2},  allowing the realization of topological phases \cite{AlexKaryn}. The stability towards disorder effects was also studied \cite{AlexKaryn}. An experimental effort to build such nano-circulators has been realized in Ref. \cite{Kamal}. In fact, inspired by the discovery of quantum Hall physics \cite{Klitzing,Laughlin,Stormer} and topological insulators \cite{Kane,Bernevig,Zhang}, experimentalists have already succeeded in realizing topological phases of light and spin-orbit type coupling \cite{MIT,Rechtsman,Hafezi1,Hafezi2,Alberto} (but not yet in microwave superconducting circuits), for example through Floquet time-periodic drive (for recent developments, see Refs. \cite{Dalibard,CayssolMoessner}); some aspects of topological phases with light have been discussed in the recent reviews \cite{Soljacic,CiutiCarusotto}. It is also important to underline the experimental progress in the field of ultra-cold atoms to realize synthetic gauge fields \cite{Gerbier,Spielman,Sengstock0,BlochHofstadter,Wolfgang,JackschZoller}. Haldane and Raghu made a pioneering contribution in this field \cite{HaldaneRaghu} that was soon confirmed experimentally \cite{MIT}; they have shown that one can realize quantum Hall phases of light through magneto-optical coupling or Faraday effect. The (chiral) transport of photons at the edges in these systems has been measured \cite{Soljacic} by analogy to quantum Hall systems \cite{Bert}. This opens the door to novel applications in quantum information and quantum computing by analogy to electron systems \cite{Stern}. The direct measurement of topological invariants \cite{Thouless} in photon lattices is in principle also feasible \cite{AlexKaryn,Carusottodissi,Hafezi3,Cooper} in particular, using the concept of anomalous quantum velocity \cite{Karplus} which has been developed further for photon systems \cite{Cominotti}. A recent work by the MIT group on photonic gyromagnetic crystals with ferrite rods reports the measurement of a Chern number with photon systems \cite{MITChern}.  Some aspects of the Berry phase \cite{Berry} and Chern number  \cite{Chern} are accessible in small superconducting circuits \cite{SchoelkopfWalraff,Martinis,Lehnert,Polkovnikov}.  Artificial graphene with Dirac photons \cite{Bellec0} and three-dimensional Weyl systems \cite{Weyl},  as well as analogues of flat band models \cite{Jacqmin}, have been realized recently. 

Other photonic arrays, including quasi-periodic arrays \cite{Patrizia} and one-dimensional geometries \cite{Sebastians,Fazio1D}, have been envisioned and some have been realized experimentally \cite{Tanese}. 

\section{Small Circuits}

This Section is organized as follows. First, we will introduce the physics of small superconducting circuits. Then, we will describe progress on quantum impurity systems, at and out of equilibrium. We will also account for dissipation effects and show that it can produce novel many-body physics, for example, based on vacuum quantum fluctuations associated with long transmission lines. We will describe the Kondo effect of light \cite{Karyn,Moshe} and address recent developments in hybrid systems \cite{MarcoKaryn}, combining circuit QED and mesoscopic electron quantum dot systems \cite{Senellart,hybrid1,hybrid2,hybrid3,hybrid4}. Note that in relation with quantum vacuum fluctuations, the dynamical Casimir effect was observed in superconducting circuits \cite{Delsing}. 

\subsection{Superconducting Elements}

These superconducting circuits embody simple building blocks for quantum simulation, based on harmonic oscillators and artificial atoms - here, two-level mesoscopic systems -  such as charge, flux or phase qubits. Nowadays, qubits in superconducting circuits yield a long decoherence time that can reach a few micro-seconds. For a recent progress on superconducting qubits, see for example Refs. \cite{Lloyd,martinis,Vion,transmon,paik,fluxonium,MartinisChen,SchmidtKoch}. Microwave photons can propagate in on-chip resonators which can be identified to Fabry-Perot type cavities made from finite sections of transmission lines. The physics of such transmission lines is reminiscent of an LC (harmonic) oscillator described by the raising and lowering operator $a^{\dagger}$ and $a$. The single-mode approximation is usually appropriate for such cavities of typical size $\sim 1cm$ (operating in the microwave domain) \cite{alexandre,wallraff}. At a simple level, $(a+a^{\dagger})$ represents the voltage fluctuations in the cavity whereas $(a-a^{\dagger})/i$ mimics the conjugate flux variable \cite{alexandre,inputoutputRMP}. Superconducting materials used in such devices are generally Niobium or Aluminum.   The propagation of micro-wave photons in the system will be described below by coupling the cavity to transmission lines (see Sec. 2.3). The input AC signal $V_0.\cos(\omega t)$ will be related to the time-averaged input power $P_{in}=V_0^2/(2 R)$ where $R$ is the resistance of the (left) transmission line. This will produce a mean number of photons $N$ such that $P_{in} = \dot{N}\hbar\omega$ and $\dot{N}=dN/dt$.  The light-matter coupling will produce non-linearities in the harmonic oscillator spectrum. The superconducting circuits allow a great flexibility afforded by the nature of the nano-fabrication system; parameters in the Hamiltonian can be widely tuned with conventional lithography. Qubit frequencies  can be set from typically 2 to 15 GHz and couplings can range from a few KiloHertz to nearly 1GHz. As the lateral dimension of the transmission lines is typically a few micrometers (which is much smaller than the wave-length), the electric field in the cavity is relatively large and a strong dipolar coupling can be achieved.  Capacitors at the ends of the cavity control the photon leakage rates and in multi-cavity systems, the hopping rates between nearest neighbor cavities \cite{Houckreview}. 

A circuit containing the cavity, artificial qubit, dissipation and AC drive effects is drawn in Fig. 1.

 \begin{figure}[t]
\center
\includegraphics[scale=0.5]{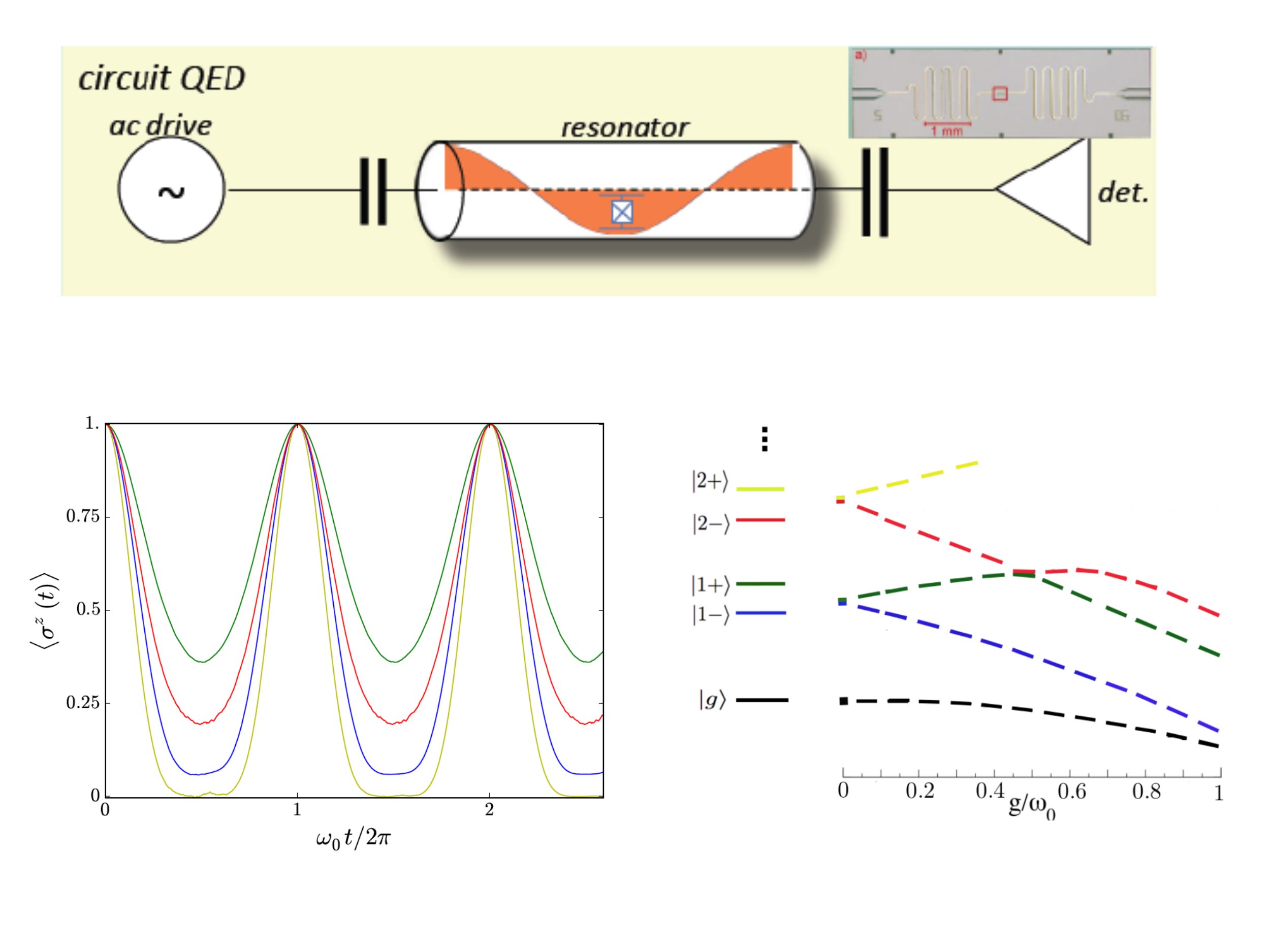}  
\vskip -1.3cm
\caption{(Color online) Example of Superconducting Circuit Quantum Electrodynamics, with an artificial atom with two levels (spin-1/2 or qubit). The electromagnetic resonator (cavity $\sim 1cm$) is described through a one-mode transmission line and is capactively coupled to an AC drive (input signal). We show an example of spin dynamics in the strong-coupling light-matter regime. We apply the methodology described in Ref. \cite{loic} and consider the far detuned or adiabatic limit of the Rabi model: $g/\omega_0$ from 0.7 (green) to 1.5 (yellow) and $\Delta/\omega_0=0.05$. The system is prepared in the initial state  $|+z\rangle\otimes|0\rangle=|0,+z\rangle$ and the revival probability of the the initial state can be computed exactly in this limit  \cite{Casanova}. The positions of the maxima only depend on the cavity frequency $\omega_0$. We also show the Jaynes-Cummings polariton ladder for $g/\omega_0=0.2$, as well as the evolution of the energy eigenstates of the Rabi model when approaching the strong coupling limit with $g\sim \omega_0$.}
\label{circuit}
\end{figure}

\subsection{Rabi, Dicke and Jaynes-Cummings Models}

The model describing the strong-coupling between a qubit and the electro-magnetic field of the cavity corresponds to the Rabi model (1936) \cite{Rabi}:
\begin{equation}
H = \frac{\Delta}{2}\sigma_z + \omega_0 \left(a^{\dagger} a+\frac{1}{2}\right) + \frac{\sigma_x}{2} g(a+a^{\dagger}).
\end{equation}
Here, $\omega_0$ represents the frequency of the quantized harmonic oscillator (quantized electro-magnetic field), $\Delta$ is the resonant frequency between the two levels forming the qubit and $\sigma_i$ where $i=x,y,z$ are the Pauli spin operators. Furthermore, $g$ represents the dipole coupling between qubit and photon. If $\sigma_x$ represents the two charge states of a Cooper pair box \cite{Devoret,Nakamura,Schonreview} then the capacitive coupling between a qubit and a resonator realizes a quantum impurity model of Rabi type \cite{Buisson}. An analogous model is obtained in the case of an inductive coupling between the qubit and the cavity. 

Although the Rabi model represents one of the simplest interacting quantum models, it does not correspond to a simple theoretical problem. Nevertheless, recently analytical solutions have been pushed forward in connection with the experimental possibilities in superconducting circuits, and more generally in nano-circuits. In particular, the limit $g/\omega_0\sim 0.1$ is commonly reached experimentally \cite{MooijSolano,solano2}. The discrete $\mathbb{Z}_2$ symmetry associated with the Hamiltonian ($a\rightarrow - a$ and $\sigma_x \rightarrow -\sigma_x$ ($\sigma_y\rightarrow -\sigma_y$)) cannot be ignored in this regime making analytical solutions more difficult to find. In this respect, it is important to emphasize the theoretical progress first realized by D. Braak in the strong-coupling limit \cite{braak,moroz,zhong,gritsev,larson,nataf,marco,Simone}, concerning the solvability of the Rabi model. It is also relevant to underline other theoretical progress on the integrability of related quantum impurity models with light \cite{Babelon,Babelon2,Faribault}. One relatively `simple' limit of the Rabi model in the strong-coupling limit is the so-called adiabatic limit \cite{Schweber,Irish}, corresponding to the highly detuned limit $\Delta\rightarrow 0$, which has been addressed for example in Ref. \cite{larson}.  In Refs. \cite{PAK1,PAK2}, in collaboration with P. P. Orth and A. Imambekov, one of the authors has developed a relatively simple stochastic scheme to address certain non-equilibrium properties and quenches on dissipative spin systems. This approach is based on a numerically exact Feynman-Vernon path integral approach \cite{Leggett,weiss} (see also Sec. 2.4). In Ref. \cite{loic}, two of the authors together with Z. Ristivojevic and P. P. Orth have extended this approach to dissipative light-matter systems beyond the weak-coupling limit. We show extra results in the strong-coupling limit (Fig. 1), in the highly detuned  limit $\Delta/\omega_0\ll 1$, in agreement with Ref. \cite{Casanova}.  The revival probability of the initial state $|+z\rangle\otimes|0\rangle=|0,+z\rangle$ (with zero photon in the cavity) can be computed exactly in the limit $\Delta\rightarrow 0$, using the shifted $b=\sigma_x a$ basis \cite{Casanova}. More precisely, this probability is of the form $\exp-|Z(t)|^2$ with $Z(t)=(g/\omega_0)(e^{-i\omega_0 t}-1)$ \cite{Casanova}, leading to periodic collapses and revivals with frequency $\omega_0/(2\pi)$. The ratio $g/\omega_0$ controls the sharpness of these revivals peaks.

The limit of large spin then referring to the Dicke Hamiltonian (1954) \cite{Dicke} is also solvable through the Holstein-Primakoff transformation, exemplifying the possibility of superradiant effects for strong light-matter  couplings \cite{LiebHepp}. As mentioned in the introduction, such a superradiant quantum phase transition has been recently observed with ultra-cold atoms in an optical cavity, in a non equilibrium setting \cite{Esslinger0,Esslinger00}. The possibility of a Dicke quantum spin glass has also been discussed \cite{Dickeglass}. Some efforts have also questioned the possibility of superradiant quantum phase transitions in mesoscopic circuits with Cooper pair boxes \cite{Nataf,MarquardtDicke}. Coupling an ensemble of atoms or artificial qubits to light is now commonly realized with various signatures of collective effects \cite{IOP,SPEC}. We have also shown that entanglement between light and matter in the Dicke model can be probed through the properties of the light field \cite{NatafMehmet,Vidal} or through bipartite fluctuations \cite{Francis,KlichLevitov}. 

In the 1960s, Jaynes and Cummings proposed an approximation to the Rabi model \cite{JaynesCummings}, referred to as the rotating-wave approximation, where the interaction part is simplified into $\frac{g}{2}(\sigma^+ a + \sigma^- a^{\dagger})$. This approximation is usually valid when $g/\omega_0\ll 1$ and at small detuning $\delta = |\omega_0 - \Delta| \ll \Delta$. The number of excitations in the system (hybrid excitations of light and matter or polaritons), the number of polaritons $N = a^{\dagger} a + \frac{1}{2}(\sigma_z + 1)$, is then conserved as a result of the emergent U(1) symmetry ($a\rightarrow a e^{-i\varphi}$ and $\sigma^+ \rightarrow \sigma^+ e^{i\varphi}$). We  can diagonalize the model in the basis $| n, + z\rangle$ and $| n+1, - z\rangle$ where $n$ refers to the number of photons in the system and $| \pm z\rangle$ refers to the qubit polarization along the $z$ axis. We have the following eigen-energies and eigenstates for $N\geq 1$ (the ground state has energy $E_0 = \delta/2$) :
\begin{eqnarray}
E_{N\pm} = N\omega_0 \pm \frac{1}{2} \sqrt{\delta^2 + N g^2},
\end{eqnarray}
and
\begin{eqnarray}
| N + \rangle &=& \alpha_N | N-1, +z\rangle +\beta_N | N, -z \rangle  \\ \nonumber
| N - \rangle &=& -\beta_N | N-1, +z\rangle +\alpha_N | N, -z \rangle  ,
\end{eqnarray}  
where 
$\beta_N = \cos(1/2 \tan^{-1} \frac{g\sqrt{N}}{\delta})$ and $\alpha_N = \sin(1/2 \tan^{-1} \frac{g\sqrt{N}}{\delta})$. We recover the usual Jaynes-Cummings (JC) ladder (see Fig. 1). The anharmonicity of the Jaynes-Cummings ladder produces exotic phenomena such as photon blockade, preventing a second photon to enter in the cavity when a first one is also present \cite{Imamoglublockade,Verger}. The $\sqrt{N}$ spacing of the JC ladder can be used to generate nonclassical states of light. Photon-photon interactions occur both in the resonance condition between cavity and qubit and in the dispersive limit, where the light frequency and the qubit level separation $\Delta$ are well distinct such that the spin remains in the ground state and the dynamics of the photons is described by an effective Hubbard interaction $U (a^{\dagger} a)^2$. More precisely, defining the small parameter $\lambda = g/| \omega_0 - \Delta|$ and performing an exact unitary transformation on the Hamiltonian results in  an effective photon-photon interaction $U = g\lambda^3$ \cite{Blais}. This also produces a shift in the resonator frequency. The photon-photon interaction certainly modifies the energy required to add a photon to the cavity depending upon the number of photons occupying the cavity. Although this is a higher order effect, the key point is that the interaction energy or Kerr energy can be made larger than the cavity line width $\kappa$ and the transport of photons through the cavity is blockaded.  This photon interaction effect has been observed in various experiments (transmission spectrum as a function of frequency or intensity correlation function), for example, in cavity \cite{Kimble} and  circuit-QED \cite{Bishop,Fink,Hofheinz,Hoffman}. 

By increasing the light-matter coupling and treating the counter-rotating wave terms perturbatively, one identifies a Bloch-Siegert shift \cite{cohenDupont} making the two polariton levels $|1-\rangle$ and $|1+\rangle$ repelling each other: when entering the strong-coupling limit the level $|1-\rangle$ becomes closer and closer to the ground state (see Fig. 1). In the strong-coupling limit, the structure of levels becomes more complicated \cite{gritsev}. In the Rabi lattice situation, features reminiscent of the Dicke model appear due to the presence of the counter-rotating wave terms \cite{marco}. To get further insights, one can project the rabi ground state in the polariton basis \cite{marco}. Due to the parity symmetry, only even polariton states contribute with different weights, which give the probability distribution of having $N$ polaritons in the ground state. This distribution is peaked around some finite value as a result of the counter-rotating terms which mix sectors at different polariton numbers, thus acting effectively as a sort of chemical potential.

In Ref. \cite{loic}, we used a relatively simple stochastic protocol to address certain non-equilibrium properties and quenches on the qubit strongly-coupled to light (see also Sec. 2.4). In particular, we show how to reach an almost pure state with one polariton in the branch $|1-\rangle$ by driving the cavity, with an almost zero detuning between light and matter. Note that in the limit of large detuning between light and matter, this protocol corresponds to make a $\pi$-pulse on the qubit \cite{HofheinzMartinis}. Below, we shall discuss further applications. 

\subsection{AC driving, Dissipation and Spin-Bath Models}

Let us now discuss dissipation and driving effects in the cavity. First, the cavity line width and driving effects from an AC source can be described in a microscopic manner by coupling the cavity to very long transmission lines \cite{Gardiner,inputoutputRMP}  which correspond to a bath of harmonic oscillators \cite{FV,Caldeira_Leggett}. 

Let us envision a configuration where a single cavity is capacitively coupled to a left and a right transmission lines (serving as photon baths). The open cavity system coupled to the transmission lines is described through the Hamiltonian:
\begin{eqnarray}
H_{lines} = \sum_{j=l,r} \sum_{k} v |k| \left( b^{\dagger}_{jk} b_{jk} + \frac{1}{2}\right) + \sum_{k} \alpha_k \left(\gamma_l(b_{lk} + b_{lk}^{\dagger}) + \gamma_r(b_{rk} + b_{rk}^{\dagger}) \right)(a+a^{\dagger}).
\end{eqnarray}
The transmission lines with dispersion $\omega_k=v|k|$ (the Planck constant $\hbar$ is normalized to unity, for simplicity) are then described by the spectral function 
\begin{equation}
J(\omega) = \pi \sum_k \lambda_k^2 \delta(\omega-\omega_k) = 2\pi\alpha \omega \exp(-\omega/\omega_c),
\end{equation}
 and $\omega_c$ is an ultraviolet cutoff for photon excitations in the environment (this corresponds to Ohmic dissipation). More precisely, defining the symmetric and anti-symmetric combinations of the two transmission lines, only the symmetric mode couples to the cavity, and the effective coupling constant $\alpha_k$ turns into $\lambda_k = \alpha_k\sqrt{\gamma_l^2 + \gamma_r^2}$. Here, $\alpha_k$ depends on the properties of the transmission lines, which are supposed to be identical, and $\gamma_{l,r}$ encode  the capacitive couplings between the cavity and the transmission lines \cite{Karyn,MarcoKaryn}.  We relate the bosonic modes in the left line $b_{lk}$ with the input signal and the input power $P_{in}$  in a usual way, and the output signal is expressed in terms of the boson $b_{rk}$ \cite{inputoutputRMP}. Using exact equations of motion and the input-output theory \cite{inputoutputRMP}, we can obtain an expression for the transmission and reflection coefficients of a photon in the system \cite{MarcoKaryn}:
\begin{eqnarray}
r(\omega) &=& 1 - 2i \frac{\gamma_l^2}{\gamma_l^2 +\gamma_r^2} J(\omega) \chi_{xx}^R(\omega) \\ \nonumber
t(\omega) &=& |t|\exp(i\varphi) = \frac{2i\gamma_r\gamma_l}{\gamma_l^2 +\gamma_r^2} J(\omega) \chi_{xx}^R(\omega).
\end{eqnarray}
In this expression, we introduce the exact retarded photon Green's function $\chi_{xx}^R(t) = -i\theta(t)\langle [x(t),x(0)]\rangle_{sys}$ describing the response of the photon displacement $x=(a+a^{\dagger})/\sqrt{2}$. The coupling with the environment then
produces a damping of Ohmic type for the photon propagator \cite{MarcoKaryn}:
\begin{equation}
\chi(\omega) = \frac{\omega_0}{\omega^2 - \omega_0^2 + i\omega_0\kappa},
\end{equation}
where $\kappa = 2\pi\alpha\omega$ is the effective dissipation induced by a coupling $\alpha$ to the modes of the transmission lines. Notice that close to resonance, the dissipation can be described through a complex frequency $\omega_0 \rightarrow
\omega_0 - i\kappa/2$. A damping rate linear with frequency is usually typical of long transmission lines which embody a Ohmic environment. Note that by treating these dissipation effects in a Hamiltonian manner through a set of bosonic variables already constitutes a first many-body effect in these systems. Some dissipation effects can be also described through the Lindblad \cite{Lindblad} or Bloch-Redfield \cite{Bloch,Redfield} type equations. Below, we show how many-body physics can be engineered from these Ohmic environments. 

Dissipation effects on the spin or the artificial two-level system already results in rich many-body quantum impurity models, such as the spin-boson Hamiltonian \cite{Leggett,weiss,Blume,KLH0,Vojta}:
\begin{equation}
H_{SB} =  \sum_{k} v |k| \left( b^{\dagger}_{k} b_{k} + \frac{1}{2}\right) - \frac{\Delta}{2} \sigma_x + \frac{h}{2}\sigma_z + \frac{1}{2}\sigma_z\sum_k  \lambda_k(b_k + b_k^{\dagger}).
\end{equation}
In the case of Ohmic dissipation (case of Eq. (5)), the physics of this model is very rich showing some exact mapping to the anisotropic Kondo model and the classical Ising model with long-range forces \cite{Spohn,AYH}; these models are now standard models in many-body physics. In particular, this Ohmic spin-boson model yields a quantum phase transition for the dissipative parameter $\alpha_c\sim 1$ by analogy to the ferromagnetic - antiferromagnetic quantum phase transition in the Kondo model \cite{Chakravarty0,Bray}; here, we use the definition of $\alpha$ given in Eq. (5).  It is important to mention recent progress in nano circuits \cite{Jezouin,Finkelstein} towards observing such dissipative quantum phase transitions \cite{KLH,SafiSaleur,Zarand,Buettiker,FurusakiMatveev}. Rabi oscillations of the qubit also show some incoherent crossover for the Toulouse limit \cite{Toulouse} corresponding to a dissipative parameter $\alpha=1/2$, where the problem becomes formally equivalent to a non-interacting resonant level model \cite{Guinea}. Similar quantum phase transitions have been predicted for a two-spin system \cite{ALJ,Garst,PeterDavid} and could be observed in light-matter systems \cite{Raftery}. One of the author and B. Coqblin have studied the problem of two-coupled spin-1 impurities \cite{KarynBernard} and extensions to lattice arrays \cite{Karynunder}.
Non-equilibrium transport at such dissipative quantum phase transitions has also been studied theoretically \cite{ChungHou} and experimentally \cite{Jezouin,Finkelstein}. 

\subsection{Dissipative Quantum Trajectories, Stochastic Dynamics and Quantum Phase Transition for two spins}

Various methods have been devised to describe quantitatively the dynamics of these spin-boson systems, from master equations \cite{Carmichael}, the non-interacting blip approximation (NIBA) and its extensions \cite{Leggett,weiss,Matteo}, stochastic wave function approaches \cite{CDM,Dum} and stochastic differential equations \cite{Gisin},  time-dependent Numerical Renormalization Group approaches \cite{AndersSchiller,Bulla,PeterDavid}, multi-polaron expansion \cite{Florenspol}, Density Matrix Renormalization Group \cite{Pollet2} and Matrix Product States \cite{Solano,Garrahan,Marconew} for example. These efforts complement other methods applied for non-equilibrium dissipative quantum systems based on Monte-Carlo techniques on the Keldysh contour \cite{MarcoMC,WernerMillis,Thomas,Waintal}. These theoretical developments are very useful owing to important progress on the experimental side both in quantum optics, ultra-cold atoms and superconducting qubits, for example, by following the real-time dynamics on the Bloch sphere. In the context of superconducting qubits, tracking single quantum trajectories has been explored experimentally with a particular focus on the competition between continuous weak measurement and driven unitary evolution \cite{Irfan,Nicolas}. Some connections also occur with the concept of open quantum walkers \cite{BauerBernard}. 

At a classical level, the dynamics of dissipative spins in a macroscopic and dissipative environment  can be described through Bloch equations $\dot{\vec{S}}(t) = \vec{H}\times \vec{S}$ where the dissipation is here included in the noisy random magnetic field $\vec{H} = h(\cos \phi(t) , \sin \phi(t), 0)$. Classically, the spin then obeys the differential equation $\dot{S}_z(t) = - h^2\int_0^t ds \cos(\phi(t) - \phi(s))S_z(s)$. At a quantum level, as a result of the Ehrenfest theorem, $\langle \sigma_z \rangle$ obeys a similar equation, where we also average over different magnetic field configurations:
\begin{equation}
\frac{d}{dt} \langle \sigma_z\rangle = - h^2 \int_0^t ds \langle \cos(\phi(t) - \phi(s)) \sigma_z(s) \rangle.
\end{equation}
A usual approximation consists in decoupling the dynamics of the stochastic field and the spin dynamics, neglecting completely the back-action of the spin dynamics on the stochastic field evolution. This leads to the well-known NIBA approximation \cite{Leggett} (where the noisy phases are then derived from the spin-boson model in Eq. (8) and are formally operators \cite{Karyn00}). Nevertheless, it is important to mention recent theoretical progress both analytical and numerical trying to go beyond these type of approximations \cite{PAK1,PAK2,Schoeller}. These developments also include applications on the driven and dissipative Jaynes-Cummings and Rabi models \cite{loic} and networks of spin-boson models, leading to many-body stochastic non-equilibrium (mean-field) theories \cite{LoicKaryn,Keeling}. Lindblad and Bloch-Redfield approaches give a complementary point of view \cite{Hakan,Vavilov} and can even include non-Markovian effects \cite{Gisin}. Similar approaches have been developed to address heat transfer \cite{Matteo}.

In Fig. 1, we show one result obtained for the driven Rabi model, by developing a stochastic Schr\" odinger equation approach. The formalism is derived on the path-integral real-time formalism \cite{FV} and allows to treat Eq. (9) in an exact manner \cite{PAK1,PAK2}. This formalism has been developed based on previous works in the field \cite{Lesovik,Demler,SG}. In Ref. \cite{loic}, the formalism was developed further  in  the context of the driven and dissipative Rabi model. In Fig. 1, we compute the spin dynamics exactly in the strong-coupling limit, by using (coupled) stochastic Gaussian variables to decouple long-range interactions in time induced by the environment (here, the cavity). The formalism has also been developed to incorporate dissipation effects, and is also inspired from Ref. \cite{SG}. In Sec. 3.1, we show one application in the driven Jaynes-Cummings lattice including dissipation in the cavities.

In Fig. 2, we show a recent extension of the stochastic Schr\" odinger equation approach in the context of two spins in a ohmic environment \cite{LoicKaryn}. The model
is identical to that in Eq. (8), but now one adds an extra direct (Ising) coupling $K\sigma_{1z}\sigma_{2z}$ between spins and the transverse field $\Delta$ acts on each spin separately \cite{PeterDavid}. There are two phases, a paramagnetic phase where the two spins are unpolarized and couple to the bath, and a polarized (ferromagnetic) phase where the spins are frozen in a $|\uparrow \uparrow\rangle$ state, and decouple from the bath: the quantum phase transition is of Kosterlitz-Thouless type by analogy to the single-spin problem. The quantum phase transition line (characterized by the parameter $\alpha_c$) is determined by considering the evolution of the system with the two spins being initially in the triplet state $|T_+\rangle=|+_z,+_z\rangle$. For $\alpha<\alpha_c/2$, we observe Rabi oscillations and a progressive damping towards the unpolarized equilibrium state. For $\alpha_c/2< \alpha<\alpha_c$, we observe an exponential relaxation towards the unpolarized equilibrium state (by analogy to the single-spin problem for $\alpha\geq 1/2$). For $\alpha>\alpha_c$, the spins stay localized in the initial state $|T_+\rangle=|+_z,+_z\rangle$. This dynamical procedure allows to determine quantitatively the position of the dissipative quantum phase transition in the two-spin system. Results for the position of the quantum phase transition are also compared with those from the Numerical Renormalization Group approach \cite{PeterDavid} and from Quantum Monte Carlo \cite{QMC2spins}. 

 \begin{figure}[t]
\center
\includegraphics[scale=0.4]{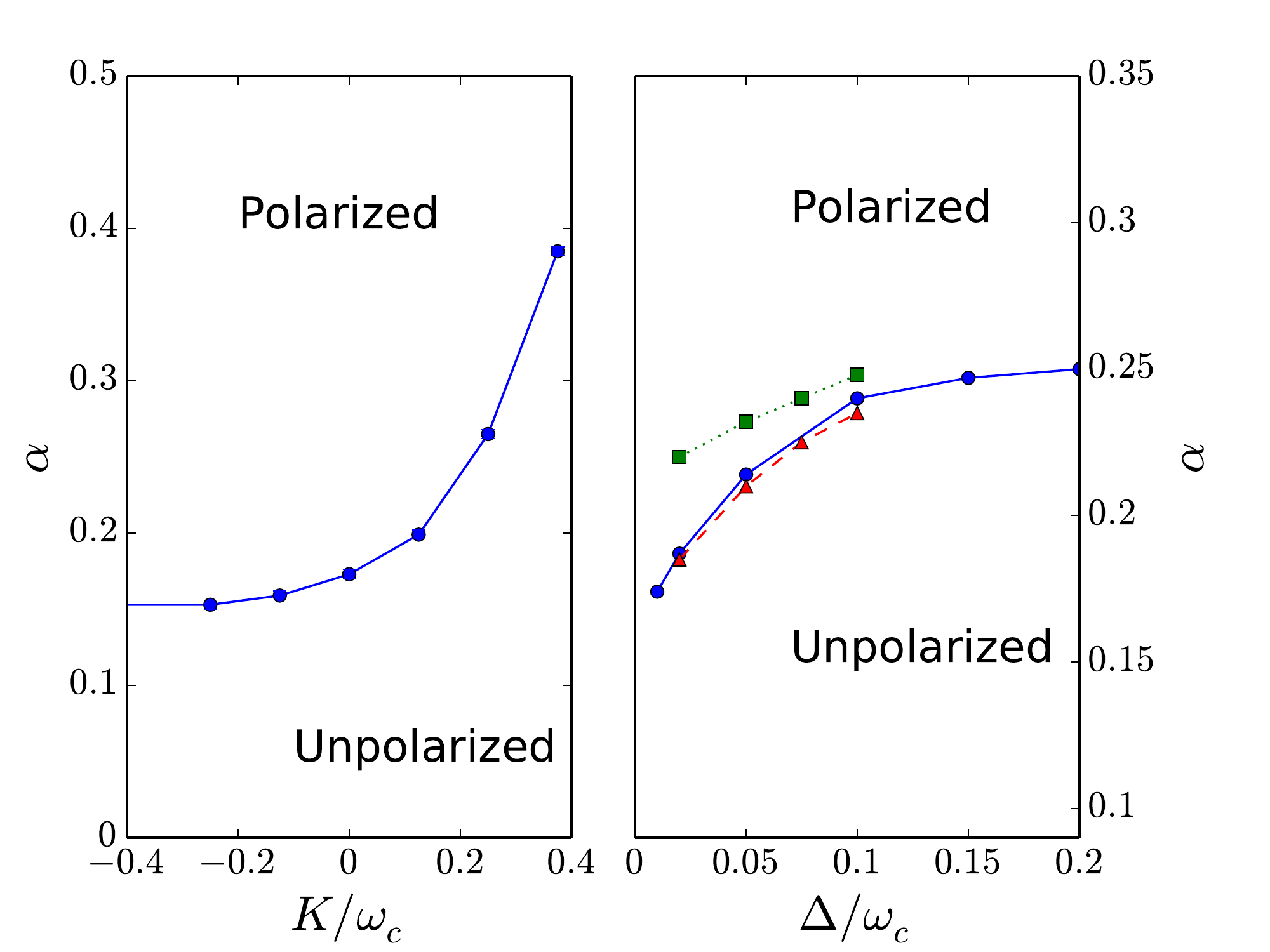}  
\caption{(Color online) (Left) Phase Diagram of the two spin system obtained from the stochastic Schr\" odinger (SSE) Equation method for two spins \cite{LoicKaryn}, in the context of 
Rabi oscillations after preparing the spin in the initial state $|T_+\rangle$. Here, $\alpha$ is the dissipative parameter of Eq. (5), $K$ represents an Ising (direct) coupling between the two spins and $\Delta$ is the transverse field acting on each spin. This model exhibits a Kosterlitz-Thouless type transition \cite{Garst,PeterDavid}, by analogy to the single-spin problem, and allows to synchronize the spins via the bath as a result of the many modes in the environment \cite{PeterDavid}. The unpolarized phase refers to a paramagnetic phase where the spins are entangled and also entangled to the bath, and the polarized phase refers to a phase where the two spins are locked into the $|T_+\rangle$ state, which does not evolve in time. This quantum phase transition could be realized by coupling the qubits (spins) to a one-dimensional transmission line, Josephson junction array, or equivalently to a one-dimensional BEC \cite{PeterIvan}. (Right) Comparison between the SSE method (blue) \cite{LoicKaryn}, the NRG approach (red, dashed) \cite{PeterDavid} and the QMC approach (green, dotted) \cite{QMC2spins} is shown. There is no direct coupling between the spins $(K=0)$.}
\label{circuit}
\end{figure}

\subsection{Exploration of Josephson-Kondo Circuits with Light}

Josephson physics \cite{Josephson,AmbegaokarBert} has been explored with superconductors \cite{superconductors,Shapiro} and quantum circuits \cite{Fulton,Devoret}. Analogous Josephson effects have also been explored with superfluids, for example, in ultra-cold atoms \cite{BECJosephson,Markus}, in superfluid He \cite{Helium} and with polariton condensates \cite{Jacqueline}. 

Here, we review small superconducting circuits in the microwave domain allowing to realize Kondo physics with one-dimensional Josephson junction arrays (mimicking long transmission lines) \cite{Karyn,Moshe} instead of electron reservoirs, as in the original Kondo Hamiltonian \cite{Kondo,Nozieres}. The Kondo effect has also been realized in quantum dot systems \cite{David,Leo,LeonidLeo}. Excitations in the long transmission lines at equilibrium are zero-point fluctuations (bosons instead of electrons in the original Kondo model). Similar geometries have been envisioned in mesoscopic rings \cite{Buettiker}, with Luttinger liquids \cite{FurusakiMatveev,Karyn} or one-dimensional BECs (Bose-Einstein condensates) \cite{Recati,PeterIvan}. In Fig. 3A, we introduce a typical Josephson-Kondo circuit where micro-wave photons are sent across the system \cite{Karyn,Moshe}. This geometry is also inspired from Refs. \cite{Meirong,jens}. Related circuits with photons have been addressed theoretically \cite{Saleur1997,Camalet,Baranger,Florensnew}. Some (fermionic) Kondo type impurity models have also been explored in ultra-cold atoms \cite{DemlerSalomon,Demler2}.

The dissipative parameter in Eq. (5) representing the one-dimensional transmission lines here takes the form $\alpha \sim 2R/R_q$, where $R$ is the resistance of each (superconducting) transmission line and $R_q=h/(2e)^2$ (where $h=2\pi$ within our units) is the quantum of resistance. With the current efforts in nanotechnology one can build transmission lines with quite large impedances \cite{Jezouin,Finkelstein,Saclay}. In particular, realizing highly dissipative transmission lines with Josephson junction arrays is also possible thanks to experimental progress \cite{fluxonium}. The artificial atom was originally considered to be a double-dot charge qubit at resonance allowing for the coherent propagation of Cooper pairs across the system in the micro-wave domain.  One could use superconducting qubits with long decoherence times \cite{Lloyd,martinis,Vion,transmon,paik,fluxonium}. It is also relevant to note recent progress in realizing such devices experimentally \cite{SolanoSB}.

The Hamiltonian representing the pseudo-spin (representing the double-island charge qubit \cite{Pashkin,Lafarge})  coupled to the left and right transmission lines takes the form \cite{Karyn}:
\begin{equation}
H = -\frac{E_J}{2} \sigma^+ e^{i(\Phi_l-\Phi_r)} + h.c. +  \sum_{j=l,r} \sum_{k} v |k| \left( b^{\dagger}_{jk} b_{jk} +\frac{1}{2}\right).
\end{equation}
The Josephson phases depict the (generalized) superfluid phases in each transmission line in the vicinity of the artificial qubit. Note that in the geometry of Fig. 3 (left), only the antisymmetric mode $(\Phi_l-\Phi_r)$ couples to the transmission line. In the underdamped limit $(0\ll\alpha\ll 0.3)$ of the spin-boson model introduced earlier, the Kondo physics here corresponds to the {\it perfect} Rayleigh transmission of a micro-wave photon at the frequency corresponding to the Kondo frequency of the system \cite{Karyn}. Such a concept of resonance fluorescence has been observed with artificial superconducting qubits \cite{Astafiev}. Note that this frequency is distinct from the bare frequency of the artificial atom (qubit) as a result of the quantum point-fluctuations in the long transmission lines. More precisely, from the two-level perspective, one can define an effective transverse field 
\begin{equation}
 E_J \langle \cos(\Phi_l - \Phi_r) \rangle. 
\end{equation}
Close to equilibrium, when the input power can be treated in linear response theory, the effective transverse field results from known results on the Ohmic spin-boson model \cite{Leggett,weiss}:
\begin{equation}
 \omega_R = E_J \left(\frac{E_J}{\omega_c}\right)^{\frac{\alpha}{1-\alpha}} \sim E_J\left(1+ \frac{\alpha}{1-\alpha}\ln\left(\frac{E_J}{\omega_c}\right)\right). 
\end{equation}
The $\ln-$corrections which stem from the many-body excitations in the long transmission lines can be understood as renormalization effects. This scenario assumes that the high-frequency cutoff $\omega_c$ of the order of the plasma frequency in the transmission lines is much larger than the Josephson frequency. The key point is that the spin susceptibility $\chi_{spin}^R(t) = -i\theta(t)\langle [\sigma_z(t),\sigma_z(0)]\rangle$ takes a form identical to Eq. (6) but with the frequency $\omega_0$ being replaced by the Kondo frequency $\omega_R$ and a damping rate $\gamma(\omega)=\omega_R J(\omega)$ which is reminiscent of the Korringa-Shiba relation in the spin-boson model \cite{spinbosonKorringa} and in the Kondo model \cite{Nozieres,Shiba,Leonid,ChristopheKaryn,Michele,MichelePRL}. Note that this damping rate can be measured directly in quantum RC circuits through the concept of quantum of resistance \cite{ButtikerRC,ButtikerRC2,Feve,Gabelli,Martin}. Multichannel effects have also been discussed theoretically \cite{Etzioni,Dutt,Texier}.

 \begin{figure}[t]
\center
\includegraphics[scale=0.3]{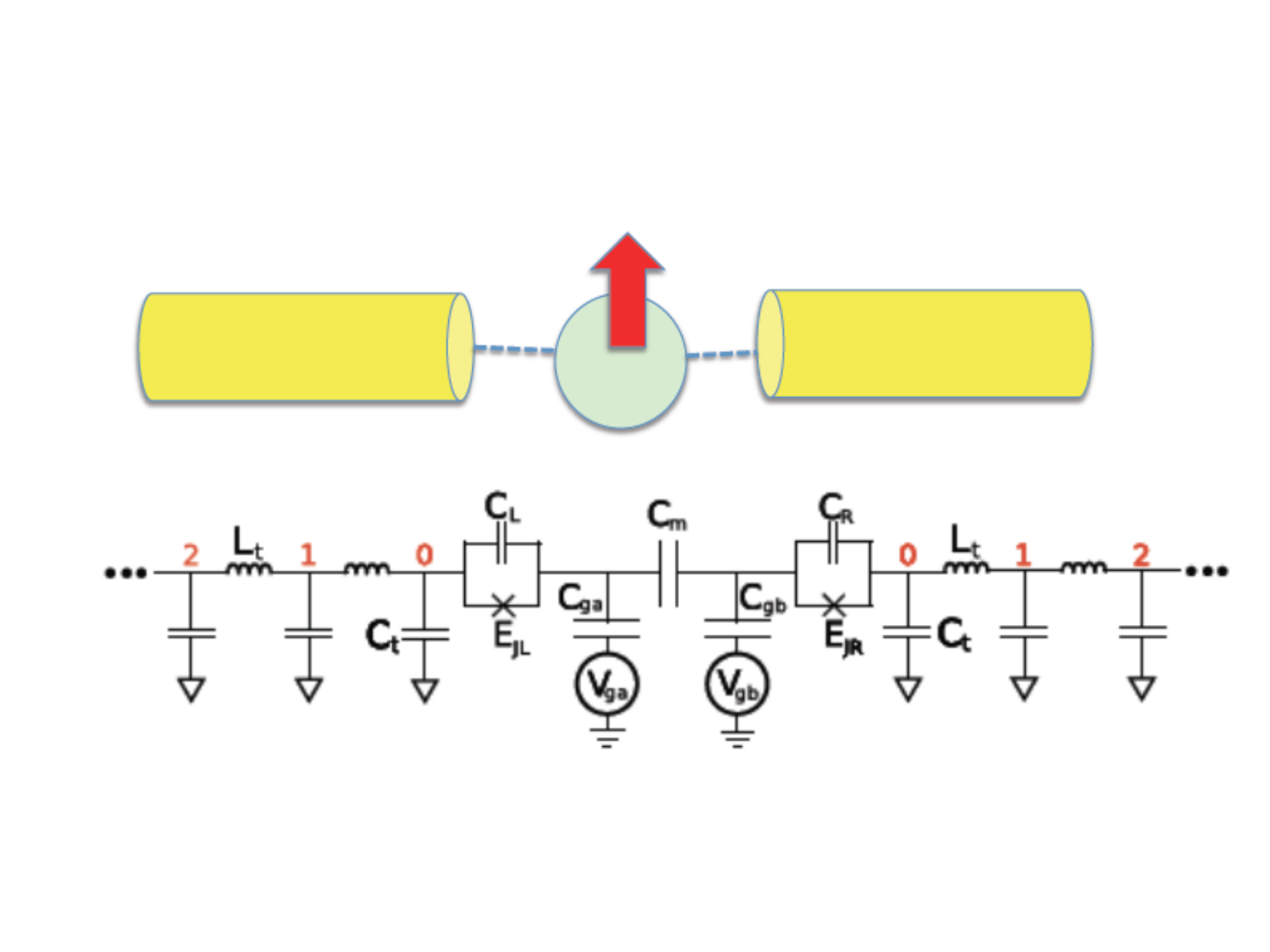}  
\includegraphics[scale=0.33]{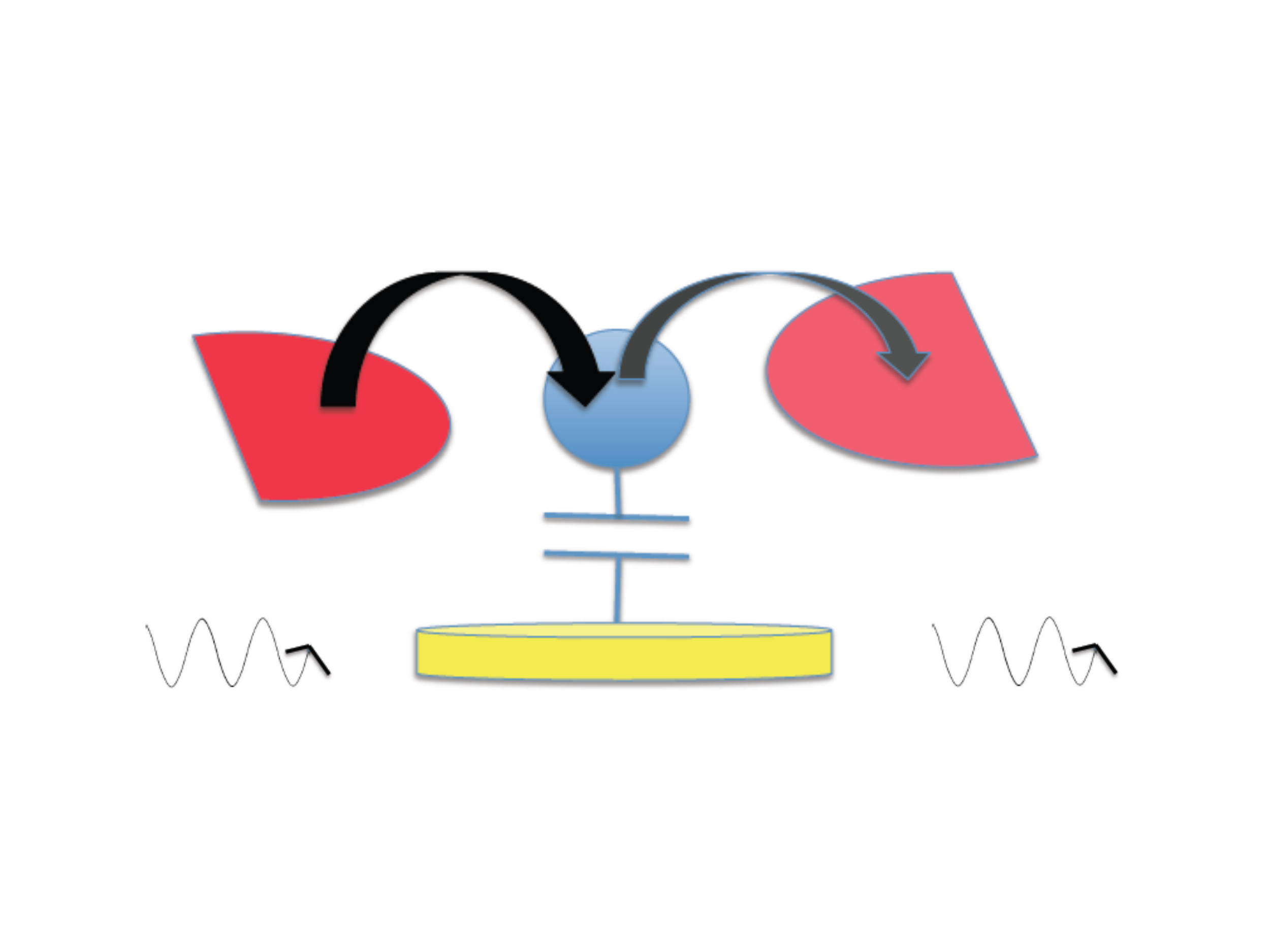}  
\vskip -0.7cm
\includegraphics[scale=0.43]{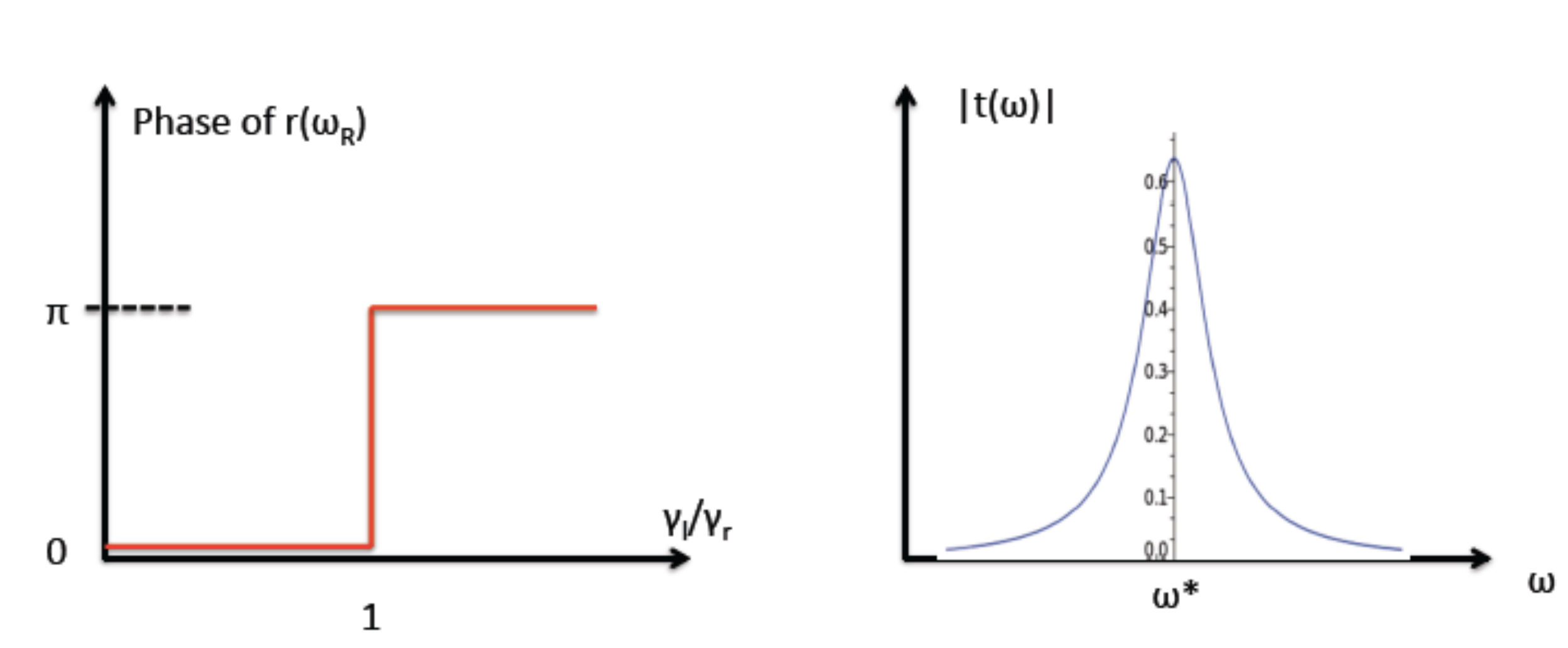}  
\caption{(Left) Typical Josephson-Kondo circuit introduced in Refs. \cite{Karyn,Moshe}. In the underdamped limit, where dissipation in the long Josephson junction chains is not so important $(0\ll \alpha \ll 0.3)$, one can observe the perfect Rayleigh transmission of a microwave photon through the system at the renormalized Kondo frequency $\omega_R$ given in Eq. (12). This many-body Lamb shift exemplifies renormalization effects. Such long transmission lines with tunable resistances can be built with current technology \cite{fluxonium}. By adjusting the ratio between the left $\gamma_l$ and right $\gamma_r$ couplings between the Josephson junction chains (transmission lines) and the artificial double-dot Cooper-pair box system \cite{Pashkin,Lafarge}, the phase of the reflected signal at the Kondo frequency passes from $0$ to $\pi$ when changing the ratio $\gamma_l/\gamma_r$. (Right) Cartoon (note the cavity in yellow is much bigger than the mesoscopic (nano-)circuit, as in Fig. 1): Hybrid system comprising a quantum dot tunnel coupled to two (red) electron reservoir leads and capacitively coupled to a one-mode (yellow) cavity via a capacitive coupling, as realized in Ref. \cite{hybrid1}. In Ref. \cite{MarcoKaryn}, we have shown that at the renormalized frequency $\omega^*$ (defined in the text), in the non-linear limit of large bias voltages applied across the quantum dot, the light is also perfectly transmitted for symmetric couplings $\gamma_l=\gamma_r$.}
\label{smallcircuits}
\end{figure}

To evaluate the elastic transport of a photon through the system, we can still use the input-output theory \cite{inputoutputRMP} and relate the reflection and transmission amplitudes of a photon \cite{Karyn}. When the averaged input power tends to zero, in the underdamped limit, we find that the scattering matrix is unitary and at the Kondo frequency the transmission is maximum as a result of $J(\omega_R)\Im  \chi_{spin}^R(\omega_R)=-1$ (using the conventions of Eq. (6)). The phase of the reflected signal also changes from zero to $\pi$ by varying the asymmetry between the left and right couplings with the transmission lines; see Fig. 3.

Note that inelastic Raman effects nevertheless become important when increasing the drive amplitude or by reaching the ultra-strong light-matter coupling (over damped regime), as described in Ref. \cite{Moshe}. Open questions concern, for example, the evolution of the Mollow triplet \cite{Astafiev} by increasing the drive amplitude and the dissipation strength $\alpha$. Simple arguments allow to conclude the disappearance of the elastic Kondo peak when increasing the amplitude of the drive \cite{Karyn}. Similar geometries could be considered to simulate the two-channel Kondo effect with light, as in electron systems \cite{NozieresBlandin,DavidPotok,Gleb,Keller,Frederic}, with finite-frequency measurements \cite{2CK1,2CK2}.

\subsection{Hybrid systems}

Coupling qubits (artificial spins) to a circuit quantum electrodynamics environment allow to explore new limits in quantum computation, where entangling distant qubits through light \cite{Senellart} has become possible for example \cite{Majer,Kontos,Deng}. One can also envision the realization of new quantum gates \cite{DiCarlo} and very robust Einstein-Poldolsky-Rosen or Bell's pairs \cite{Michel} by analogy to photon systems \cite{Aspect} and Cooper pairs \cite{CooperBCS}. Coupling a real spin-1/2 object to light has also been realized experimentally \cite{PascaleLoic,Takis2}. Extending the concepts and techniques of electrons to quantum optics has also resulted in new developments in quantum opto-electronics \cite{LPA}. 

Recently, experimentalists have accomplished another step further by designing hybrid systems, containing electrons with a circuit quantum electrodynamics environment. For example, coupling a quantum dot to an electromagnetic resonator has been achieved experimentally \cite{hybrid1,hybrid2,hybrid3,hybrid4}. Similar systems have been built with Nitrogen Vacancy centers \cite{SPEC}. More precisely, the coupling between a quantum dot, a microwave resonator and two biased electronic reservoirs was built on chip and signatures of many-body correlations  have been observed, for example, in the phase of the microwave transmitted photon as a function of the bias voltage $V$ applied across the leads. The system can be modeled by an Anderson-Holstein type Hamiltonian \cite{Anderson,Holstein}:
\begin{equation}
H_{matter} = \sum_{k\sigma\alpha} \epsilon_{k\alpha} c^{\dagger}_{k\alpha\sigma} c_{k\alpha\sigma} + \sum_{k\alpha\sigma} V_{k\alpha}(c^{\dagger}_{k\alpha\sigma} d_{\sigma} +h.c.) + \epsilon \sum_{\sigma} d^{\dagger}_{\sigma} d_{\sigma} + U n_{\uparrow}n_{\downarrow} + \lambda (a+a^{\dagger}) \sum_{\sigma} d^{\dagger}_{\sigma} d_{\sigma} + \omega_0 a^{\dagger} a.
\end{equation} 
Here, $c_{k\alpha\sigma}$ and $d_{\sigma}$ are the annihilation operators of the leads and of the dot respectively with spin $\sigma=\uparrow,\downarrow$. In addition, $\epsilon_{k\alpha}=\epsilon_k \pm eV/2$ and $n_{\sigma} = d^{\dagger}_{\sigma} d_{\sigma}$. The cavity is also coupled to two long transmission lines, similar to Eq. (4), that serve to send and receive the micro-wave signals. A similar Hamiltonian has been addressed to study transport through molecules with particular modes of vibration \cite{Mitra,Grempel,Vinkler}.
 
One natural question concerns the phase of the transmitted photon signal as a function of the bias voltage $V$ across the two electronic reservoirs \cite{MarcoKaryn}. Let us consider the quantum limit in the cavity with a few photon number. The phase of the transmitted (photon) signal can be computed through the relation $\tan \varphi(\omega) = -\Re \chi_{xx}^R(\omega)/\Im \chi_{xx}^R(\omega)$ in Eq. (6). By including the cavity-dot coupling, the (light) susceptibility becomes modified as $\chi_{xx}^R(\omega) = \omega_0/(\omega^2 - \omega_0^2 + i\omega_0\kappa -\omega_0 \Pi^R(\omega))$. We have treated the light-matter coupling to second order in perturbation theory. Interestingly, in the limit of large bias voltage applied across the quantum dot, the damping of the light field is only weakly affected by the (fast) motion of the electrons $(\Im \Pi^R(\omega_0)\ll \Re\Pi^R(\omega_0))$, then resulting in a phase $\varphi(\omega_0)$ which tends to a universal value $\pi/2$ in the limit of very small dissipation from the cavity environment $(\kappa\rightarrow 0)$ \cite{MarcoKaryn}. Note that this phenomenon occurs both when the quantum dot is at resonance and in the Kondo regime. Note that this phenomenon occurs both when the quantum dot is at resonance and in the Kondo regime. It is also relevant to note that at the renormalized frequency $\omega^*$ defined through the equation $\omega^2 - \omega_0^2 -\omega_0 \Re \Pi^R(\omega)=0$, the transmission coefficient exhibits a pronounced peak  (see Fig. 2). At large bias voltages $\omega^*$ converges towards $\omega_0$ and the transmission peak approaches unity for symmetric couplings with the left and right transmission lines (in Eq. (6)). In addition, the phase of the reflected signal is $\pi$ by analogy to Fig. 3 (bottom left) and has been recently experimentally measured in double-quantum-dot geometries in graphene settings at low temperatures \cite{Deng2}. It is also relevant to mention other recent efforts from the theoretical point of view to quantitatively describe transport in these systems \cite{Audrey}, in particular at low-frequency in relation with the Korringa-Shiba relation \cite{Olesia}. The strong-coupling limit between light and matter has been also achieved experimentally \cite{Hennessy}.

The effect of the electronic non-equilibrium environment on the cavity photon is to induce non linearity, damping/friction term and noise term. These terms in general are independent due to current flow and non-equilibrium effects and can be fully tuned by changing bias and gate voltages. In the limit of large bias, we can formally integrate out all the fermionic modes to obtain an explicit expression for the cavity photon effective action \cite{MarcoKaryn}. From this we can see how these hybrid systems can also serve to simulate novel regimes of the Langevin equation and some chaotic properties of the Duffing oscillator \cite{MarcoKaryn}, which has been realized experimentally \cite{Kozinsky}. Other recent theoretical developments can be found in Refs. \cite{Audrey,Samuelsson,Jordan,Nanoengine,H1,H2}. Similar ideas to build dissipative arrays have been addressed with hybrid system technology \cite{Berkeleypolaron}. It is also relevant that coupling topological systems such as (one-dimensional) topological superconductors \cite{Alicea,Kitaev,Read} to circuit QED has also attracted some attention, both for simulation and measurement point of view \cite{Schmidt,Nunnenkamp,Cottet,Trif,Simon,GinossarGrosfeld}. AC Josephson effects in these systems have also attracted some attention \cite{ManuelJulia}. Some theoretical efforts have also been done to engineer topological Kondo ground states with Majorana quantum boxes \cite{Beri,Altland,Erik} and multichannel Kondo models \cite{Tsvelikmajo}. Ideas to implement fermionized photons have also been considered \cite{CarusottoF}. Engineering Majorana fermions with photons has also engendered some attention in the community \cite{Atac}. 

\section{Large Circuits}

\subsection{Circuit QED array: Interaction Effects, Effective Chemical Potential and Driving}

In the last few years, some experimental progress were also accomplished towards realizing larger arrays in cQED \cite{Houckreview,experimentHouck}. Experimental efforts have been realized in various geometries \cite{RochLPA}. Two cavities can be coupled through a capacitive or Josephson coupling \cite{Raftery,Benjamin}. This point has also been carefully addressed in Ref. \cite{SchmidtKoch}. In particular, a cavity-dimer was realized experimentally \cite{Raftery} allowing to investigate novel non-dynamical steady states and entanglement features \cite{Aron}. 

 \begin{figure}[t]
\center
\includegraphics[scale=0.2]{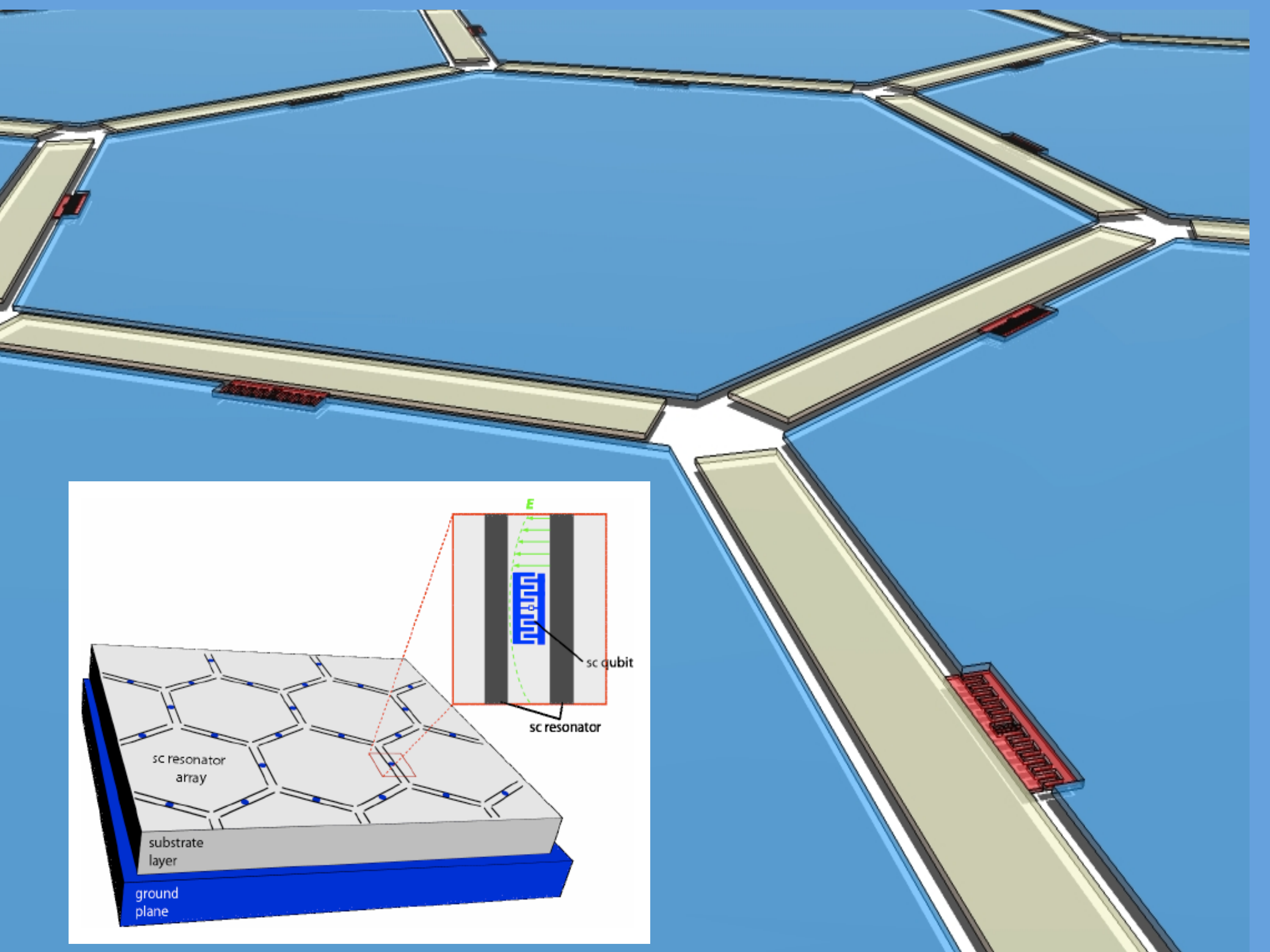}  
\includegraphics[scale=0.25]{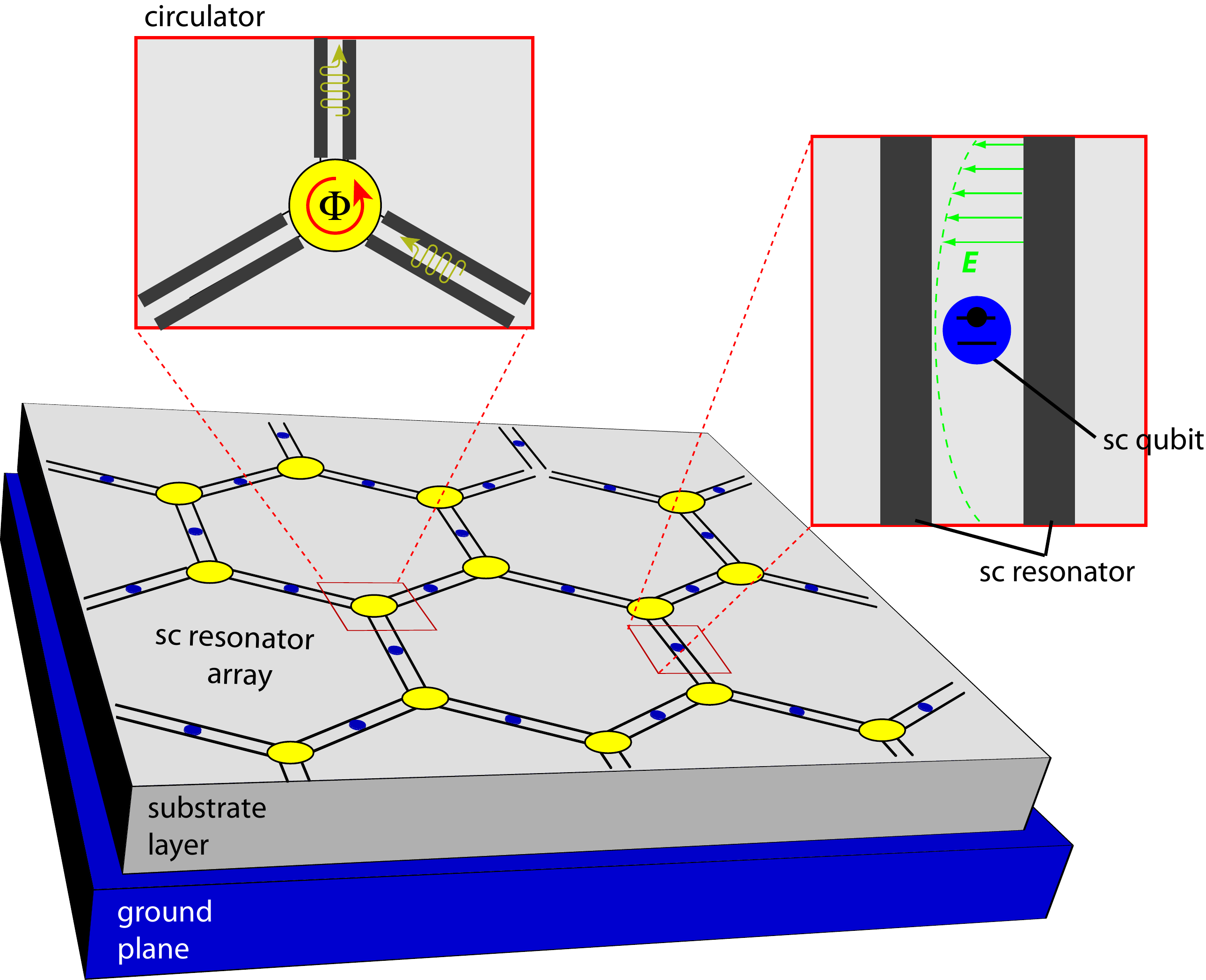}   
 \includegraphics[scale=0.35]{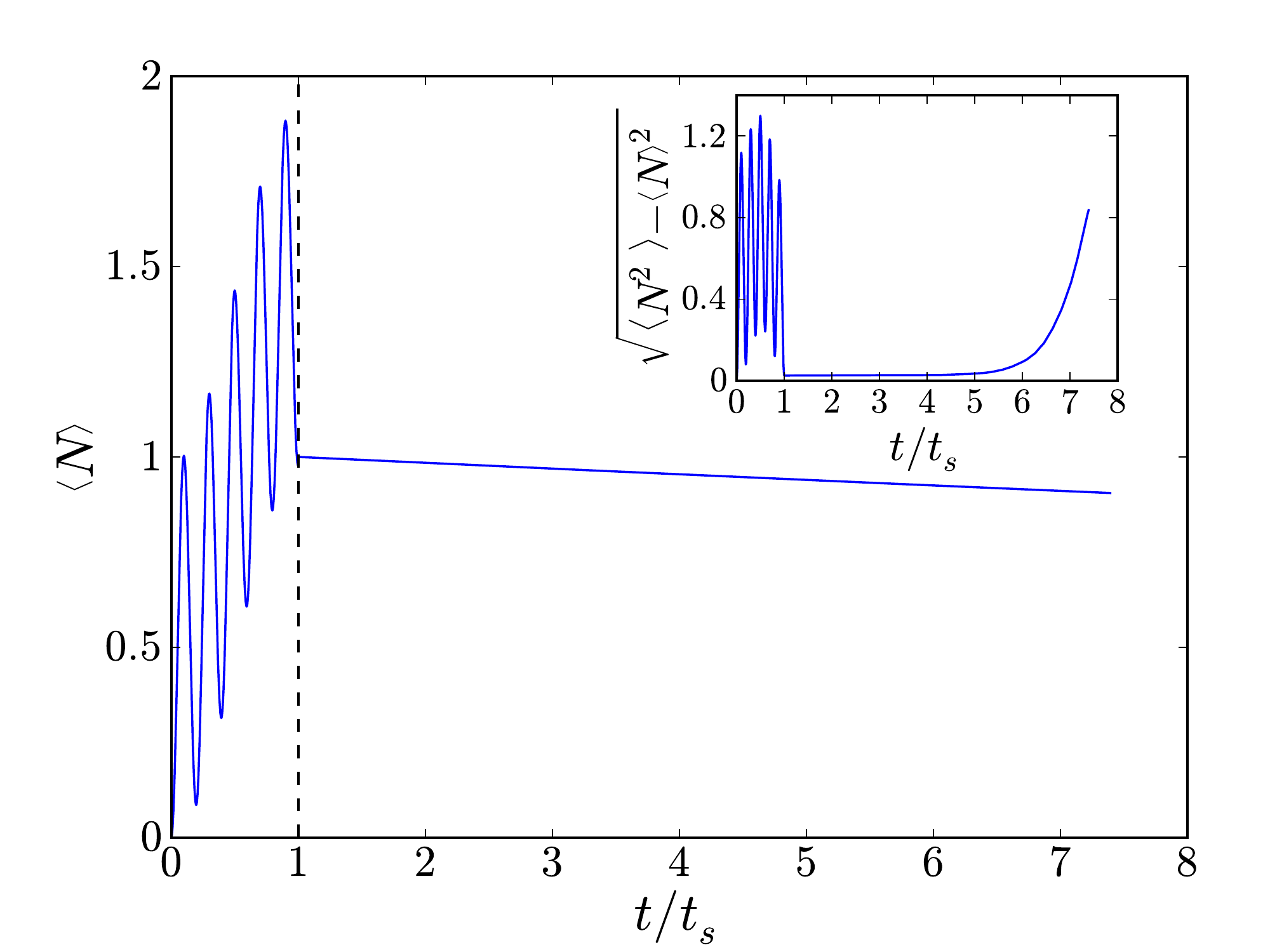}  
 \vskip -0.3cm
\caption{(Color online) (Left) Example of Superconducting Quantum Electrodynamics Networks, from Refs. \cite{Houckreview,KochHur}. Some experimental efforts are currently realized to engineer these artificial photonic arrays operating in the microwave domain \cite{experimentHouck}. Some other photonic (polaritonic) lattices have been built \cite{Jacqmin,Bellec0}. (Right) Synthetic gauge fields on photon systems  can be generated through nano-circulators or Josephson junction rings at each Y-junction, on the honeycomb lattice \cite{KochT}. Efforts are currently developed to realize such nano-circulators on-chip \cite{Kamal}. This honeycomb lattice with artificial gauge fields is also dual to a Kagome lattice with particular flux configurations \cite{KochT,AlexKaryn}, as described below. Below, we also show how these gauge fields can be realized through the Floquet theory. (Bottom) We apply the methodology described in Ref. \cite{loic} and include dissipation effects in each cavity $(\omega_0\rightarrow \omega_0-i\kappa/2)$. The protocol is as follows: we drive the cavities to reach a polarition state $|1-\rangle$ in each cavity at a (sufficiently short) time scale $t_s$. Then, we increase the coupling between cavities and observe an intermediate state in time, where the standard deviation  of the polariton number $N = a^{\dagger} a + \frac{1}{2}(\sigma_z + 1)$ is negligible (similar to a Mott phase). At long times, we identify a dynamical `transition' towards a delocalized state. The parameters are: $g/\omega_0=0.02$, $\Delta/\omega_0=0.9$, $V_0/\omega_0=0.1$ ($V_0$ is the amplitude of the AC signal on each cavity), $\kappa/\omega_0=10^{-6}$, $J=v_J(t-t_s)$ (for $t\geq t_s$) with $v_J=10^{-2}$ (in units of $\omega_0 t_s^{-1}$) and $\omega_0 t_s=311$. A similar procedure has been considered to compute the $g_2$ correlation function in time \cite{Tomadin}. }
\label{largecircuits}
\end{figure}

Assuming that the coupling between cavities (usually in the MHz range) is much smaller than the internal frequencies of the cavities, the coupling between cavities turns into an effective hopping term of bosons (photons) from one cavity to the neighboring ones.  For a large quality factor, we may restrict our treatment to a single photon mode, similar to the Jaynes-Cummings, Rabi and Dicke Hamiltonians. One model that has been investigated a lot theoretically in the literature
is the Jaynes-Cummings lattice model, which takes the form \cite{Houckreview}:
\begin{equation}
H = \sum_i H_i^{JC} - J\sum_{\langle i;j\rangle} \left(a^{\dagger}_i a_j +  a^{\dagger}_j a_i\right)-\mu_{eff}\sum_i \left(a^{\dagger}_i a_i + \sigma_i^+ \sigma_i^-\right),
\end{equation}
where $H_i^{JC}$ describes the Jaynes-Cummings Hamiltonian in each cavity, introduced earlier, and the bracket notation $\langle i;j\rangle$ denotes summation over nearest neighbors (see Fig. 4).  One can realize experimentally complex hopping tunneling elements $J$ at each junction, by inserting a Josephson junction ring, which allows to break-time reversal symmetry and to realize topological phases \cite{KochT,Andreas,AlexKaryn,Kamal}, as described below. Similar quantum nano-circulators have also been designed in cQED architectures \cite{Lehnert2}. This model has attracted some attention, for example, in the light of realizing analogues of Mott insulators with polaritons \cite{Greentree,KochHur,SchmidtBlatter,SchmidtBlatter1}. More precisely, one observes two competing effects in these arrays. For large values of $J$, formally the system would tend to form a wave delocalized over the full lattice, by analogy to a polariton superfluid \cite{CiutiCarusotto}, whereas for very small values of $J$, the photon blockade in each cavity addressed earlier in Sec. 2.2 will play some role.  Theoretical works \cite{Greentree,SchmidtBlatter,SchmidtBlatter1}, including ours \cite{KochHur,Karyn00}, have focused their attentions on solving the phase diagram at equilibrium in the presence of a tunable chemical potential $\mu_{eff}$. For $\mu_{eff}=0$, the ground state at small $J$ would correspond to the vacuum in each cavity. By driving the system with a quite strong drive amplitude, one can eventually reach for intermediate times, a state with one polariton in average in each cavity \cite{loic}, by analogy to applying a $\pi$ pulse on the qubits for a large detuning between light and matter \cite{HofheinzMartinis}. By driving cavities individually, an intermediate state analogous to a Mott state, was also identified \cite{loic}. In Fig. 4, we apply the methodology of Ref. \cite{loic}, and apply for the following protocol: we prepare the cavity in a $|1-\rangle$ polariton state by driving each cavity, then we stop the drive and we switch on the coupling between cavities. First, we observe that the effect of dissipation on each cavity described by the modification $\omega_0\rightarrow \omega_0 - i\kappa/2$ (see Eq. (7)) produces some small leakage on the mean polariton number per cavity $\langle N\rangle= \langle a^{\dagger} a + \frac{1}{2}(\sigma_z + 1)\rangle$, compared to Ref. \cite{loic}. At longer times, the coupling between cavities increases substantially the fluctuations, producing dynamically an extended state for the photons (polaritons). The coupling between cavities here is treated at a mean-field level. A possible extension could use the dynamical mean-field theory \cite{DMFT}.

By changing $\mu_{eff}$ and keeping $J$ small, one could eventually turn the vacuum in each cavity into a polariton state in the branch $|1-\rangle$ (by analogy to Fig. 1). Simple energetic arguments predict that this change would occur when $E_{1-} - \mu_{eff} = E_0$ (we use the notations of Sec. 2.2). This result can be made more formal by using a mean-field theory and a strong-coupling expansion \cite{Greentree,KochHur,SchmidtBlatter,SchmidtBlatter1}. In the atomic limit where  $J$ is small, one then predicts the analogue of Mott-insulating incompressible phases, as observed in ultra-cold atoms \cite{Greiner}, where it costs a finite energy to change the polariton number \cite{Greentree,Hartmann}. This point has been summarized in various recent reviews \cite{Houckreview,Fazio}. By increasing the hopping $J$, one can build an equivalent of the $\psi^4$  theory, where $\psi\sim \langle a_i\rangle$, in order to describe the second-order quantum phase transition between the Mott region of polaritons and the superfluid limit \cite{KochHur}. These results, which make a connection with the Bose-Hubbard model \cite{Fisher,GiamarchiSchulz}, have been confirmed by numerical methods such as Quantum Monte Carlo \cite{Pollet}. The effect of disorder has also been addressed using the Density Matrix Renormalization Group, leading potentially to glassy phases \cite{Fazio2}. Disorder can also result in Coddington-Chalker type models for light \cite{pasek}. Weak-localization effects have been observed in superconducting circuits \cite{MartinisChen}.These spin-boson networks can also serve to emulate various quantum spin models \cite{KochHur,Angelakis,Camille}.  It is also relevant to stress the appearance of novel phases in the limit of Rabi lattices, beyond the weak Jaynes-Cummings coupling limit; the presence of the counter-rotating wave terms somehow acts as an effective chemical potential \cite{marco}. The limit of strong light-matter coupling has been achieved with two stripline resonators \cite{Baust}. Dissipation effects in the two resonator problem have been addressed in Ref. \cite{Hanggi}.

Similar to ultra-cold atoms \cite{coldatomreview,coldatom2}, it seems important to be able to tune the effective chemical potential of photons to develop further quantum simulation proposals and observe these analogues of Mott phases in the long time limit. Recently, an idea to simulate an effective chemical potential for photons was pushed forward in Ref. \cite{Hafezi}, based on a parametric coupling with a bath of the form $\lambda\cos(\omega_p t)H_{SB}$ where $H_{SB}=\sum_j(a_j+a^{\dagger}_j)B_j$ where $B_j$ is a bath operator. Superfluid-Mott transitions using this scheme were also addressed in finite lattices. The Bose-Hubbard model was considered which can be achieved for highly detuned cavities \cite{Blais}.  At a more general level, considerable attention has been turned towards describing driven cavity arrays in order to realize novel steady states. In this context, it is certainly important to develop analytical and numerical tools \cite{Nissen,Leboite,Leboite2}, from stochastic non-equilibrium mean-field theories \cite{LoicKaryn,Keeling} to Matrix Product States for example \cite{Marconew,Joshi,Biella}. We have also done some progress to tackle the real-time dynamics of a spin dimer and the quantum Ising model in a transverse field coupled to a bosonic environment out of equilibrium \cite{LoicKaryn}. See also recent Refs. \cite{Smitha,CamilleChamon}. Features of the quantum Ising model has been recently observed in cQED \cite{Wallraffnew}. Note that the time-dependent numerical renormalization group is also powerful to describe the dissipative two spin system \cite{PeterDavid}.

Based on all these theoretical developments, we expect more experimental developments in cQED multi-cavity systems \cite{Houckreview,experimentHouck,Baust}.

\subsection{Introducing Topological Phases}

It is also important to emphasize the recent realizations of topological phases with photons in artificial systems together with developments in ultra-cold atoms. While in nature this is achieved by a magnetic field coupled to charged particles, artificial gauge fields have been realized with ultracold atoms in optical lattices \cite{Gerbier,Spielman} and photonic systems \cite{Soljacic}. We may think of a topological phase of noninteracting particles as characterized by a bulk topological invariant which necessarily implies the existence of edge states protected against backscattering at the system boundary. The lure of optical lattices and photonic systems is that band topology, which underlies the characterization of quantum Hall-like phases, as well as edge state transport, can be probed. Moreover, a strong magnetic field at a suitably chosen filling in the presence of interactions leads to the fractional quantum Hall effect. This phenomenon, specific to two-dimensional electron gases,  underlies multiple theoretical proposals for ultracold atoms \cite{A,B,C,D,E,F,G} or photons \cite{Sougato,Greentree2,CarusottoHall,Lukin}, but no realization to date.

Defining (time-dependent) topological invariants in relation with observables is certainly timely for Floquet theories \cite{Lyon,netanel}. Measuring topological invariants in quantum fluids can also be done exploiting the Karplus-Luttinger anomalous velocity \cite{Karplus,Cooper,AlexKaryn}. Another important progress concerns the first realization of a photonic topological Floquet insulator \cite{Rechtsman}. Rechtsman et al. translated the modulation from the time domain to the spatial domain, leading to the experimental demonstration of the photonic analogue of the quantum Hall effect at optical frequencies (633nm). Similar developments have been done in micro-cavities where a spin-orbit coupling for polaritons has been engineered \cite{Alberto}. Concerning quantum circuits, a topological insulator has been realized based on meta-materials, composed of capacitively coupled high-Q inductors \cite{simon2}, and other suggestions have been addressed \cite{Macdonald}. Other quantum circuit architectures have been proposed \cite{Liang}. The effect of dissipation in driven lattices has also been considered theoretically, in particular, on honeycomb lattice systems \cite{Carusottodissi}. Topological edge states have also been observed on photonic honeycomb lattices \cite{Plotnik, Segev,LPN}. Novel ways of pumping a non-equilibrium photonic system were also discussed in Refs. \cite{Kapit,Lebreuilly}. In fact, these developments are not restricted to two dimensions in these quantum fluids by analogy to the topological properties of Bismuth materials \cite{Kane,Bernevig,Zhang}.

First, we will review a generic model, the Haldane model, on the honeycomb lattice and its extension to the Kagome lattice. Then, we will describe how to implement a Peierls phase for neutral particles based on the Floquet mechanism. We will continue with quasi-one-dimensional ladder-type geometries described by a Bose-Hubbard type model, related to experiments in ultra-cold atoms and Josephson junction arrays. These systems are also related to circuit-QED systems since the Jaynes-Cummings lattice model is described by the same $\psi^4$ theory as the Bose-Hubbard model \cite{KochHur,Karyn00}.

\subsection{Artificial Graphene, Kagome lattice and Haldane Model}

Some experimental works have also succeeded in realizing topological phases through the Haldane model and analogous ideas. Here, we review these developments starting from the artificial graphene on the honeycomb lattice \cite{Jacqmin,Bellec,Pellegrini}. The Haldane model takes the form (here, hopping elements are denoted $t_1$ and $t_2$ by analogy to condensed-matter systems) \cite{Haldane1988}:
\begin{equation}
H = -t_1\sum_{\langle i;j\rangle} c_i^{\dagger} c_j - t_2\sum_{\langle\langle i;j\rangle\rangle} e^{i\phi_{ij}}c^{\dagger}_i c_j +h.c. + \sum_i (-1)^i M c^{\dagger}_i c_i. 
\end{equation}
We fix the effective chemical potential of photons to be zero here. The presence of the term $t_1 $ will produce the Dirac particles on the graphene lattice and the term $t_2$ which couples the next nearest neighbor sites corresponds to the Haldane term. We also add the Semenoff mass $M$ which has been introduced in high-energy physics \cite{Semenoff}; for cQED arrays, it corresponds to an adjustment of frequencies in each cavity and in ultra-cold atoms it can be engineered through magnetic field gradients, for example. In the continuum limit, we will make an analogy with the equation $(i\gamma^{\mu}\partial_{\mu}-M)\psi=0$ where $\psi$ will be a two-component spinor and the Clifford algebra will be generated from the Pauli matrices.The configuration of fluxes for the Haldane model is shown in Fig. 5 and $\phi_{ij}=\phi$. The key point here is that the $t_2$ term opens a gap at the Dirac points by breaking the time-reversal symmetry: this produces the emergence of chiral edge modes at a single-particle level. In fact, the system exhibits a non-trivial topology as long as $M<3\sqrt{3}t_2\sin\phi$ \cite{Haldane1988}. 

 \begin{figure}[t]
\center
\includegraphics[scale=0.55]{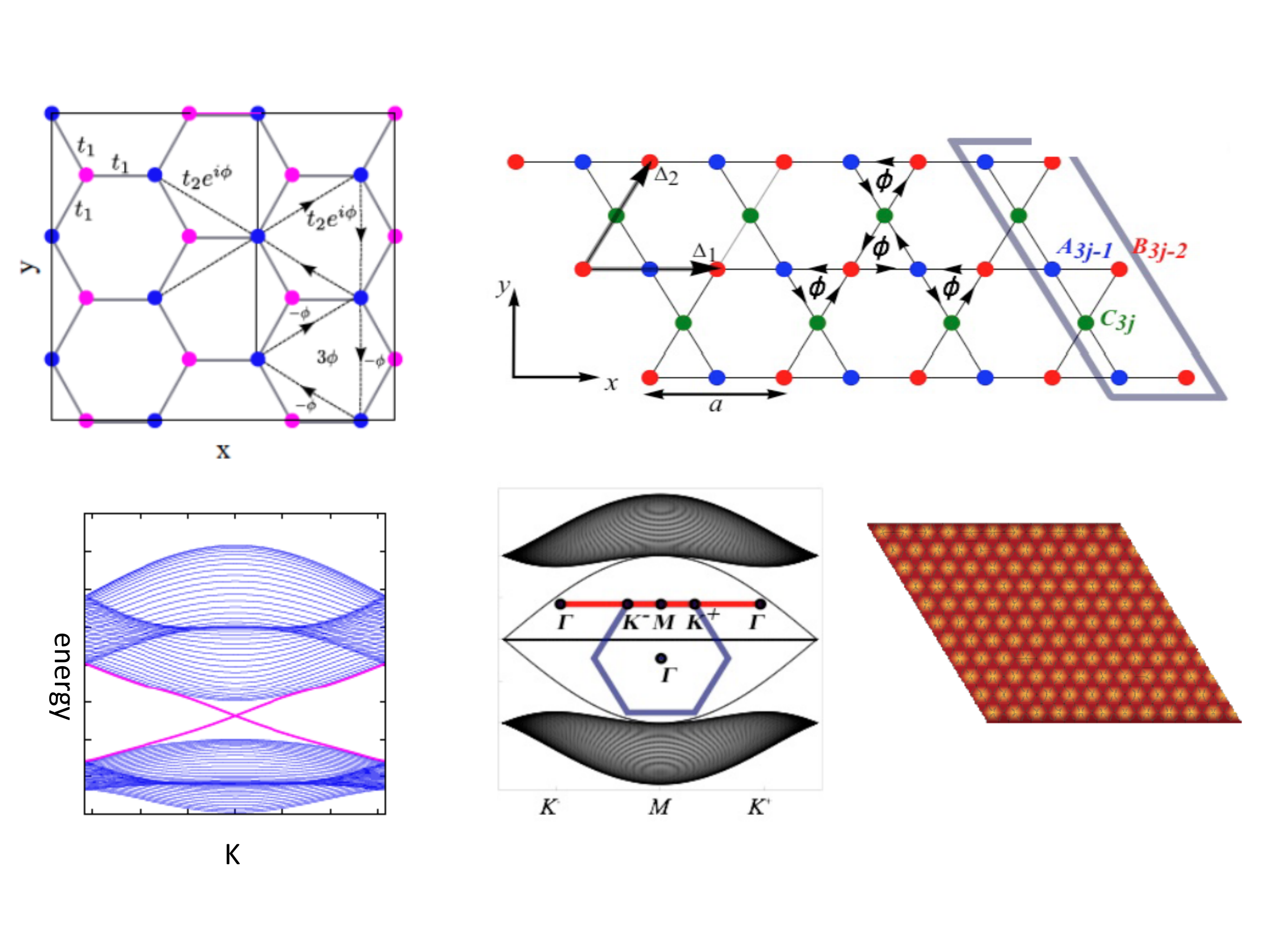}  
\vskip -1cm
\caption{Left: Representation of the Haldane model on the honeycomb lattice \cite{Haldane1988} (top) and spectrum of the system in the topological phase (bottom). Right: Kagome lattice with artificial fluxes \cite{AlexKaryn}, dual to the honeycomb lattice with Josephson junction rings introduced in Ref. \cite{KochT}. This can be seen as a realization of the Haldane model on the Kagome lattice with no net flux on a parallelogram unit cell. Bottom, middle: spectrum of the Kagome lattice for one particular flux configuration $\phi=\pi/6$ (with the notations of Ref. \cite{AlexKaryn}), showing a flat band in the middle of the energy spectrum in addition to chiral edge modes. Right: the local density of states then also shows states trapped on hexagonal cells \cite{AlexKaryn}, as observed in the experimental setup of Ref. \cite{Jacqmin}. }
\end{figure}

The precise analogy with the Dirac equation in two spatial dimensions can be seen by introducing the spinor $\psi_i=(a_i,b_i)$, where $a$ and $b$ correspond to the two sub-lattices \cite{Annica}. When $t_2=0$ two inequivalent Dirac points $K$ and $K'$ naturally emerge in the band structure \cite{Wallace} and are protected by the PT symmetry (Time-reversal and inversion symmetry); for an introduction, see Refs. \cite{Annica,Cayssol}. The dispersion relation close to the Dirac points in that case obeys $\pm \sqrt{(v_x k_x)^2 + (v_y k_y)^2}$, where $v_i$ are the group velocities (below, $v_i=v)$, and one can check that $(PT)H({\bf k})(PT)^{-1}=H({\bf k})^*$. The implementation of Dirac particles has also been discussed in the cQED context \cite{Solano2}, in relation with existing experiments. Two linear dispersion bands in the three-dimensional momentum space can also intersect at a single degenerate point -- the Weyl point. Recently, these three-dimensional Weyl fermions have also been identified in photonic crystals \cite{Weyl}. 

The total Hamiltonian can also be written as an effective spin-1/2 particle subject to a magnetic field in the wave-vector space: $H=\int_{BZ} d{\bf k} \psi^{\dagger}({\bf k}) H_H({\bf k}) \psi({\bf k})$ where $H_H = - {\bf h}({\bf k})\cdot \hat{\mathbf{\sigma}}$, and in the basis  of the Pauli matrices $(\sigma_x,\sigma_y,\sigma_z)$ the effective magnetic field takes the form:
\begin{equation}
{\bf h}({\bf k}) = (t_1\sum_i \cos({\bf k}.{\bf a}_i), t_1\sum_i \sin({\bf k}.{\bf a}_i), +M-2t_2\sum_i \sin({\bf k}.{\bf b}_i)).
\end{equation}
Here, ${\bf a}_i$ and ${\bf b}_i$ are the vectors connecting the nearest-neighbors and the next-nearest neighbors respectively and we assume that the phase $\phi=\pi/2$ such that the $t_2$ term in Eq. (15) is purely imaginary. The presence of the $t_2$ opens a gap and produces a quantized Chern number similar to the quantum Hall case \cite{Haldane1988}. In the presence of $P$ ($T$ broken), the Berry curvature satisfies ${\cal F}({\bf k})={\cal F}(-{\bf k})$ and the total Berry flux adds up to $2\pi$ and the Chern number equals one. By analogy with a spin-1/2 particle, the Chern number of the lowest band can also be written as: $C_- = (1/4\pi)\int_{BZ} d{\bf k} \tilde{\bf h}\cdot (\partial_x \tilde{\bf h}\times \partial_y \tilde{\bf h})$, where $\tilde{\bf h}$ is the normalized ${\bf h}$ vector (here, ${\bf k}$ lives on the Brillouin zone torus whereas $\tilde{\bf h}$ lives on the Bloch sphere). A pedagogical derivation is done in Ref. \cite{Cayssol}. An important point is that the $t_2$ contribution is odd under changing ${\bf k}$ in -${\bf k}$. The bulk-edge correspondence ensures the presence of a chiral edge mode in this case similar to the quantum Hall situation \cite{Bert,Thouless} (see Fig. 5). A similar quantum Hall state without Landau levels can be generated through (long-range) interaction effects \cite{RaghuSC,Tianhan}.

 \begin{figure}[t]
\center
\includegraphics[scale=0.6]{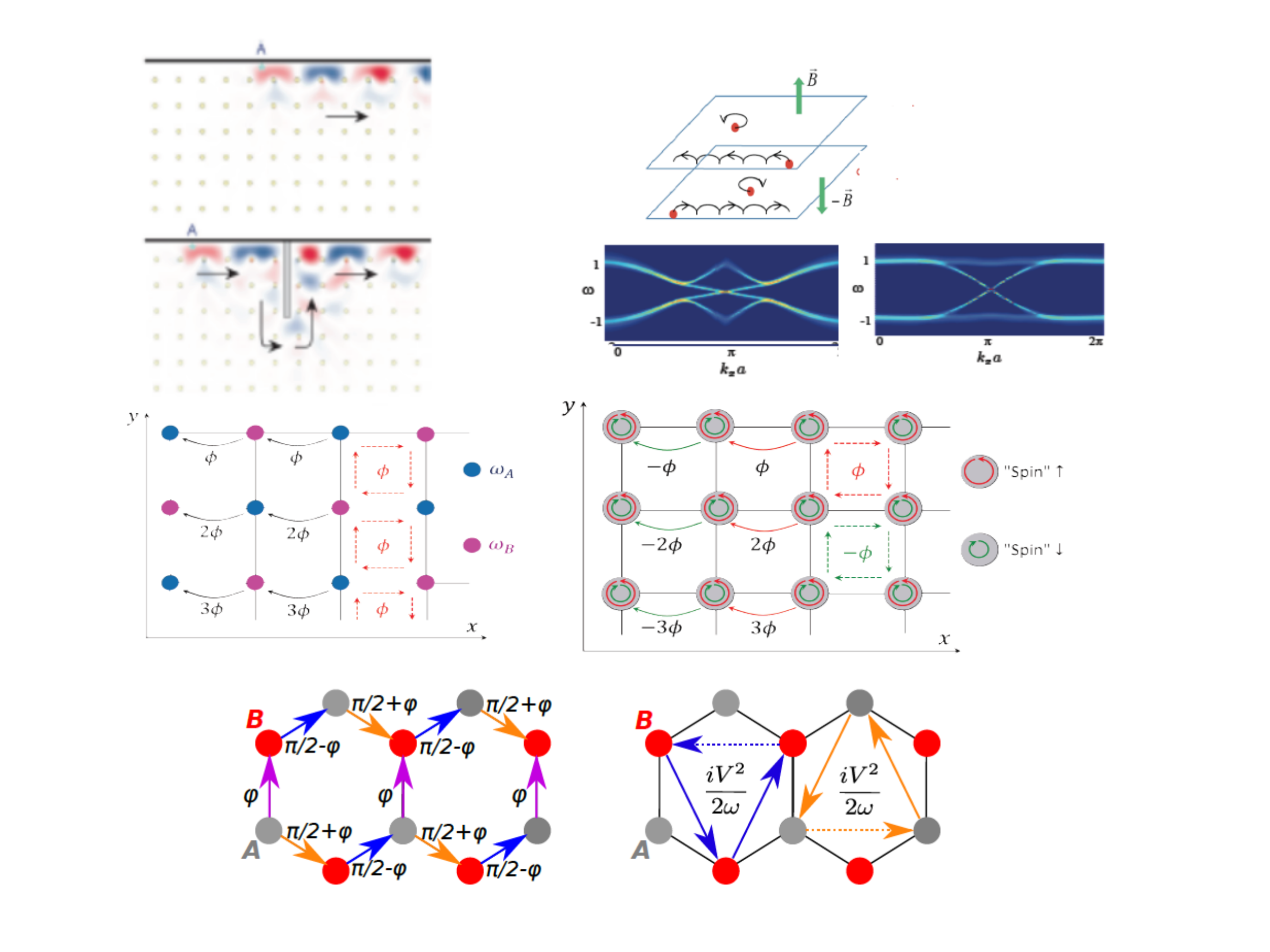}  
\vskip -1cm
\caption{Top, Left: first experimental observation of chiral edge modes and quantum Hall phases of photons in gyromagnetic photonic crystals \cite{MIT,Soljacic}. Middle, Left: possible Floquet scheme suggested in Ref. \cite{Stanfordlight} to realize Peierls phases and quantum Hall phases for photon systems without the Faraday effect. Middle Right: Generalization to a quantum Spin Hall state, as observed experimentally in silicon photonics at University of Maryland \cite{Hafezi1,Hafezi2}. Top, Right: Spectral function from Ref. \cite{Wei} computed with CDMFT \cite{DMFT,RMP} exemplifying that the helical edge modes of the Kane-Mele model are robust towards interaction effects. Bottom: Possible phase configuration using the Floquet scheme of Eq. (17), producing an (anisotropic) Haldane state with anisotropic couplings (see Eq. (19); here $\omega=\Omega$).}
\label{}
\end{figure}

In fact, Haldane and Raghu have extended this scheme to the field of gyromagnetic photonic crystals with ferrite rods and the magneto-optical Faraday effect then mimics the $t_2$ term of the Haldane Hamiltonian \cite{HaldaneRaghu}. This theoretical idea has been first realized at MIT in 2009, where a photonic band gap has been observed as well as robust chiral edge modes for light in the microwave regime \cite{MIT}.  Note that the experiment has been realized with ferrite rods forming a square lattice instead of the honeycomb lattice (see Fig. 6). Photon systems are particularly appropriate to produce chiral edge modes, where selecting the right-frequency of the incoming wave and observing the local density of states is relatively straightforward \cite{Soljacic}.  The Haldane Hamiltonian has also been achieved in 2014 at ETH Zuerich \cite{Esslinger}, with fermions in optical lattices, where the $t_2$ term has been simulated using the Floquet point of view and the shaking protocol \cite{Dalibard,CayssolMoessner}.  Similar experiments have also been realized at Hamburg \cite{Sengstock}. Preceding experiments on ultra-cold Dirac fermions have managed to move the Dirac points by tuning the hopping amplitudes \cite{Tarruell,Montambaux}. The Chern number has also been measured in ultra-cold atoms \cite{Bloch1}. The Zak phase has been experimentally observed \cite{Monica,Delplace}. A similar quantum Hall state without a uniform magnetic flux was also observed in Bismuth materials \cite{QAH1}.  

It is also important to note that the Haldane model can be extended to more complicated lattices such as the Kagome lattice, with three sites (atoms) per unit cell. Following the protocol of Refs. \cite{KochT,Andreas,AlexKaryn}, summarized in Figs. 4 and 5, one can realize an analogous situation with chiral edge modes of light in cQED. One can map the honeycomb lattice with Josephson rings at each link onto a ``dual'' Kagome lattice with fluxes in the ``bow ties''. The Kagome model contains particles hopping between nearest-neighbor sites with a complex integral $|t|e^{\pm i \phi}$. Photons then acquire a phase $3\phi$ around a triangular plaquette and a phase $-6\phi$ on a honeycomb cell, amounting to a zero net flux in a parallelogram unit cell. By adjusting the flux in each Josephson junction ring, one can realize a quantum Hall effect without Landau levels \cite{AlexKaryn} (see Fig. 4). This model with three sites per unit cell also allows the occurrence of flat bands which corresponds to wave-functions localized on each hexagonal cell (see Fig. 5) \cite{AlexKaryn,Chamon}. The search of flat bands has a long history in condensed-matter physics \cite{Lieb,Mielke,Kagomedot}. Note that flat bands have been recently observed at LPN Marcoussis in artificial honeycomb systems \cite{Jacqmin}. Other efforts in polariton systems to engineer and observe flat bands have been done recently both theoretically \cite{Baboux} and experimentally \cite{Huber}. Similar lattices (with fluxes) have been widely investigated theoretically and some have been realized in quantum circuits and ultra-cold atoms \cite{Pannetier,HuseChaikin,VidalMosseriBenoit,Berkeley}.  The Kagome lattice is also a candidate for the exploration of quantum spin liquids (non-ordered N\' eel states) \cite{Fak}. Other correlated states have been predicted on the Kagome lattice \cite{Lecheminant,Lecheminant2,Balents,White,Schollwock,Messio,AntoineS,Cecile}.

\begin{figure}[t]
\center
\includegraphics[scale=0.7]{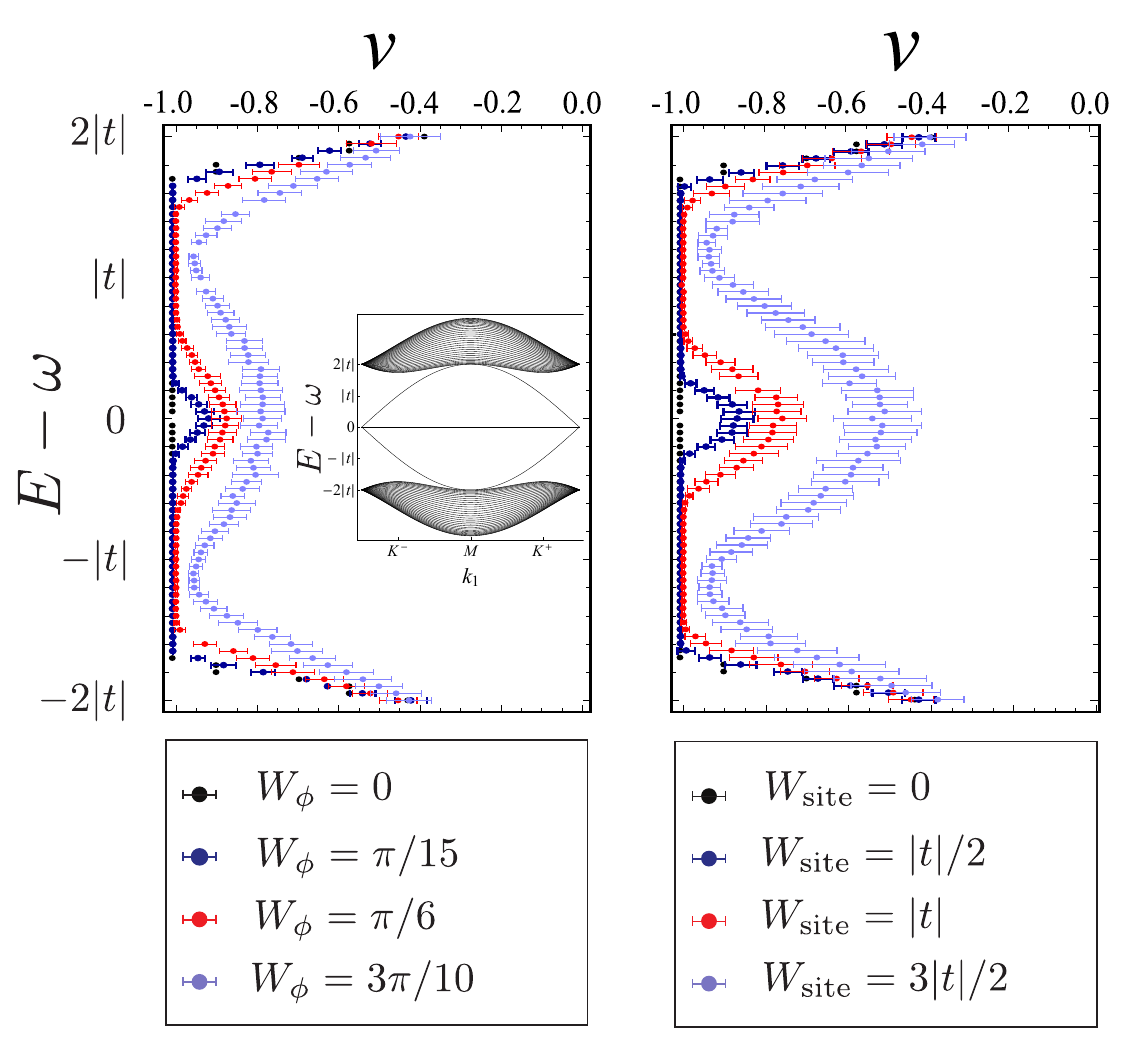}  
\caption{Disorder effects on the Kagome lattice. Calculation of the Chern number $\nu$ in real space at a given energy and at $\phi=\pi/6$ with increasing amounts of phase and on-site disorders. Details of the calculations follow the procedure shown in Ref. \cite{AlexKaryn}. The disorder is a white noise (uncorrelated from site to site) disorder. For a disorder amplitude $W$, a random number between $-W/2$ and $W/2$ is selected. For photonic networks, the left (right) panel depicts the effect of increasing phase (resonator frequency) disorder. Here, $\omega$ represents the cavity frequency. The inset shows the reference bandstructure for the clean translationally invariant system, where Chern numbers corresponding to bottom and top bands are -1 and 1, respectively, and the flat middle band has vanishing Chern number.}
\end{figure}

In the presence of disorder, there is however a clear difference between the honeycomb and Kagome lattice. In the case of the Kagome lattice, disorder will spread the middle flat band, making it possible for intragap edge states to scatter into the bulk. This is consistent with the fact that the Chern number at energies energies close to the intragap flat band energy loses its quantization as disorder strength increases \cite{AlexKaryn}. In Fig. 7, we illustrate such a difference by computing the Chern number $\nu$ at a given energy on the disordered Kagome lattice with translational symmetry breaking (using non-commutative rules and averaging over disorder configurations) \cite{AlexKaryn}. The common characteristic of the flat middle band and the lower band is that as soon as disorder is turned on $|\nu|$ starts decreasing from $1$ in the close vicinity of the band edge, as indicated in the plots of Fig. 7.  We also notice that disorder effects have already been controlled and adjusted in superconducting circuits \cite{Martinisdisorder}.

Very recently, another progress was achieved in cQED \cite{Martinis,Lehnert}, by observing similar topological effects and simulating directly the Haldane-type Hamiltonian $H(k_x,k_y)=v(k_x\sigma_x+k_y\sigma_y)+(M-m_t)\sigma_z$, where $m_t$ corresponds to the second-neighbor hopping (tunneling) in a local magnetic field. This Hamiltonian has been implemented with one or two superconducting qubits on the Bloch sphere: the BZ on a torus is then simply replaced by an integration over the polar and azimuthal angles on the Bloch sphere. A path from north to south pole on the Bloch sphere was achieved in time, through microwave pulses. Note that these measurements constitute a dynamical (direct) measurement of the Berry phase. These measurements use the general scheme proposed in Ref. \cite{Polkovnikov}. Other measurements of the Berry phase were also done in Ref. \cite{SchoelkopfWalraff}. Motion in a curved space will be deflected from a straight trajectory and the state of the system feels a force leading to deviations from the adiabatic path. The experiments have directly measured the Chern number $C = \int_0^{\pi} F_{\theta\phi}d\theta$ where the Berry curvature component $F_{\theta\phi}$ encodes the information of the force and is related to the spin observable $\langle \sigma_y\rangle$ (the experiments work at constant speed allowing a linear response). The two experiments report a quantized Chern number $1$ for $M<m_t$ and a topological phase transition for $M=m_t$ where the Chern number drops to zero.   

\subsection{Topological Quantum Fluids and Spin-Orbit Coupling}

The goal to achieve arbitrary control of photon flows has motivated an increasing recent research on photonic crystals, metamaterials and artificial quantum circuits. Here, we first review the theoretical proposal of Ref. \cite{Stanfordlight} and describe the emergence of protected one-way photon edge states that are robust against disorder, without the use of magneto-optical effects \cite{MIT,Soljacic} (see also Fig. 6, top, left). Using simple protocols, based on particular choices of spatial distribution and modulation phases, one can produce an effective magnetic field for photons, leading to a Lorentz force. The Hamiltonian takes the form (see Figure 6, middle left)
\begin{eqnarray}
H &=& \omega_A\sum_i a^{\dagger}_i a_i + \omega_B\sum_i b^{\dagger}_i b_i \\ \nonumber
&+& \sum_{\langle i;j\rangle} V\cos(\Omega t+\phi_{ij})(a^{\dagger}_i b_j +h.c).
\end{eqnarray}
In this scheme, two neighboring resonators have different frequencies. One can simulate various situations by adjusting the modulation frequency $\Omega$. Let us assume that we are close to resonance $\Omega=\omega_A-\omega_B$, then we can build a simple physical argument based on the rotating-wave approximation. Assuming that $V\ll \Omega$, the evolution in time of the photons is controlled by $\omega_A$ and $\omega_B$ mostly. It is then justified to define the operators $c_{i(j)}=e^{[i\omega_{A(B)}t c^{\dagger}_{i(j)} c_{i(j)}]}a_i(b_j)$. 

In the rotating frame, the Hamiltonian is equivalent to charged particles subject to a magnetic field \cite{Harper,Hofstadter}: 
\begin{equation}
H_{eff} = \sum_{\langle i;j\rangle} \frac{V}{2}(e^{-i\phi_{ij}}c^{\dagger}_i c_j + e^{i\phi_{ij}}c^{\dagger}_j c_i). 
\end{equation}
Using this resonant scheme, then one can produce an equivalent of Peierls phases on a given lattice, 
for neutral particles; here, the nearest-neighbor coupling is dominant. In particular, by adjusting experimentally the phases $\phi_{ij}$ (see Fig. 6, middle left) one can realize for example a uniform magnetic field related to the effective vector potential ${\bf A}_{eff}$ defined as $\int_i^j {\bf A}_{eff}\cdot d{\bf l} = \phi_{ij}$.  An idea to realize the dynamical coupling in photon (quantum circuit) systems has been suggested in Ref.  \cite{Stanfordlight}. A similar scheme has been used in ultra-cold atoms, based on resonant laser-assisted tunneling and tilted lattices \cite{BlochHofstadter,Wolfgang,JackschZoller}. In the context of ultra-cold atoms, tilts can be achieved in various ways, from magnetic to gravitational gradients. 

In the case of a uniform net flux, the effective Hamiltonian then exhibits  the same properties as a two-dimensional gas in the presence of a uniform magnetic field. If the flux is rational, one observes a Hofstadter butterfly and chiral edge modes, as achieved in ultra-cold atoms for example \cite{BlochHofstadter}. To show this, Fang et al. applied a simple Floquet argument \cite{Stanfordlight}. The original Hamiltonian is periodic in time $H(t)=H(t+\frac{2\pi}{\Omega})$ and the eigenstates then take the form $|\psi(t)\rangle = e^{-i\epsilon t}|\chi(t)\rangle$ where $\epsilon$ represents the ``quasi-energy'' which can be found using the equation $(i\partial_t - H(t))|\chi(t)\rangle = -\epsilon|\chi(t)\rangle$. The time-dependent Hamiltonian here takes the particular form: $H(t) = H_0 + H_1e^{i\Omega t} + H_{-1}e^{-i\Omega t}$, where $H_0=H_{0,a}+H_{0,b}=\sum_{i} \omega_i c^{\dagger}_i c_i$. The problem can be diagonalized using the decomposition $|\chi(t) \rangle = \sum_{n=-\infty}^{+\infty} |\chi_n\rangle e^{i n \Omega t}$ and using the $(a,b)$ basis, by writing $|\chi_n\rangle = |\chi_{n,a}\rangle +  |\chi_{n,b}\rangle$ \cite{Fang2}. Outside the resonant condition, one can apply the Floquet theory, a multi-phase expansion and a ${\cal O}(1/\Omega)$ expansion \cite{Dalibard,CayssolMoessner}. This allows to achieve various flux configurations and spin-orbit couplings \cite{Kitagawa,Lindner}. In particular, one can realize an effective Haldane-type model, as follows. We expand the effective Floquet Hamiltonian in powers of $1/\Omega$. To first order in $1/\Omega$, away from the resonance condition, one obtains
\begin{equation}
H = H_{graphene} + 
\frac{i V^2}{2 \Omega} 
\left(
\sum\limits_{\Braket{\Braket{i;k}}}
	\sin \left( \phi_{ij} + \phi_{jk} \right)
	{a}^\dag_i {a}_k  +
\sum\limits_{\Braket{\Braket{i;k}}}
	\sin \left( \phi_{kj} + \phi_{ji} \right) 
	{b}^\dag_i {b}_k + h.c.
\right).
\end{equation}
We assume that the Hamiltonian also contains a constant nearest-neighbor hopping term (which can be generated through a perturbation of the form $V\cos^2(\Omega t +\phi_{ij})$). Here, $j$ is the unique nearest neighbor of both sites $i$ and $k$. In Fig. 6 (bottom), we show one application of Floquet theory to realize an anisotropic Haldane model (in agreement with the relative signs generated in the terms for the a and b sub-lattices). The generated $t_2$ terms are indeed anisotropic: the $t_2$ term parallel to the horizontal axis is formally zero. We have numerically checked that the Chern number is still quantized and currents propagate at the edges despite the anisotropic choice of phases.

A generalization of the Haldane model on the honeycomb lattice to spin-1/2 particles has been made by Kane and Mele in 2005 \cite{KaneMele1} (see also Refs. \cite{Kane,Bernevig,Zhang}):
\begin{equation}
H = -\sum_{\langle i;j\rangle} t_1 c^{\dagger}_{i\alpha}c_{j\alpha} + \sum_{\langle\langle i;j\rangle\rangle \alpha\beta} i t_2 \nu_{ij} s^z_{\alpha\beta} c^{\dagger}_{i\alpha}c_{j\beta} +h.c.
\end{equation} 
The $i t_2$ term (here, purely imaginary) mimics an atomic spin-orbit coupling $s^z L^z$. More precisely, on a given pair $i$ and $j$ of next-nearest-neighbor sites within the same sub-lattice, we must satisfy $\nu_{ij}=-\nu_{ji}=\pm 1$.
Similar to the Haldane model, the spin-orbit $t_2$ term opens a gap in the system. The main difference is that now the system respects the T-symmetry and is characterized by a different topological invariant. A simple picture of the edge states
is the Kramers pair (Fig. 6, top right): a particle (electron) with spin-up  propagates to the left whereas a particle (electron) with spin-down propagates to the right. This leads to a spin current at the edges and to a $\mathbb{Z}_2$ type topological invariant \cite{KaneMele2,JoelMoore} and a spin Chern number \cite{Donna}. The $\mathbb{Z}_2$ invariant can also be expressed in terms of the four time-reversal invariant momenta (points) in the Brillouin zone \cite{FuKane}. Progress have been accomplished in ultra-cold atoms, with magnetic field gradients, to observe such a quantum spin Hall state \cite{BlochHofstadter}.  Another important progress from the experimental point of view comes from the discovery of Mercury Telluride materials which have allowed to observe a similar physics in two dimensions, leading to the discovery of two-dimensional topological insulators \cite{Molenkamp,ZhangHughesBernevig}. In Refs. \cite{Stephan,Wei}, together with S. Rachel, W. Wu and W.-M. Liu, we have shown that the helical edge modes in the Kane-Mele model are robust towards finite to moderate interactions. For free particles, the spectrum of the edge modes can be computed analytically \cite{TianhanSpiral}. For recent reviews on interaction effects in topological spin-orbit systems and iridate materials, consult for example, Refs. \cite{Assaad,BalentsKim}. In Fig. 6, we show the spectral functions in the presence of a Hubbard interaction obtained with a real-space extension \cite{Wei} of the Cluster Dynamical Mean-Field Theory (CDMFT) \cite{DMFT,RMP}. We have reached a similar conclusion using the Slave-Rotor approach \cite{Stephan,SergeAntoine}. It is also interesting to note the possibility of chiral bogoliubons in a BEC system \cite{Bardyn}.

Recently, a spinful version has been realized in photon systems by Hafezi et al. \cite{Hafezi1,Hafezi2} based on a Silicon architecture (see Fig. 6, middle): the two spin-polarizations correspond to the two possible circulations (clockwise or anti-clockwise) in a given resonator. The phase between resonators are carefully adjusted to follow the scheme of Fig. 6. The main difference with the usual quantum spin Hall effect is that here the two spin polarizations couple when T is not broken. To decouple the two spin polarizations, one requires to break the T symmetry; in this sense, one realizes two independent copies of a Harper-Hofstadter model; as a result, topological protection is only partial and limited to devices with no back-scattering. The measurement of a topological invariant in these systems has been suggested \cite{Carusottodissi,Hafezi3}. An optical spin-Hall effect in photonic graphene and other spin-orbit topological phenoma have also been suggested \cite{Nalitov,Karzig}.

\subsection{Josephson Effect and Chiral Bosonic Phases}

Here, we discuss interaction effects on topological phases. Then, we give a specific example related to bosons systems which can be realized in Josephson junctions, ultra-cold atoms or photon systems.

The effect of interactions on topological phases have been addressed in various contexts \cite{Joel,Gurarie,Levin}. The stability of topological insulators, for example, towards disorder has been thoroughly analyzed from a theoretical point of view \cite{Beenakker,Prodan,AlexKaryn}. For problems such as the quantum spin Hall state, one can apply various analytical arguments to answer these types of questions \cite{KaneMele1,Stephan}. The stability of topological insulators and superconductors is in general ensured perturbatively due to the gap protection, for fermion systems. Extending the notion of topological invariants to interacting fermion systems has also attracted some attention in the community in relation with numerical developments \cite{ZhangQi,Kim,Budich,Gurarie2}. 

For boson systems, the situation is different since the bosons can condense in a Bose-Einstein type manner and occupy macroscopically the bottom of the lowest Bloch band. Defining topology in these many-body boson systems has attracted attention recently. Here, we review two examples of non-trivial effects in chiral bosonic systems, in relation with current experiments \cite{Esslinger,Bloch2}. The breaking of time-reversal symmetry via the spontaneous formation of chiral order has been observed in Ref. \cite{Hemmerich}. The physics here is associated  with Meissner physics. A connection
to quantum Hall phases in ladder systems will be addressed.

The first effect concerns a situation related to Eqs. (17) and (18), which is obtained from the time-dependent Hamiltonian at the resonance condition. 
Here we follow the recent experiment realized at Muenich on a ladder-type geometry \cite{Bloch2}, and focus on the effective (time-independent) Hamiltonian:
\begin{eqnarray}
H &=& -t\sum_{\langle i;j\rangle \alpha} e^{i l A_{ij}^{\alpha}} a_{i\alpha}^{\dagger} a_{j\alpha} - t_{\perp}\sum_{i} e^{-i l A_{\perp i}} a_{i2}^{\dagger} a_{i 1} + h.c. \\ \nonumber
&+& \frac{U}{2}\sum_{i\alpha} n_{i\alpha}(n_{i\alpha}-1) + V_{\perp}\sum_i n_{i1}n_{i2} - \mu\sum_{i\alpha} n_{i\alpha}.
\end{eqnarray}
$\alpha=1,2$ denotes the two chain index, $l$ is the lattice spacing in both directions (parallel and perpendicular to the chains) and we have considered a general gauge field acting on the boson operators of each chain. We have also added a chemical potential which is justified in ultra-cold atoms, Josephson junction arrays and can be simulated in photon networks, as discussed in Sec. 3.1. Here, we also consider the effect of on-site repulsion and repulsion between chains, $n_{i\alpha}=a^{\dagger}_{i\alpha} a_{i\alpha}$. The $t_{\perp}$ term represents a Josephson coupling between chains favoring (superfluid) phase-locking between the chains and the occurrence of a superfluid. The artificial flux phases can be explicitly derived from a Floquet approach \cite{Polkovnikov2}. The application of a small (synthetic) magnetic field perpendicular to the ladder plane then will result in the Meissner effect as in superconductors. We can describe this Meissner effect as follows \cite{OrignacGiamarchi,Crepin}. Let us introduce the explicit bosonic representation $a_{i\alpha}=\sqrt{n}e^{i\theta_{i\alpha}}$, where $n$ represents the mean density in each chain, describing the (quasi-)superfluid in each chain. The Josephson term can be re-written as $-t_{\perp}n\cos(lA_{\perp i}+\theta_{1i}-\theta_{2i})$. The current (density) operators at point $i$ take the form: $j_{\parallel}=it(-e^{i l A_{ij}^1} a^{\dagger}_{i1} a_{j1} + e^{i l A_{ij}^2} a^{\dagger}_{i2} a_{j2})+h.c.$ and $j_{\perp} = -i t_{\perp} a^{\dagger}_{i1} a_{i2} e^{i l A_{\perp i}} +h.c.$. The key point is that for strong values of $t_{\perp}$, the superfluid phases are pinned such that $\theta_{1i}-\theta_{2i}+ l A_{\perp i}=0$ and therefore we obtain a Meissner effect, meaning that $\langle j_{\perp}\rangle=0$ in the bulk (see Fig. 6). Furthermore, in the small field limit, we may expand to get
\begin{equation}
\langle j_{\parallel} \rangle = \langle j_1\rangle - \langle j_2\rangle = -2 t n\ {\varphi}_{ij}.
\end{equation}
 \begin{figure}[t]
\center
\includegraphics[scale=0.4]{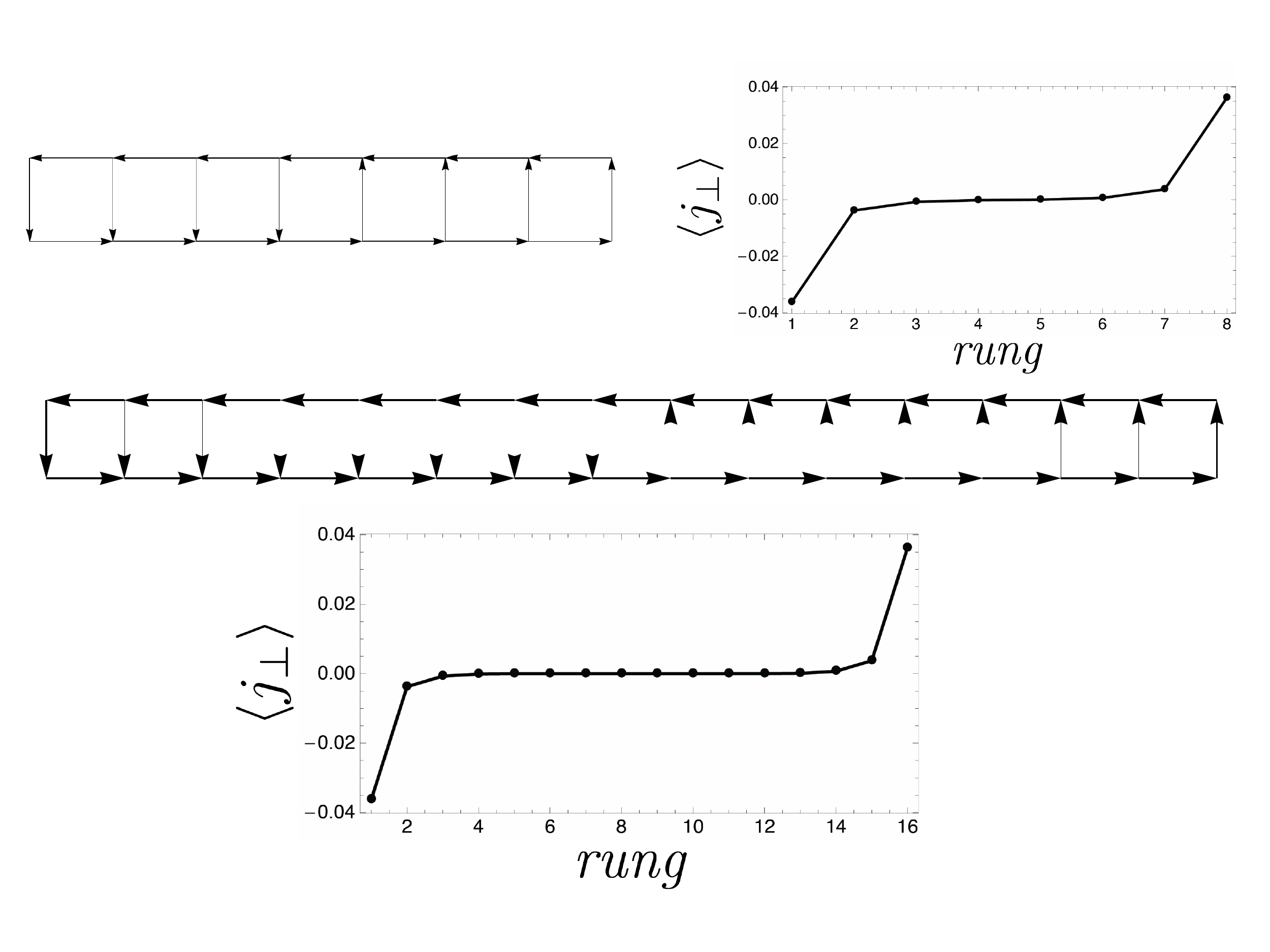}  
\vskip -0.7cm
\caption{DMRG and Exact Diagonalization results on small ladders with bosons, with 8 and 16 rungs, showing that the Meissner currents subsist in a Mott phase with total density $1$, as predicted by the theory \cite{AlexKarynMeissner1,AlexKarynMeissner2}. Here, the parameters are fixed to: formally $U\rightarrow +\infty$ (hard-core bosons), $V_{\perp}=4t$, $t=1$, $t_{\perp}=1$ and the flux is fixed to $\chi=\pi/25$ using the notations of Ref. \cite{AlexKarynMeissner2}. Other careful DMRG analyses of this model
have been performed in Refs. \cite{Marie,Fabian,Dhar}.}
\label{circuit}
\end{figure}
Here, we have defined the phase around a plaquette, $\varphi_{ij}=(A_{ij}^2-A_{ij}^1)l + (A_{\perp i} - A_{\perp j})l$. Then, this corresponds to the usual Meissner current, here defined on a lattice: a chiral current moves in opposite
direction than the applied external flux. This is an example of chiral superfluid phase in low dimensions which has been observed experimentally \cite{Bloch2}. This argument is in fact general since the $t_{\perp}$ hopping term is a relevant perturbation under Renormalization Group arguments in the superfluid phase; one can also apply bosonization techniques explicitly \cite{Giamarchi}. By increasing the magnetic flux, one sees the emergence of vortices; similarly to type II superconductors, there is a critical field strength. Another nice phenomenon is that these Meissner currents can subsist when entering into a Mott phase. For example, for one boson per rung in average, as shown in Refs. \cite{AlexKarynMeissner1,AlexKarynMeissner2}, by adjusting the parameters $U$ and $V_{\perp}$, one can enter a Mott state where the total density in each rung is fixed but still Meissner currents develop: this is an example of chiral Mott phase which breaks time-reversal symmetry. In Fig. 8, we show that the Meissner currents
in the Mott phase can be already observed for relatively small systems (with 8 or 16 rungs for example; these results are obtained with the Density matrix Renormalization Group (DMRG) and Exact Diagonalization). More results have been confirmed independently based on the DMRG \cite{Marie,Fabian}.  Other chiral phases have been predicted for other filling factors \cite{Dhar,Mueller,AkiyukiAntoine}. Note that Laughlin type phases in ladder systems \cite{TeoKane,KaneLubensky,BertStern} characterized by chiral edge states are also possible by adjusting densities and the magnetic flux, as shown in Ref. \cite{AlexKarynMeissner2}. These correlated phases can also be realized in quantum circuits such as Josephson junction ladders \cite{AlexKarynMeissner2}. Here, it is also important to emphasize other experimental effort in ultra-cold atoms, with boson and fermion systems \cite{Maryland,Florence}.

Another example of chiral bosonic phases has been be identified through the Haldane model on the honeycomb lattice discussed in Sec. 3.2, with bosons. Extensive results have been obtained through a combination of bosonic DMFT, Exact Diagonalization and analytical arguments \cite{IvanaAlex}. Here, we review simple physical results. One can easily identify the presence of two superfluid phases in the phase diagram. For $t_2=0$ (here, we follow the notations of Sec. 3.2) all the bosons condense at the center of the Brillouin zone, at zero momentum. On the other hand, for $t_1=0$, the model turns into two decoupled triangular lattices, and the bosons condense at the two inequivalent corners of the Brillouin zone $K$ and $K'$. Whereas the superfluid phase at $t_2=0$ will exhibit Meissner physics in response to the Haldane phases $\phi_{ij}=\pi/2$ between the next-nearest-neighbor bonds, the phase at large $t_2$ constitutes a novel chiral superfluid phase. More precisely, as an order parameter, one can define the average current $\langle j_{ij}\rangle$ coupling a pair of next-nearest-neighbors which belong to the same sub-lattice. In the superfluid phase with $t_2=0$, the Meissner response to the phase $\phi_{ji}=-\phi_{ij}=-\pi/2$ is $\langle j_{ij}\rangle = -2n t_2 \Im(\exp(-i\pi/2))=2 n t_2$.  In the chiral superfluid phase, one rather observes a sign change of the response $\langle j_{ij} \rangle = - n t_2$. This implies that the quantum phase transition between the two phases is a first-order-phase transition. A thorough analysis has been performed to build the complete phase diagram \cite{IvanaAlex}. The properties of the Mott phase, emerging for one particle per site, are also interesting; the Mott phase exhibits at high energy topological quasiparticle and quasi hole excitations visible in the spectral function. Similar chiral superfluid phases have been predicted on the square lattice \cite{Cristiane}.

We note that the emergence of chiral currents have been predicted in other Hamiltonians, for example, in doped Mott insulators \cite{AffleckMarston,Chakravarty}, high-Tc superconductors \cite{Bourges} and other ladder models \cite{ladder1,ladder2,ladder3}.

\subsection{Quantum Link Model in superconducting circuits}

Substantial theoretical effort has gone into implementing gauge theories in ultra cold atoms and quantum circuits  \cite{FisherZoller}. Superconducting circuits have also attracted some attention in the context of topological quantum computing \cite{DoucotIoffe} and Quantum Error Correction codes \cite{Barbarareview}, with some experimental efforts in superconducting circuits \cite{JohnMartinis}. There are also current theoretical efforts to suggest Majorana surface codes with superconducting wires (circuits) \cite{Fu,AltlandEgger}. One can provide a useful scheme to think about these gauge fields with potential realizations in superconducting circuits, using loops of Josephson junctions as described in Ref. \cite{ZollerMarcos}. 
One can build a Quantum link model. Let us consider a square lattice and two nearest neighboring sites connected with the phase $\varphi_{ij}=\int_i^j d{\bf l}\cdot {\bf A}$. The flux variable is then defined as $U_{ij}=\exp(i\varphi_{ij})\in U(1)$, and the electric field (conjugate) variable connecting this link is then defined as $E_{ij}=-i\partial/\partial\varphi_{ij}$. These variables are operators satisfying the angular momentum type algebra $[E_{ij},U_{ij}]=U_{ij}$ and $[E_{ij},U^{\dagger}_{ij}]=-U_{ij}^{\dagger}$; $U_{ij}$ and $U_{ij}^{\dagger}$ then act as lowering and raising operators of the electric field $e_{ij}$, such
that
\begin{equation}
E_{ij} |e_{ij} \rangle = e_{ij} |e_{ij}\rangle.
\end{equation}
Here, eigenvalues of the electric field are integer or half-integer. One can write down the Hamiltonian, by introducing the flux per plaquette. For example, for a square type geometry, the flux per plaquette reads $U_{\square} = (U_{ij}U_{jk} U_{kl} U_{li})=\exp(i\Phi)$ where
the magnetic flux per plaquette is $\Phi=\varphi_{ij}+\varphi_{jk}+\varphi_{kl}+\varphi_{li}$. The dynamics of the gauge fields here is described by the Hamiltonian \cite{ZollerMarcos}:
\begin{equation}
H = \frac{g^2}{2} \sum_{\langle i;j\rangle} E_{ij}^2 - \frac{1}{4g^2} \sum_{\square} (U_{\square} + U_{\square}^{\dagger}).
\end{equation}
This Hamiltonian corresponds to a Maxwell-type Hamiltonian on a lattice (the second term in $\cos\Phi$ is related to Josephson physics). 
Lattice gauge transformations keeping the Hamiltonian invariant have been discussed explicitly in Ref. \cite{ZollerMarcos}. One can envision different Quantum link Models. A class of models making links with spin-1/2 Hamiltonians and dimer models is defined as follows. If $e_{ij}=\pm 1/2$, then one can define spin-up and spin-down states through $(E_{ij}=S_{ij}^z,U_{ij}=S^-_{ij})$ and  $(E_{ij}=-S_{ij}^z,U_{ij}=S^+_{ij})$, where $E_{ij}^2=(S_{ij}^z)^2=1/4$. In this language, the $U_{\square}$ represent ring-exchange terms $S_{ij}^+ S_{jk}^- S_{kl}^+ S_{li}^-$. In a neutral sub-space  of the Hilbert space we have the same number of spin-up and spin-down states. Then, one can also engineer a Rokhsar-Kivelson type interaction between dimers \cite{RK}
\begin{equation}
H = -J\sum_{\square} \left(U_{\square} + U^{\dagger}_{\square} - \lambda(U_{\square} + U_{\square}^{\dagger})^2 \right).
\end{equation}
This model has attracted some attention in condensed-matter physics, in relation with confined-deconfined phase transitions. Theoretical questions about gauge theories are also relevant \cite{MoessnerSondhi}. Generalizations to non-Abelian situations might lead to more complex gauge theories. Some applications of quantum link models to quantum chromodynamics have been suggested \cite{Wiese}. Other gauge theories may have applications for quantum computing and the toric code \cite{Kitaev2}. In fact, theorists have recently shown that linear circuits can already mimic non-trivial hopping matrix elements, in relation with current technology, allowing potentially to realize non-Abelian gauge fields. In particular, loops of ``triangle-like capacitors'' have been considered as a potential simulator of a spin-doubled Azbel-Hofstadter \cite{Hofstadter,Azbel} model, with 1/3 magnetic flux per plaquette \cite{Liang}. This model is related to the time-reversal invariant Hofstadter model \cite{GoldmanSpielman,Cocks,Peter,StephanNature}. Non-abelian models have also been considered in ladder systems \cite{Piraud}.

\section{Conclusion}

We have summarized several recent developments, both theoretical and experimental, to realize many-body physics with photons in superconducting Quantum Electrodynamics (QED) Networks. At the theoretical level, we have outlined analytical and numerical approaches to describe quantitatively these open, dissipative and many-body quantum systems beyond the equilibrium limit. We hope that this review will also serve to inspire novel experimental developments, related for example to the realization of Kondo physics with light in Josephson junction arrays \cite{Karyn,Moshe} or to the realization of the Jaynes-Cummings lattice \cite{Houckreview,Greentree,Angelakis,Hartmann,KochHur,SchmidtBlatter,SchmidtBlatter1} or Rabi model \cite{marco,Marconew} with superconducting circuits. Some experimental efforts already exist for a few cavities \cite{Raftery,RochLPA,Benjamin}. At a more general level, one can also anticipate closer links  with propagating light in bulk nonlinear media \cite{pomeau} and with phononics \cite{huberpho}. Dissipation has also been shown to be efficient to induce topological states of matter \cite{diehl}.

We have presented results on the driven Jaynes-Cummings lattice from a stochastic Schr\" odinger equation approach \cite{loic,PAK1,PAK2,Lesovik,Demler} including dissipation effects in the cavities. The coupling between cavities is treated at a mean-field level. A more rigorous dynamical-mean field theory approach could be, in principle, built \cite{DMFT} by analogy with recent results on dissipative spin chains with long-range forces \cite{LoicKaryn}. We have also extended stochastic approaches for two-spin dimer problems \cite{LoicKaryn}, which have attracted some attention both theoretically \cite{Aron} and experimentally \cite{Raftery}. On the experimental side, hybrid mesoscopic circuits  \cite{Senellart,hybrid1,hybrid2,hybrid3,hybrid4} comprising quantum dots  and photons will hopefully confirm our theoretical predictions \cite{MarcoKaryn}. Regarding the implementation of synthetic gauge fields and spin-orbit couplings in photon systems, experimentalists have already managed to observe topological phases of light \cite{MIT,Rechtsman,Hafezi1,Hafezi2,Alberto,Soljacic}, in synchronization with the developments in ultra-cold systems \cite{Gerbier,Spielman,Sengstock0,BlochHofstadter,Wolfgang,JackschZoller}. Using the Floquet approach, we have shown that the Haldane model on the honeycomb lattice can be engineered from a perturbation periodic in time. The Haldane model has been recently implemented with ultra-cold fermions \cite{Esslinger}. A similar scheme has been considered to simulate a Kane-Mele model for photons \cite{Hafezi1,Hafezi2}. Josephson junction loops can  serve to realize artificial gauge fields and break time-reversal symmetry \cite{KochT,Andreas,AlexKaryn,Lehnert2,Kamal}.  Applying the Density Matrix Renormalization Group and Exact Diagonalization (on small systems), we have checked that Meissner currents can develop in a Mott insulating phase, in a ladder geometry with synthetic gauge fields and Josephson physics \cite{AlexKarynMeissner1,AlexKarynMeissner2,AkiyukiAntoine}. Such ladder geometries have been built with ultra-cold bosonic atoms \cite{Bloch2} based on the theoretical work of Ref. \cite{OrignacGiamarchi}. Experimental accomplishments with lumped elements \cite{simon2} and small superconducting circuits \cite{Martinis,Lehnert} pave the way towards observing topological
phases in superconducting Quantum Electrodynamics (QED) networks. Entanglement properties in such quantum Hall phases could be probed through charge noise \cite{Francis,KlichLevitov,JSTATAlex,Glattli,Heiblum}. 

{\it Acnowledgements}: We would like to acknowledge A. Amo, C. Aron, I. Bloch, J. Bloch, S. Bose, I. Carusotto, J. Cayssol, C. Ciuti, G. Deng, M. Dogan, B. Dou\c{c}ot, P. Dutt, E. Eriksson, J. Esteve, M. Filippone, S. Florens, J. Gabelli, T. Giamarchi, S. M. Girvin, A. Georges, M. Goerbig, T. G\" oren, G. Guo, M. Hafezi, L. Herviou, W. Hofstetter, A. Houck, B. Huard, A. Jordan, J. Koch, T. Kontos, Ph. Lecheminant, M.-R. Li, T. Liu, V. Manucharyan, C. Mora, P. Nataf, E. Orignac, F. Pistolesi, C. Repellin, N. Roch, Z. Ristivojevic, P. P. Orth, M. Piraud, S. Rachel, N. Regnault, P. Senellart, S. Schmidt, P. Simon, H.-F. Song, A. Sterdyniak, J. Taylor, H. T\"ureci, I. Vasic and W. Wu, for discussions or collaborations related to this work. This work has benefitted from the financial support by DOE, under the grant DE-FG02-08ER46541, and by the Labex PALM Paris-Saclay ANR-10-LABX-0039. This work has also benefitted from discussions at the CIFAR meeting in Canada, at the memorial symposia in honor of Adilet Imambekov at Harvard and Bernard Coqblin in Orsay, at conferences in Bordeaux, Jouvence (Quebec), KITP Santa Barbara, Natal, Nordita, Paris, Trento, Trieste for example.

\vskip 1cm

\bibliographystyle{CRAS_with_doi_eprint}

\end{document}

%% file: macrosNotations.inc
\newcommandx{\Iverson}[1]{\ensuremath{\left[ #1 \right] }}

\newcommandx\PermutationGroup[1]{\ensuremath{\mathfrak{S}_{#1}}}

\newcommandx\GeneralLinearGroup[2][2={}]{
\ifthenelse{\equal{#2}{}}{
\ensuremath{\text{GL}(#1)}
}{
\ensuremath{\text{GL}(#1,#2)}
}\xspace
}
\newcommandx\SpecialLinearGroup[2][2={}]{
\ifthenelse{\equal{#2}{}}{
\ensuremath{\text{SL}(#1)}
}{
\ensuremath{\text{SL}(#1,#2)}
}\xspace
}

\newcommandx\ContinuityClass[3][2={},3={}]{
\ifthenelse{\equal{#2}{}}{
    \ensuremath{\mathcal{C}^{#1}}
  }{
    \ifthenelse{\equal{#3}{}}{
      \ensuremath{\mathcal{C}^{#1}(#2)}
    }{
      \ensuremath{\mathcal{C}^{#1}(#2,#3)}
  }
}\xspace
}

\newcommandx\IntegerPart[1]{\ensuremath{\text{E}\left[ #1 \right]}}

\newcommandx\PauliMatrix[2][1={\sigma}]{\xspace\ensuremath{
\ifthenelse{\equal{#2}{}}{
#1
}{
#1_{#2}
}
}\xspace}
\newcommandx\ExteriorAlgebra[2][1={}]{\xspace\ensuremath{
\ifthenelse{\equal{#1}{}}{
\Lambda #2
}{
\Lambda^{#1} #2
}
}\xspace}

\newcommandx\Norm[1]{\ensuremath{\left\lVert #1 \right\rVert}}
\newcommandx\norm[1]{\ensuremath{\lVert #1 \rVert}}

\renewcommand{\Re}{\operatorname{Re}}
\renewcommand{\Im}{\operatorname{Im}}



%% file: specificNotations.inc
\NewDocumentCommand{\GammaMatrix}{mg}{
\IfNoValueTF{#2}{
\Gamma_{#1}
}{
\Gamma_{#1 #2}
}
}

\NewDocumentCommand{\TRIClosedCurve}{O{} m}{\ensuremath{\mathcal{C}_{#2}^{#1}}}

%% file: Circuit.Photon.summary.FINAL2.bbl
\begin{thebibliography}{99}

\bibitem{cohen}
C. Cohen-Tanoudji, J. Dupont-Roc and G. Grynberg, Photons and atoms, introduction to quantum electrodynamics, Wiley (1997).

\bibitem{raimond}
J.-M. Raimond, M. Brune and S. Haroche, Manipulating quantum entanglement with atoms and photons in a cavity, Rev. Mod. Phys. 73, 565 (2001).

\bibitem{haroche}
S. Haroche and J.-M. Raimond, Exploring the Quantum: Atoms, Cavities, and Photons, Oxford University Press (2006).

\bibitem{Wineland}
D. Leibfried, R. Blatt, C. Monroe and D. Wineland, Quantum dynamics of single trapped ions, Rev. Mod. Phys. 75, 281, (2003).

\bibitem{Esslinger0}
H. Ritsch, P. Domokos, F. Brennecke, and T. Esslinger, Cold atoms in cavity-generated dynamical optical potentials, Rev. Mod. Phys. 85, 553-601 (2013).

\bibitem{Esslinger00}
K. Baumann, R. Mottl, F. Brennecke and T. Esslinger, Exploring Symmetry Breaking at the Dicke Quantum Phase Transition, Phys. Rev. Lett. 107, 140402 (2011).

\bibitem{SchoelkopfGirvin}
R. J. Schoelkopf and S. M. Girvin, Wiring up quantum systems, Nature 451, 664 (2008). 

\bibitem{Circuits}
M. H. Devoret, in {\it Quantum Fluctuations}, edited by S. Reynaud, E. Giacobino, and J. Zinn-Justin (Elsevier, 1995) Chap. 10.

\bibitem{braak}
D. Braak, Integrability of the Rabi Model, Phys. Rev. Lett. 107, 100401 (2011).

\bibitem{Houckreview}
A. A. Houck, H. E. T\"{u}reci and J. Koch, On-chip quantum simulation with superconducting circuits, Nature Phys. 8, 292-299 (2012).

 \bibitem{Fazio}
A. Tomadin and R. Fazio, Many-body phenomena in QED-cavity arrays, J. Opt. Soc. Am. 27, A130 (2010).

\bibitem{experimentHouck}
D. L. Underwood, W. E. Shanks, J. Koch and A. A. Houck, Low-Disorder Microwave Cavity Lattices for Quantum Simulation with Photons, Phys. Rev. A 86, 023837 (2012).

\bibitem{Wallraffnew}
Y. Salath\' e {\it et al.}, Digital quantum simulation of spin models with circuit quantum electrodynamics, arXiv:1502.06778.

\bibitem{Martinis1}
R. Barends {\it et al.}, Digital quantum simulation of fermionic models with a superconducting circuit,  arXiv:1501.07703.

\bibitem{Martinis2}
Y. Chen {\it et al.}, Simulating weak localization using superconducting quantum circuits,  Nature Communications 5, 5184 (2014).

\bibitem{Irfan}
S. J. Weber, A. Chantasri, J. Dressel, A. N. Jordan, K. W. Murch, and I. Siddiqi Mapping the optimal route between two quantum states, Nature 511, 570-573 (2014).

\bibitem{Nicolas} 
N. Roch {\it et al.}, Observation of measurement-induced entanglement and quantum trajectories of remote superconducting qubits, Phys. Rev. Lett. 112, 170501 (2014).

\bibitem{coldatomreview}
I. Bloch, J. Dalibard and W. Zwerger, Many-Body Physics with Ultracold Gases, Rev. Mod. Phys. 80, 885 (2008).

\bibitem{coldatom2}
I. Bloch, J. Dalibard and S. Nascimb\`ene, Quantum simulations with ultracold quantum gases,  Nature Physics 8, 267-276 (2012).

\bibitem{Optique}
M. Viteau, P. Huillery, M. G. Bason, N. Malossi, D. Ciampini, O. Morsch, E. Arimondo, D. Comparat and P. Pillet, Cooperative excitation and many-body interactions in a cold Rydberg gas, Phys. Rev. Lett. 109, 053002 (2012). 

\bibitem{AntoineThierry}
H. Labuhn, S. Ravets, D. Barredo, L. B\' eguin, F. Nogrette, T. Lahaye and A. Browaeys, Single-Atom Addressing in Microtraps for Quantum-State Engineering using Rydberg Atoms, Phys. Rev. A 90, 023415 (2014).

\bibitem{IOP}
V. Parigi, E. Bimbard, J. Stanojevic, A. J. Hilliard, F. Nogrette, R. Tualle-Brouri, A. Ourjoumtsev, and P. Grangier, Observation and Measurement of Interaction-Induced Dispersive Optical Nonlinearities in an Ensemble of Cold Rydberg Atoms, Phys. Rev. Lett. 109, 233602 (2012).

\bibitem{SPEC}
Y. Kubo, C. Grezes, A. Dewes, T. Umeda, J. Isoya, H. Sumiya, N. Morishita, H. Abe, S. Onoda, T. Ohshima, V. Jacques, A. Dr\' eau, J.-F. Roch, I. Diniz, A. Auffeves, D. Vion, D. Esteve, P. Bertet, Hybrid quantum circuit with a superconducting qubit coupled to a spin ensemble, Phys. Rev. Lett. 107, 220501 (2011).

\bibitem{Ludwig1}
M. Ludwig and F. Marquardt, Quantum many-body dynamics in optomechanical arrays, Phys. Rev. Lett. 111, 073603 (2013).

\bibitem{Marquardt0}
M. Schmidt, V. Peano and F. Marquardt, Optomechanical Dirac Physics, arXiv:1410.8483.

\bibitem{Kondo}
 J. Kondo, Resistance Minimum in Dilute Magnetic Alloys, Progress of Theoretical Physics 32 37 (1963).
 
 \bibitem{Anderson0}
 P. W. Anderson, A poor man's derivation of scaling laws for the Kondo problem, J. Phys. C: Solid St. Phys. 3 2436-2441 (1970).

\bibitem{Nozieres}
Ph. Nozi\`eres, A Fermi-liquid description of the Kondo model at low temperatures, Journ. of Low Temperature Physics 17, 31 (1974).

\bibitem{Wilson}
K. Wilson, The renormalization group: Critical phenomena and the Kondo problem, Rev. Mod. Phys. 47 (4): 773-840 (1975).

\bibitem{Affleck}
For a review: I. Affleck, Conformal Field Theory Approach to the Kondo Effect, Acta Phys. Polon. B26 1869-1932 (1995). 

\bibitem{WiegmannTsvelik}
A. M. Tsvelick and P. Wiegmann, Exact results in the theory of magnetic alloys, Adv. Phys. 32 , 453 (1983).

\bibitem{Karyn}
K. Le Hur, Kondo Resonance of a Microwave Photon, Phys. Rev. B 85, 140506(R) (2012).

\bibitem{Moshe}
M. Goldstein, M. H. Devoret, M. Houzet, and L. I. Glazman, Inelastic Microwave Photon Scattering off a Quantum Impurity in a Josephson-Junction Array, Phys. Rev. Lett. 110, 017002 (2013).

\bibitem{Saleur1997}
A. Leclair,  F. Lesage, S. Lukyanov and H. Saleur, The Maxwell-Bloch Theory in Quantum Optics and the. Kondo Model, Phys. Lett. A 235 203-208 (1997).

\bibitem{Camalet}
S. Camalet, J. Schriefl, P. Degiovanni and F. Delduc, Quantum Impurity Approach to a coupled Qubit Problem, Europhysics Letters 68 37 (2004).

\bibitem{loic}
L. Henriet, Z. Ristivojevic, P. P. Orth and K. Le Hur, Quantum Dynamics of the Driven and Dissipative Rabi Model, Phys. Rev. A 90, 023820 (2014).

\bibitem{PAK1}
P. P. Orth, A. Imambekov, and K. Le Hur, Universality in dissipative Landau-Zener transitions, Phys. Rev. A 82, 032118 (2010).

\bibitem{PAK2}
P. P. Orth, A. Imambekov, and K. Le Hur, Non-perturbative stochastic method for driven spin-boson model,  Phys. Rev. B 87, 014305 (2013).

\bibitem{Lesovik}
G. B. Lesovik, A. O. Lebedev and A. O. Imambekov, Dynamics of Two-Level System Interacting with Random Classical Field, JETP Lett, 75 474 (2002).
    
  \bibitem{Demler}
A. O. Imambekov, V. Gritsev and E. Demler, Proceedings of the 2006 Enrico Fermi Summer School on ``Ultracold Fermi gases'' , Varenna, 2006 edited by M. Inguscio,
W. Ketterle, and C. Salomon (IOS Press, Amsterdam) 2008.

 \bibitem{Greentree}
 A. D. Greentree, C. Tahan, J. H. Cole, and L. C. L. Hollenberg, Simulating quantum fields with cavity QED, Nat. Phys. 2, 856 (2006).
 
 \bibitem{Angelakis1}
 D. G. Angelakis, M. F. Santos and S. Bose, Photon-blockade-induced Mott transitions and XY spin models in coupled cavity arrays, Phys. Rev. A {\bf 76}, 031 (2007).
 
 \bibitem{Angelakis}
J. Cho, D. G. Angelakis and S. Bose, Simulation of high-spin Heisenberg models in coupled cavities,  Phys. Rev. A 78, 062338 (2008).
 
  \bibitem{Hartmann}
 M. J. Hartmann, F. G. S. L. Brandao, and M. B. Plenio, Quantum many-body phenomena in coupled cavity arrays, Nat. Phys. 2, 849 (2006).
 
 \bibitem{KochHur}
 J. Koch and K. Le Hur, Superfluid-Mott Insulator Transition of Light in the Jaynes-Cummings Lattice,  Phys. Rev. A 80, 023811 (2009).
 
 \bibitem{SchmidtBlatter}
 S. Schmidt and G. Blatter, Strong coupling theory for the Jaynes-Cummings-Hubbard model, Phys. Rev. Lett. 103, 086403 (2009).
 
 \bibitem{SchmidtBlatter1}
 S. Schmidt and G. Blatter, Excitations of strongly correlated polaritons, Phys. Rev. Lett. 104, 216402 (2010).
 
 \bibitem{marco}
M. Schir\' {o}, M. Bordyuh, B. \"{O}ztop and H. E. T\"{u}reci, Phase Transition of Light in Cavity QED Lattices, Phys. Rev. Lett. 109, 053601 (2012).

\bibitem{Marconew}
M. Schir\' o, C. Joshi, M. Bordyuh, R. Fazio, J. Keeling, and H. E. T\" ureci, Exotic attractors of the non-equilibrium Rabi-Hubbard model,  arXiv:1503.04456.
 
\bibitem{Hafezi}
M. Hafezi, P. Adhikari and J. M. Taylor, A chemical potential for light, arXiv:1405.5821.
 
\bibitem{Karyn00}
K. Le Hur, Quantum Phase Transitions in Spin-Boson Systems: Dissipation and Light Phenomena in the book ``Understanding Quantum Phase Transitions'', edited by Lincoln D. Carr (Taylor and Francis, Boca Raton, 2010); see also arXiv:0909.4822.

\bibitem{Leggett}
A. J. Leggett, S. Chakravarty, A. T. Dorsey, M. P. A. Fisher, A. Garg and W. Zwerger, Dynamics of the dissipative two-state system, Rev. Mod. Phys, 59, 1 (1987).

\bibitem{weiss}
U. Weiss, Quantum Dissipative Systems, World Scientific, Singapore (2008).

\bibitem{moroz}
A. Moroz, On solvability and integrability of the Rabi model, Ann. Phys. 338, 319-340 (2013).

\bibitem{zhong}
H. Zhong, Q. Xie, M. Batchelor and C. Lee, Analytical eigenstates for the quantum Rabi model, J. Phys. A: Math. Theor. 46 415302 (2013). 

\bibitem{gritsev}
M. Tomka, O. El Araby, M. Pletyukhov and V. Gritsev, Exceptional and regular spectra of a generalized Rabi model,  Phys. Rev. A 90, 063839 (2014).

\bibitem{larson}
J. Larson, Dynamics of the Jaynes-Cummings and Rabi models: old wine in new bottles, Phys. Scr. 76, 146 (2007).

\bibitem{nataf}
P. Nataf and C. Ciuti, Vacuum Degeneracy of a Circuit QED System in the Ultrastrong Coupling Regime, Phys. Rev. Lett. 104 023601 (2010). 

\bibitem{Simone}
Simone de Liberato, Light-Matter Decoupling in the Deep Strong Coupling Regime: The Breakdown of the Purcell Effect, Phys. Rev. Lett. 112, 016401 (2014).

\bibitem{braak2}
F. A. Wold, F. Vallone, G. Romero, M. Kollar, E. Solano and D. Braak, Dynamical correlation functions and the quantum Rabi model, Phys. Rev. A 87, 023835 (2013).

\bibitem{KochT}
J. Koch, A. A. Houck, K. Le Hur and S. M. Girvin, Time-reversal symmetry breaking in circuit-QED based photon lattices, Phys. Rev. A 82, 043811 (2010).

\bibitem{Andreas}
A. Nunnenkamp, J. Koch and S. M. Girvin, Synthetic gauge fields and homodyne transmission in Jaynes-Cummings lattices, New J. Phys. 13, 095008 (2011).

\bibitem{Lehnert2}
J. Kerckhoff, K. Lalumi\`ere, B. J. Chapman, A. Blais and K. W. Lehnert, On-chip superconducting microwave circulator from synthetic rotation, arXiv:1502.06041.

\bibitem{AlexKaryn}
A. Petrescu, A. A. Houck and K. Le Hur, Anomalous Hall Effects of Light and Chiral Edge Modes on the Kagome Lattice, Phys. Rev. A 86, 053804 (2012).

\bibitem{Kamal}
A. Kamal, J. Clarke and M. Devoret, Noiseless nonreciprocity in a parametric active device, Nature Physics 7, 311-315 (2011).

\bibitem{Klitzing}
K. von Klitzing, G. Dorda and M. Pepper, New Method for High-Accuracy Determination of the Fine-Structure Constant Based on Quantized Hall Resistance, Phys. Rev. Lett. 45, 494 (1980).

\bibitem{Laughlin}
R. B. Laughlin, Anomalous Quantum Hall Effect: An Incompressible Quantum Fluid with Fractionally Charged Excitations, Phys. Rev. Lett. 50, 1395 (1983). 

\bibitem{Stormer}
D. C. Tsui, H. L. Stormer, and A. C. Gossard, Two-Dimensional Magnetotransport in the Extreme Quantum Limit, Phys. Rev. Lett. 48, 1559 (1982).

\bibitem{Kane}
M. Z. Hasan and C. L. Kane, Topological Insulators, Rev. Mod. Phys. 82 3045 (2010).

\bibitem{Bernevig}
B. A. Bernevig with T. L. Hughes, Topological Insulators and Topological Superconductors, Princeton University Press (2013).

\bibitem{Zhang}
Xiao-Liang Qi and Shou-Cheng Zhang, Topological insulators and superconductors, Rev. Mod. Phys. 83, 1057 (2011).

\bibitem{MIT}
Z. Wang, Y. Chong, J. D. Joannopoulos and M. Soljacic, Observation of unidirectional backscattering-immune topological electromagnetic states Nature 461, 772-775 (2009).

\bibitem{Rechtsman}
M. C. Rechtsman, J. M. Zeuner, Y. Plotnik, Y. Lumer, D. Podolsky, F. Dreisow, S. Nolte, M. Segev and A. Szameit, Photonic Floquet Topological Insulators, Nature 496, 196-200 (2013).

\bibitem{Hafezi1}
M. Hafezi, J. Fan, A. Migdall, and J. Taylor, Observation of photonic edge states in a versatile Silicon platform, Nature Photonics 7, 1001 (2013).

\bibitem{Hafezi2}
M. Hafezi, E. Demler, M. Lukin and J. Taylor, Robust optical delay lines via topological protection, Nature Physics 7, 907--912 (2011).

\bibitem{Alberto}
V. G. Sala, D. D. Solnyshkov, I. Carusotto, T. Jacqmin, A. Lema\^itre, H. Ter\c{c}as, A. Nalitov, M. Abbarchi, E. Galopin, I. Sagnes, J. Bloch, G. Malpuech and A. Amo, Engineering spin-orbit coupling for photons and polaritons in microstructures,
 arXiv:1406.4816
 
 \bibitem{Dalibard}
N. Goldman and J. Dalibard, Periodically-driven quantum systems: Effective Hamiltonians and engineered gauge fields, Phys. Rev. X 4, 031027 (2014).

\bibitem{CayssolMoessner}
J. Cayssol, B. D\' ora, F. Simon and R. Moessner, Floquet topological insulators, Phys. Status Solidi RRL, 7, 101-108 (2013).

\bibitem{Soljacic}
L. Lu, J. D. Joannopoulos and M. Soljacic, Topological photonics, Nature Photonics 8, 821-829 (2014).

\bibitem{CiutiCarusotto}
I. Carusotto and C. Ciuti,  Quantum fluids of light, Rev. Mod. Phys. 85, 299 (2013).

\bibitem{Gerbier}
J. Dalibard, F. Gerbier, G. Juzeli\" unas, Patrik \"Ohberg, Artificial gauge potentials for neutral atoms, Rev. Mod. Phys. 83, 1523 (2011).

\bibitem{Spielman}
N. Goldman, G. Juzeliunas, P. Ohberg, I. B. Spielman, Light-induced gauge fields for ultracold atoms,  Rep. Prog. Phys. 77 126401 (2014).

\bibitem{Sengstock0}
Philipp Hauke, Olivier Tieleman, Alessio Celi, Christoph \" Olschl\" ager, Juliette Simonet, Julian Struck, Malte Weinberg, Patrick Windpassinger, Klaus Sengstock, Maciej Lewenstein, Andr\'e Eckardt, Non-Abelian gauge fields and topological insulators in shaken optical lattices,  Phys. Rev. Lett. 109, 145301 (2012).

\bibitem{BlochHofstadter}
M. Aidelsburger, M. Atala, M. Lohse, J. T. Barreiro, B. Paredes and I. Bloch, Realization of the Hofstadter Hamiltonian with ultracold atoms in optical lattices,  Phys. Rev. Lett. 111, 185301 (2013).

\bibitem{Wolfgang}
H. Miyake, G. A. Siviloglou, C. J. Kennedy, W. C. Burton, and W. Ketterle, Realizing the Harper Hamiltonian with Laser-Assisted Tunneling in Optical Lattices, Phys. Rev. Lett. 111, 185302 (2013).

\bibitem{JackschZoller}
D. Jaksch and P. Zoller, Creation of effective magnetic fields in optical lattices: The Hofstadter butterfly for cold neutral atoms, New J. Phys. 5 56 (2003).

\bibitem{HaldaneRaghu}
F. D. M. Haldane and S. Raghu, Possible Realization of Directional Optical Waveguides in Photonic Crystals with Broken Time-Reversal Symmetry, Phys.Rev.Lett. 100, 013904 (2008).

\bibitem{Bert}
B. I. Halperin, Quantized Hall conductance, current-carrying edge states, and the existence of extended states in a two-dimensional disordered potential, Phys. Rev. B 25, 2185 (1982).

\bibitem{Stern}
Chetan Nayak, Steven H. Simon, Ady Stern, Michael Freedman, Sankar Das Sarma, Non-Abelian Anyons and Topological Quantum Computation, Rev. Mod. Phys. 80, 1083 (2008).

\bibitem{Thouless}
D. J. Thouless, M. Kohmoto, M. P. Nightingale and M. den Nijs, Quantized  Hall  Conductance  in  a  Two-Dimensional  Periodic Potential. Phys. Rev. Lett. 49, 405-408 (1982).

 \bibitem{Carusottodissi}
 T. Ozawa and I. Carusotto, Phys. Rev. Lett. 112, 133902 (2014).

\bibitem{Hafezi3}
M. Hafezi, Measuring topological invariants in photonic systems, Phys. Rev. Lett. 112, 210405 (2014).

\bibitem{Cooper}
H. M. Price and N. R. Cooper, Mapping the Berry Curvature from Semiclassical Dynamics in Optical Lattices, Phys. Rev. A 85, 033620 (2012).

\bibitem{Karplus}
R. Karplus and J. M. Luttinger, Hall Effect in Ferromagnetics, Phys. Rev. 95, 1154 (1954).

\bibitem{Cominotti}
Marco Cominotti and Iacopo Carusotto, Berry curvature effects in the Bloch oscillations of a quantum particle under a strong (synthetic) magnetic field, EPL 103 (2013) 10001.

\bibitem{MITChern}
S. A. Skirlo, L. Lu, Y. Igarashi, J. Joannopoulos and M. Soljacic, Experimental Observation of Large Chern numbers in Photonic Crystals,  arXiv:1504.04399.

\bibitem{Berry}
M. V. Berry, Quantal Phase Factors Accompanying Adiabatic Changes, Proceedings of the Royal Society of London. Series A, Mathematical and Physical Sciences, Volume 392, Issue 1802, pp. 45-57 (1984).

\bibitem{Chern}
S. S. Chern, Characteristic classes of Hermitian Manifolds, Annals of Mathematics 47, 1 (1948).

\bibitem{SchoelkopfWalraff}
P. J. Leek, J. M. Fink, A. Blais, R. Bianchetti, M. G\"oppl, J. M. Gambetta, D. I. Schuster, L. Frunzio, R. J. Schoelkopf and A. Wallraff, Observation of Berry's Phase in a Solid State Qubit, Science 318, 1889 (2007).

\bibitem{Martinis}
P. Roushan {\it et al.}, Observation of topological transitions in interacting quantum circuits,  Nature 515 241-244 (2014).

\bibitem{Lehnert}
M. D. Schroer, M. H. Kolodrubetz, W. F. Kindel, M. Sandberg, J. Gao, M. R. Vissers, D. P. Pappas, Anatoli Polkovnikov and K. W. Lehnert, Measuring a topological transition in an artificial spin 1/2 system, Phys. Rev. Lett. 113, 050402.

\bibitem{Polkovnikov}
V. Gritsev, A. Polkovnikov, Dynamical Quantum Hall Effect in the Parameter Space,  PNAS, 109, 6457 (2012).

\bibitem{Bellec0}
M. Bellec, U. Kuhl, G. Montambaux and Fabrice Mortessagne, Tight-binding couplings in microwave artificial graphene, Phys. Rev. B 88, 115437 (2013).

\bibitem{Weyl}
 L. Lu, Z. Wang, D. Ye, L. Ran, L. Fu, J. D. Joannopoulos, M. Solja\c{c}ic, Experimental observation of Weyl points,  arXiv:1502.03438.

\bibitem{Jacqmin}
Th. Jacqmin,  I. Carusotto, I. Sagnes, M. Abbarchi, D. Solnyshkov, G. Malpuech, E. Galopin, A. Lema\^itre, J. Bloch and A. Amo, Direct observation of Dirac cones and a flatband in a honeycomb lattice for polaritons, Phys. Rev. Lett. 112 116402 (2014).

\bibitem{Patrizia}
P. Vignolo, M. Bellec, J. Boehm, A. Camara, J.-M. Gambaudo, U. Kuhl, and F. Mortessagne,  Energy landscape in two-dimensional Penrose-tiled quasicrystal, arXiv:1411.1234.

\bibitem{Sebastians}
M. Biondi, E. P. L. van Nieuwenburg, G. Blatter, S. D. Huber, S. Schmidt, Incompressible polaritons in a flat band,  arXiv:1502.07854.

\bibitem{Fazio1D}
Feng Mei, Jia-Bin You, Wei Nie, R. Fazio, Shi-Liang Zhu, L. C. Kwek, Simulation and Detection of Photonic Chern Insulators in One-Dimensional Circuit Quantum Electrodynamics Lattice,  arXiv:1504.05686.

\bibitem{Tanese}
D. Tanese, E. Gurevich, F. Baboux, T. Jacqmin, A. Lema\^itre, E. Galopin, I. Sagnes, A. Amo, J. Bloch and E. Akkermans, Fractal energy spectrum of a polariton gas in a Fibonacci quasi-periodic potential, Phys. Rev. Lett. 112, 146404 (2014).

\bibitem{MarcoKaryn}
M. Schir\' o and K. Le Hur, Tunable Hybrid Quantum Electrodynamics from Non-Linear Electron Transport, Phys. Rev. B 89, 195127 (2014).

\bibitem{Senellart}
A. Dousse,  L. Lanco, J. Suffczynski, E. Semenova, A. Miard, A. Lema\^itre, I. Sagnes, C. Roblin, J. Bloch, P. Senellart, Controlled light-matter coupling for a single quantum dot embedded in a pillar microcavity using far-field optical lithography, Phys. Rev. Lett. 101, 267404 (2008).

\bibitem{hybrid1}
T. Frey, P. J. Leek, M. Beck, J. Faist, A. Wallraff, K. Ensslin, T. Ihn, M. B\" uttiker, Quantum dot admittance probed at microwave frequencies with an on-chip resonator, Phys. Rev. B 86, 115303 (2012).

\bibitem{hybrid2}
M. R. Delbecq,  V. Schmitt, F.D. Parmentier, N. Roch, J.J. Viennot, G. F\`eve, B. Huard, C. Mora, A. Cottet, T. Kontos, Coupling a quantum dot, fermionic leads and a microwave cavity on-chip,  Phys. Rev. Lett. 107, 256804 (2011).

\bibitem{hybrid3}
K. D. Petersson,  L. W. McFaul, M. D. Schroer, M. Jung, J. M. Taylor, A. A. Houck, J. R. Petta, Circuit Quantum Electrodynamics with a Spin Qubit, Nature 490, 380 (2012).

\bibitem{hybrid4}
Z.-R. Lin, G.-P. Guo, T. Tu, F.-Y. Zhu, and G.-C. Guo, Generation of quantum-dot cluster states with a superconducting transmission line resonator, Phys. Rev. Lett. 101, 230501 (2008).

\bibitem{Delsing}
 C. M. Wilson, G. Johansson, A. Pourkabirian,	M. Simoen, J. R. Johansson, T. Duty, F. Nori and P. Delsing, Observation of the Dynamical Casimir Effect in a Superconducting Circuit, Nature 479, 376-379 (2011).

\bibitem{Lloyd}
J. E. Mooij, T. P. Orlando, L. Levitov, L. Tian, C. H. Van der Wal and S. Lloyd, Josephson persistent-current qubit, Science 285, 1036-1039, (1999).

\bibitem{martinis}
J. M. Martinis, S. Nam, J. Aumentado and C. Urbina, Rabi Oscillations in a Large Josephson-Junction Qubit, Phys. Rev. Lett. 89, 117901 (2002).

\bibitem{Vion}
D. Vion, A. Aassime, A. Cottet, P. Joyez, H. Pothier, C. Urbina, D. Esteve and M. H. Devoret, Manipulating the Quantum State of an Electrical Circuit, Science 296, 886 (2002).

\bibitem{transmon}
J. Koch, T. M. Yu, J. Gambetta, A. A. Houck, D. I. Schuster, J. Majer, A. Blais, M. H. Devoret, S. M. Girvin and R. J. Schoelkopf, Charge-insensitive qubit design derived from the Cooper pair box, Phys. Rev. A 76, 042319 (2007).

\bibitem{paik}
H. Paik {\it et al.}, Observation of High Coherence in Josephson Junction Qubits Measured in a Three-Dimensional Circuit QED Architecture, Phys. Rev. Lett. 107, 240501 (2011).

\bibitem{fluxonium}
V. E. Manucharyan, J. Koch, L. Glazman and M. Devoret, Fluxonium: Single Cooper-Pair Circuit Free of Charge Offsets, Science 326, 113-116 (2009).

\bibitem{MartinisChen}
Y. Chen {\it et al.}, qubit architecture with High Coherence and Fast Tunable Coupling, Phys. Rev. Lett. 113, 220502 (2014).

\bibitem{SchmidtKoch}
S. Schmidt and J. Koch, Circuit QED lattices: towards quantum simulation with superconducting circuits, Annalen der Physik, 525, 395-412 (2013).

\bibitem{alexandre}
A. Blais, R.-S. Huang, A. Wallraff, S. M. Girvin and R. J. Schoelkopf, Cavity quantum electrodynamics for superconducting electrical circuits: An architecture for quantum computation, Phys. Rev. A 69, 062320 (2004).

\bibitem{wallraff}
A. Wallraff, D. I. Schuster, A. Blais, L. Frunzio, R.- S. Huang, J. Majer, S. Kumar, S. M. Girvin and R. J. Schoelkopf, Circuit Quantum Electrodynamics: Coherent Coupling of a Single Photon to a Cooper Pair Box, Nature 431,162-167 (2004).

\bibitem{inputoutputRMP}
A. A. Clerk, M. H. Devoret, S. M. Girvin, F. Marquardt, and R. J. Schoelkopf, Introduction to Quantum Noise, Measurement and Amplification,  Rev. Mod. Phys. 82, 1155 (2010).  

\bibitem{Rabi}
I. I. Rabi, On the Process of Space Quantization, Phys. Rev. 49, 324 (1936);  I. I. Rabi, Space Quantization in a Gyrating Magnetic Field, Phys. Rev. 51, 652 (1937).

\bibitem{Devoret}
V. Bouchiat, D. Vion, Ph. Joyez, D. Esteve and M. H. Devoret, Quantum Coherence with a Single Cooper Pair, Physica Scripta, Volume T76, pp. 165-170 (1998). 

\bibitem{Nakamura}
Y. Nakamura, Yu. A. Pashkin, and J. S. Tsai, Coherent control of macroscopic quantum states in a single-Cooper-pair box, Nature 398, 786-788 (1999).

\bibitem{Schonreview}
G. Ithier, E. Collin, P. Joyez, P. J. Meeson, D. Vion, D. Esteve, F. Chiarello, A. Shnirman, Y. Makhlin, J. Schriefl, and G. Sch\" on, Decoherence in a superconducting quantum bit circuit, Phys. Rev. B 72, 134519 (2005).

\bibitem{Buisson}
O. Buisson and F. W. J. Hekking, Entangled states in a Josephson charge qubit coupled to a
superconducting resonator, proceedings of the conference on Macroscopic Quantum Coherence and Computing, Naples, Italy, June 2000.

\bibitem{MooijSolano}
P. Forn-Diaz, J. Lisenfeld, D. Marcos, J. J. Garcia-Ripoll, E. Solano, C. J. P. M. Harmans, J. E. Mooij,  Observation of the Bloch-Siegert Shift in a Qubit-Oscillator System in the Ultrastrong Coupling Regime, Phys. Rev. Lett. 105, 237001 (2010).

\bibitem{solano2}
T. Niemczyk, F. Deppe,  F. Deppe, H. Huebl, E. P. Menzel, F. Hocke, M. J. Schwarz, J. J. Garcia-Ripoll, D. Zueco, T. H\" ummer, E. Solano, A. Marx, R. Gross, Circuit quantum electrodynamics in the ultrastrong-coupling regime, Nature Physics
    6, 772-776 (2010).
  
\bibitem{Babelon}
O. Babelon, L. Cantini and B. Dou\c{c}ot, A semiclassical study of the Jaynes-Cummings Model, J. Stat. Mech. (2009) P07011.

\bibitem{Babelon2}
O. Babelon and B. Dou\c{c}ot, Classical Bethe Ansatz and Normal Forms in the Jaynes-Cummings Model, arXiv:1106.3274.

\bibitem{Faribault}
H. Tschirhart and A. Faribault, Algebraic Bethe Ans\" atze and eigenvalue-based determinants for Dicke-Jaynes-Cummings-Gaudin quantum integrable models,  J. Phys. A: Math. Theor. 47 405204 (2014).

\bibitem{Schweber}
S. Schweber, On the application of Bargmann Hilbert spaces to dynamical problems, Ann. Phys. (N.Y.) 41, 205 (1967).

\bibitem{Irish}
E. K. Irish,  J. Gea-Banacloche, I. Martin and K. C. Schwab, Dynamics of a two-level system strongly coupled to a high-frequency quantum oscillator, Phys. Rev. B 72, 195410 (2005). 

\bibitem{Casanova}
J. Casanova, G. Romero, I. Lizuain, J. J. Garcia-Ripoll, and E. Solano, Deep Strong Coupling Regime of the Jaynes-Cummings model, Phys. Rev. Lett. 105, 263603 (2010).

\bibitem{Dicke}
R. H. Dicke, Coherence in Spontaneous Radiation Processes, Phys. Rev. 93 99-110 (1954). 

\bibitem{LiebHepp}
K. Hepp and E. H. Lieb, On the superradiant phase transition for molecules in a quantized radiation field: the dicke maser model, Ann. Phys. 76 360-404, 1973.

\bibitem{Dickeglass}
Philipp Strack, Subir Sachdev, Dicke quantum spin glass of atoms and photons, Phys. Rev. Lett. 107, 277202 (2011).

\bibitem{Nataf}
P. Nataf and C. Ciuti, Is there a no-go theorem for superradiant quantum phase transitions in cavity and circuit QED?, Nat. Commun. 1 72 (2010).

\bibitem{MarquardtDicke}
O. Viehmann, J. von Delft and F. Marquardt, Superradiant Phase Transitions and the Standard Description of Circuit QED, Phys. Rev. Lett. 107, 113602 (2011). 

\bibitem{NatafMehmet}
P. Nataf, M. Dogan and K. Le Hur, Heisenberg uncertainty principle as a probe of entanglement entropy: Application to superradiant quantum phase transitions, Phys. Rev. A, 86, 043807 (2012).

\bibitem{Vidal}
 S. Dusuel and J. Vidal, Finite-Size Scaling Exponents of the Lipkin-Meshkov-Glick Model, Phys. Rev. Lett. 93, 237204 (2004).
 
 \bibitem{Francis}
 H. Francis Song, Stephan Rachel, Christian Flindt, Israel Klich, Nicolas Laflorencie, and Karyn Le Hur, Bipartite Fluctuations as a Probe of Many-Body Entanglement, Phys. Rev. B 85, 035409 (2012), Editors' Suggestion.
 
 \bibitem{KlichLevitov}
 Israel Klich, Leonid Levitov, Quantum Noise as an Entanglement Meter, Phys. Rev. Lett. 102, 100502 (2009).

\bibitem{JaynesCummings}
E. T. Jaynes and F. W. Cummings, Comparison of quantum and semiclassical radiation theories with application to the beam maser, Proc. IEEE 51 89, 1963.

\bibitem{Imamoglublockade}
A. Imamoglu, H. Schmidt, G. Woods, and M. Deutsch, Strongly Interacting Photons in a Nonlinear Cavity, Phys. Rev. Lett. 79, 1467 (1998).

\bibitem{Verger}
A. Verger, C. Ciuti and I. Carusotto, Polariton quantum blockade in a photonic dot, Phys. Rev. B 73, 193306 (2006).

\bibitem{Blais}
M. Boissonneault, J. Gambetta and A. Blais, Dispersive Regime of CQED: photon-dependent qubit dephasing and relaxation rates,  Phys. Rev. A 79, 013819 (2009).

\bibitem{Kimble}
K. M. Birnbaum {\it et al.}, Photon blockade in an optical cavity with one trapped atom, Nature 436, 87 (2005).

\bibitem{Bishop}
L. S. Bishop {\it et al.}, Nonlinear response of the vacuum Rabi resonance, Nature Phys. 5, 105-109 (2008).

\bibitem{Fink}
J. M. Fink {\it et al.}, Climbing the Jaynes-Cummings ladder and observing its $\sqrt{n}$ nonlinearity in a cavity QED system, Nature 454, 315-8 (2008).

\bibitem{Hofheinz}
M. Hofheinz {\it et al.}, Generation of Fock states in a superconducting quantum circuit, Nature 454, 310-4 (2008).

\bibitem{Hoffman}
A. J. Hoffman, S. J. Srinivasan, S. Schmidt, L. Spietz, J. Aumentado, H. E. T\" ureci, and A. A. Houck, Dispersive Photon Blockade in a Superconducting Circuit, Phys. Rev. Lett. 107, 053602 (2011)

 \bibitem{cohenDupont}
C. Cohen-Tannoudji, J. Dupont-Roc and C. Fabre, A quantum calculation of the higher order terms in the Bloch-Siegert shift, J. Phys. B: Atom. Molec. Phys., Vol. 6, August 1973. 

\bibitem{HofheinzMartinis}
M. Hofheinz {\it et al.}, Synthesizing arbitrary quantum states in a superconducting resonator, Nature 459, 546-549 (28 May 2009).

\bibitem{Gardiner}
C. W. Gardiner and M. J. Collett, Input and ouptut in damped quantum systems: Quantum Stochastic Differential Equations and the Master Equation, Phys. Rev. A {\bf 31}, 3761 (1985).

\bibitem{FV}
 R. P. Feynman and F. L. Vernon, The Theory of a  General Quantum System Interacting with a Linear Dissipative System, Ann. Phys. (N.Y.), 24 118 (1963).

\bibitem{Caldeira_Leggett}
A. O. Caldeira and A. J. Leggett, Path integral approach to quantum Brownian motion, Physica 121A: 587 (1983).

\bibitem{Lindblad}
G. Lindblad, On the generators of quantum dynamical semigroups, Commun. Math. Phys. 48, 119 (1976).

\bibitem{Bloch}
F. Bloch, Generalized Theory of Relaxation, Physcal Review 105, 1206 (1957).

\bibitem{Redfield}
A. G. Redfield, On the theory of relaxation processes, IBM Journal of Research and Development 1, 19 (1957).

\bibitem{Blume}
M. Blume, V. J. Emery and A. Luther, Spin-Boson Systems: One-Dimensional Equivalents and the Kondo Problem, Phys. Rev. Lett. 25, 450 (1970).

\bibitem{KLH0}
K. Le Hur, Entanglement entropy, decoherence, and quantum phase transitions of a dissipative two-level system, Annals Phys. (NY) 323, 2208 (2008).

\bibitem{Vojta}
M. Vojta, Impurity Quantum Phase Transitions, Phil Mag 86, 1807 (2006).

\bibitem{Spohn}
R. D\" umcke and H. Spohn, Quantum tunneling with dissipation and the Ising model over R, J. Stat. Phys. 41, 389 (1985).

\bibitem{AYH}
P. W. Anderson, G. Yuval and D. R. Hamann, Exact Results in the Kondo Problem. II. Scaling Theory, Qualitatively Correct Solution, and Some New Results on One-Dimensional Classical Statistical Models, Phys. Rev. B 1, 4464 (1970).

\bibitem{Chakravarty0}
S. Chakravarty, Quantum Fluctuations in the Tunneling between Superconductors, Phys. Rev. Lett. 49, 681 (1982).

\bibitem{Bray}
A. J. Bray and M. A. Moore, Influence of Dissipation on Quantum Coherence, Phys. Rev. Lett. 49, 681 (1982).

\bibitem{Jezouin}
S. Jezouin, M. Albert, F. D. Parmentier, A. Anthore, U. Gennser, A. Cavanna, I. Safi, F. Pierre, Tomonaga-Luttinger physics in electronic quantum circuits, Nat. Commun. 4 1802 (2013).

\bibitem{Finkelstein}
H. T. Mebrahtu, I. V. Borzenets, D. E. Liu, H. Zheng, Y. V. Bomze, A. I. Smirnov, H. U. Baranger, G. Finkelstein, Quantum Phase Transition in a Resonant Level Coupled to Interacting Leads, Nature 488, p. 61 (2012).

\bibitem{KLH}
K. Le Hur, Coulomb Blockade of a Noisy Metallic Box: A Realization of Bose-Fermi Kondo Models, Phys. Rev. Lett. 92 196804 (2004); K. Le Hur and M.-R. Li, Unification of electromagnetic noise and Luttinger liquid via a quantum dot,
Phys. Rev. B 72, 073305 (2005); M.-R. Li, K. Le Hur and W. Hofstetter, Hidden Caldeira-Leggett dissipation in a Bose-Fermi Kondo model, Phys. Rev. Lett. 95, 086406 (2005).

\bibitem{SafiSaleur}
See also, I. Safi and H. Saleur, A one-channel conductor in an ohmic environment: mapping to a TLL and full counting statistics, Phys. Rev. Lett. 93, 126602 (2004).

\bibitem{Zarand}
L. Borda,  G. Zarand and P. Simon, Dissipation-induced quantum phase transition in a quantum box, Phys. Rev. B 72, 155311 (2005).

\bibitem{David}
D. Goldhaber-Gordon, H. Shtrikman, D. Mahalu, D. Abusch-Magder, U. Meirav and M. A. Kastner, Nature 391, 156-159 (1998).

\bibitem{Leo}
L. P. Kouwenhoven and C. M. Marcus, Quantum dots, Physics World 11 35-39 (1998).

\bibitem{LeonidLeo}
Leo Kouwenhoven, Leonid Glazman, Revival of the Kondo effect, Physics World, 14 33-38 (2001).

\bibitem{Buettiker}
P. Cedraschi and M. B\"uttiker, Quantum Coherence of the Ground State of a Mesoscopic Ring, Annals of Physics (NY) 289, 1 - 23 (2001).

\bibitem{FurusakiMatveev}
A. Furusaki and K. Matveev, Occupation of a resonant level coupled to a chiral Luttinger liquid, Phys. Rev. Lett. 88, 226404 (2002).

\bibitem{Toulouse}
G. Toulouse, Expression exacte de l'\' energie de l'\' etat de base de l'hamiltonien de Kondo pour une valeur particuli\`ere de Jz, C. R. Acad. Sci. Paris 268 1200 (1969).

\bibitem{Guinea}
F. Guinea, V. Hakim and A. Muramatsu, Bosonization of a two-level system with dissipation, Phys. Rev. B 32, 4410 (1985).

\bibitem{ALJ}
I. Affleck, A. A. Ludwig and B. A. Jones, Conformal-field-theory approach to the two-impurity Kondo problem: Comparison with numerical renormalization-group results, Phys. Rev. B 52 9528 (1995). 

\bibitem{Garst}
M. Garst,  S. Kehrein, T. Pruschke, A. Rosch and M. Vojta, Quantum phase transition of Ising-coupled Kondo impurities, Phys. Rev. B 69, 214413 (2004).

\bibitem{PeterDavid}
P. P. Orth, D. Roosen, W. Hofstetter and K. Le Hur, Dynamics, Synchronization and Quantum Phase Transitions of Two Dissipative Spins, Phys. Rev. B 82, 144423 (2010).

\bibitem{Raftery}
J. Raftery, D. Sadri, S. Schmidt, H. E. T\"{u}reci and A. A. Houck, Observation of a Dissipation-Induced Classical to Quantum Transition, Phys. Rev. X 4, 031043.

\bibitem{KarynBernard}
Karyn Le Hur and Bernard Coqblin, The underscreened Kondo effect: a two S=1 impurity model, Phys. Rev. B, 56 668 (1997).

\bibitem{Karynunder}
Karyn Le Hur, The underscreened Kondo effect in ladder systems, Phys. Rev. Lett., 83 848 (1999).

\bibitem{ChungHou}
C.-H. Chung, K. Le Hur, M. Vojta and P. W\" olfle, Non-equilibrium transport at a dissipative quantum phase transition, Phys. Rev. Lett. 102, 216803 (2009).

\bibitem{Carmichael}
H. Carmichael, An open system approach to Quantum Optics, (Springer, Berlin) (1994).

\bibitem{Matteo}
M. Carrega, P. Solinas, A. Braggio, M. Sassetti and U. Weiss, Functional Integral approach to time-dependent heat exchange in open quantum systems: general method and applications, arXiv:1412.6991

\bibitem{CDM}
J. Dalibard, I. Castin and K. Molmer, Wave-function approach to dissipative processes in quantum optics, Phys. Rev. Lett. 68, 580 (1992).

\bibitem{Dum}
R. Dum, P. Zoller, H. Ritsch, Monte Carlo simulation of the atomic master equation for spontaneous emission, Phys. Rev. A 45, 4879 (1992).

\bibitem{Gisin}
W. T. Strunz, L. Diosi and N. Gisin, Open System Dynamics with Non-Markovian Quantum Trajectories, Phys. Rev. Lett. 82, 1801 (1999).

\bibitem{AndersSchiller}
F. B. Anders and A. Schiller, Spin Precession and Real Time Dynamics in the Kondo Model: A Time-Dependent Numerical Renormalization-Group Study,  Phys. Rev. B 74, 245113 (2006).

\bibitem{Bulla}
A. B. Anders, R. Bulla and M. Vojta, Equilibrium and non-equilibrium dynamics of the sub-ohmic spin-boson model, Phys. Rev. Lett. 98, 210402 (2007).

\bibitem{Florenspol}
Soumya Bera, Ahsan Nazir, Alex W. Chin, Harold U. Baranger, Serge Florens, A generalized multi-polaron expansion for the spin-boson model: Environmental entanglement and the biased two-state system, Phys. Rev. B 90, 075110 (2014).

\bibitem{Pollet2}
Z. Cai, U. Schollwoeck and L. Pollet, Identifying a bath-induced Bose liquid in interacting spin-boson models, Phys. Rev. Lett. 113, 260403 (2014). 

\bibitem{Solano}
E. Sanchez-Burillo, D. Zueco, J. J. Garcia-Ripoll, and L. Martin-Moreno, Scattering in the Ultrastrong Regime: Nonlinear Optics with One Photon, Phys. Rev. Lett. 113, 263604 (2014).

\bibitem{Garrahan}
Igor Lesanovsky, Merlijn van Horssen, Madalin Guta, Juan P. Garrahan, Characterization of dynamical phase transitions in quantum jump trajectories beyond the properties of the stationary state, Phys. Rev. Lett. 110, 150401 (2013).

\bibitem{MarcoMC}
Marco Schir\' o, Michele Fabrizio, Real-Time Diagrammatic Monte Carlo for Nonequilibrium Quantum Transport, Phys. Rev. B 79, 153302 (2009).

\bibitem{WernerMillis}
Philipp Werner, Takashi Oka, Andrew J. Millis, Diagrammatic Monte Carlo simulation of non-equilibrium systems, Phys. Rev. B 79, 035320 (2009).

\bibitem{Thomas}
T. L. Schmidt, P. Werner, L. Muehlbacher, A. Komnik, Transient dynamics of the Anderson impurity model out of equilibrium, Phys. Rev. B 78, 235110 (2008).

\bibitem{Waintal}
Rosario E. V. Profumo, Christoph Groth, Laura Messio, Olivier Parcollet, Xavier Waintal, Quantum Monte-Carlo for correlated out-of-equilibrium nanoelectronics devices, arXiv:1504.02132.

\bibitem{BauerBernard}
M. Bauer, D. Bernard and A. Tilloy, The Open Quantum Brownian Motion, J. Stat. Mech. P09001 (2014).

\bibitem{Schoeller}
D. M. Kennes, O. Kashuba, M. Pletyukhov, H. Schoeller and V. Meden, {Oscillatory dynamics and non-markovian memory in dissipative quantum systems}  Phys. Rev. Lett. 110, 100405 (2013). 

\bibitem{LoicKaryn}
Loic Henriet and Karyn Le Hur, Many-Body Stochastic Dynamics: Quenches in Dissipative Quantum Spin Arrays,  arXiv:1502.06863.

\bibitem{Keeling}
G. Kulaitis, F. Kr\" uger, F. Nissen and J. Keeling, Disordered driven coupled cavity arrays: Non-equilibrium stochastic mean-field theory, Phys. Rev. A 87, 013840 (2013).

\bibitem{Hakan}
S. Mandt, D. Sadri, A. A. Houck and H. T\"{u}reci, Stochastic Differential Equations for Quantum Dynamics of Spin-Boson Networks, arXiv:1410.3142.

\bibitem{Vavilov}
C. Xu, A. Poudel and M. G. Vavilov, Nonadiabatic Dynamics of a Dissipative Two-level System, Phys. Rev. A 89, 052102 (2014). 
   
\bibitem{SG}
 J. T. Stockburger and H. Grabert,  Exact c-number Representation of Non-Markovian Quantum Dissipation, Phys. Rev. Lett. 88, 170407 (2002).
 
 \bibitem{QMC2spins}
 A. Winter and H. Rieger, Quantum phase transition and correlations in the multi-spin-boson model, Phys. Rev. B 90, 224401 (2014).
 
\bibitem{Josephson}
B. D. Josephson, Possible new effects in superconductive tunnelling, Physics Letters 1, 251 (1962).

\bibitem{AmbegaokarBert}
V. Ambegaokar and B. I. Halperin, Voltage Due to Thermal Noise in the dc Josephson Effect, Phys. Rev. Lett. 22, 1364 (1969).

\bibitem{superconductors} 
P. W. Anderson and J. M. Rowell, Probable Observation of the Josephson Tunnel Effect, Phys. Rev. Letters 10 230 (1963).

\bibitem{Shapiro}
S. Shapiro, Josephson Currents in Superconducting Tunneling: The Effect of Microwaves and Other Observations, Phys. Rev. Lett. 11, 80 (1963).

\bibitem{Fulton}
T. A. Fulton {\it et al.}, Observation of combined Josephson and charging effects in small tunnel junction circuits, Phys. Rev. Lett. 63, 1307 (1989).

\bibitem{BECJosephson}
M. R. Andrews, C. G. Townsend, H. J. Miesner, D. S. Durfee, D. M. Kurn, and W. Ketterle, Observation of interference between two Bose condensates, Science 275, 637 (1997).

\bibitem{Markus}
M. Albiez {\it et al.} Direct Observation of Tunneling and Nonlinear Self-Trapping in a single Bosonic Josephson Junction, Phys. Rev. Lett. 95, 010402 (2005).

\bibitem{Helium}
K. Sukhatme {\it et al.}, Observation of the ideal Josephson effect in superfluid $4^{He}$, Nature 411, 280-283 (2001). 

\bibitem{Jacqueline}
M. Abbarchi {\it et al.} Macroscopic quantum self-trapping and Josephson oscillations of exciton-polaritons, Nature Physics 9, 275 (2013).

\bibitem{Recati}
A. Recati, P.O. Fedichev, W. Zwerger, J. von Delft, and P. Zoller, Atomic quantum dots coupled to BEC reservoirs, Phys. Rev. Lett. 94, 040404 (2005).

\bibitem{PeterIvan}
Peter P. Orth, Ivan Stanic, Karyn Le Hur, Dissipative Quantum Ising model in a cold atomic spin-boson mixture, Phys. Rev. A 77, 051601(R) (2008).

\bibitem{Meirong}
M.-R Li and K. Le Hur, Double-dot charge qubit and transport via dissipative cotunneling, Phys. Rev. Lett. 93, 176802 (2004).

\bibitem{jens}
J. Koch and K. Le Hur, Discontinuous current-phase relations in small 1D Josephson junction arrays, Phys. Rev. Lett. 101, 097007 (2008).

\bibitem{Baranger}
H. Zheng, D. J. Gauthier and H. Baranger, Waveguide QED: Many-Body Bound State Effects on Coherent and Fock State Scattering from a Two-Level System, Phys. Rev. A 82, 063816 (2010). 

\bibitem{Florensnew}
Izak Snyman, Serge Florens, Josephson-Kondo screening cloud in circuit quantum electrodynamics,  arXiv:1503.05708.

\bibitem{DemlerSalomon}
Johannes Bauer, Christophe Salomon, Eugene Demler, Realizing a Kondo-correlated state with ultracold atoms, Phys. Rev. Lett. 111, 215304 (2013).

\bibitem{Demler2}
Michael Knap, Dmitry A. Abanin, Eugene Demler, Dissipative dynamics of a driven quantum spin coupled to a bath of ultracold fermions,  Phys. Rev. Lett. 111, 265302 (2013).

\bibitem{Saclay}
C. Altimiras, O. Parlavecchio, Ph. Joyez, D. Vion, P. Roche, D. Esteve and F. Portier, Fluctuation-dissipation relations of a tunnel junction driven by a quantum circuit, Applied Physics Letters, 103 212601 (2013).

\bibitem{SolanoSB}
Max Haeberlein, Frank Deppe, Andreas Kurcz, Jan Goetz, Alexander Baust, Peter Eder, Kirill Fedorov, Michael Fischer, Edwin P. Menzel, Manuel J. Schwarz, Friedrich Wulschner, Edwar Xie, Ling Zhong, Enrique Solano, Achim Marx, Juan-Jos\' e Garcia-Ripoll, and Rudolf Gross, Spin-boson model with an engineered reservoir in circuit quantum electrodynamics,  arXiv:1506.09114.

\bibitem{Pashkin}
Y. A. Pashkin,  T. Yamamoto, O. Astafiev, Y. Nakamura, D. V. Averin, J. S. Tsai, Quantum oscillations in two coupled charge qubits, Nature (London) 421, 823 (2003).

\bibitem{Lafarge}
E. Bibow, P. Lafarge and L. Levy, Resonant Cooper Pair Tunneling through a Double-Island Qubit, Phys. Rev. Lett. 88, 017003 (2001).

\bibitem{Astafiev}
O. Astafiev, A. M. Zagoskin, A. A. Abdumalikov Jr, Yu. A. Pashkin, T. Yamamoto, K. Inomata, Y. Nakamura and J.-S. Tsai, Resonance fluorescence of a single artificial atom, Science 327, 840 (2010).

\bibitem{spinbosonKorringa}
M. Sassetti and U. Weiss, Universality in the dissipative two-state system, Phys. Rev. Lett. 65, 2262 (1990).

\bibitem{Shiba}
H. Shiba, The Korringa Relation for the Impurity Nuclear Spin-Lattice Relaxation in Dilute Kondo Alloys, Prog. Theor. Phys. 54, 967 (1975).

\bibitem{Leonid}
M. Garst, P. Wolfle, L. Borda, J. von Delft, and L. Glazman, Energy-resolved inelastic electron scattering off a magnetic impurity, Phys. Rev. B 72, 205125 (2005).

\bibitem{ChristopheKaryn}
C. Mora and K. Le Hur, Universal Resistances of the Quantum RC circuit, Nature Physics 6, 697 (2010).

\bibitem{MichelePRL}
M. Filippone, K. Le Hur and C. Mora, Giant Charge Relaxation Resistance in the Anderson Model, Phys. Rev. Lett. 107, 176601 (2011).

\bibitem{Michele}
M. Filippone and C. Mora, Fermi liquid approach to the quantum RC circuit: renormalization-group analysis of the Anderson and Coulomb blockade models, Phys. Rev. B 86, 125311 (2012).

\bibitem{ButtikerRC}
M. B\" uttiker, A. Pr\^etre, and H. Thomas, Dynamic Conductance and the Scattering Matrix of small Conductors, Phys. Rev. Lett. 70, 4114 (1993); 
M. B\" uttiker, H. Thomas, and A. Pr\^etre, Mesoscopic capacitors, Phys. Lett. A 180, 364 (1993).

\bibitem{ButtikerRC2}
S. E. Nigg, R. Lopez and M. B\" uttiker, Mesoscopic Charge Relaxation, Phys. Rev. Lett. 97, 206804 (2006).

\bibitem{Feve}
J. Gabelli et al., Violation of Kirchoff's Laws for a Coherent RC Circuit, Science 313, 499 (2006); G. F\`eve et al., An On-Demand Coherent Single Electron Source, Science 316, 1169 (2007).

\bibitem{Gabelli}
J. Gabelli, G. F\`eve, J.-M. Berroir and B. Pla\c{c}ais, A Coherent RC Circuit, Rep. Prog. Phys. 75 126504 (2012).

\bibitem{Martin}
Y. Hamamoto, T. Jonckheere, T. Kato and T. Martin, Dynamic response of a mesoscopic capacitor in the presence of strong electron interactions, Phys. Rev. B 81, 153305 (2010).

\bibitem{Etzioni}
Y. Etzioni, B. Horovitz and P. Le Doussal, Rings and boxes in dissipative environments,  Phys. Rev. Lett. 106, 166803 (2011) and Phys. Rev. B 86, 235406 (2012).

\bibitem{Dutt}
P. Dutt, T. L. Schmidt, C. Mora and K. Le Hur, Strongly correlated dynamics in multichannel quantum RC circuits, Phys. Rev. B 87, 155134 (2013).

\bibitem{Texier}
C. Texier, Wigner time delay and related concepts - Application to transport in coherent conductors,  arXiv:1507.00075.

\bibitem{NozieresBlandin}
P. Nozi\`eres and A. Blandin,  Kondo effect in real metals, J. Phys. 41, 193 (1980).

\bibitem{DavidPotok}
R. M. Potok, I. G. Rau, H. Shtrikman, Y. Oreg and D. Goldhaber-Gordon, Observation of the two-channel Kondo effect, Nature 446, 167 (2007).
 
\bibitem{Gleb}
H. T. Mebrahtu {\it et al.}, Observation of Majorana Quantum Critical Behavior in a Resonant Level Coupled to a Dissipative Environment, Nature Physics 9, 732 (2013). 

\bibitem{Keller}
A. J. Keller {\it et al.}, Universal Fermi liquid crossover and quantum criticality in a mesoscopic device,  Nature 526, 237-240 (2015).

\bibitem{Frederic}
Z. Iftikhar {\it et al.}, Two-channel Kondo effect and renormalization flow with macroscopic quantum charge states, Nature 526, 233-236 (2015). 

\bibitem{2CK1}
C. Mora and K. Le Hur, Probing dynamics of Majorana fermions in quantum impurity systems, Phys. Rev. B 88, 241302 (2013).

\bibitem{2CK2}
P. Dutt, T. L. Schmidt, C. Mora and K. Le Hur, Strongly correlated dynamics in multichannel quantum RC circuits, Phys. Rev. B 88, 241302 (2013).

\bibitem{Majer}
J. Majer {\it et al.} Coupling Superconducting Qubits via a Cavity Bus,  Nature 449, 443-447 (2007). 

\bibitem{Kontos}
M. R. Delbecq, L. E. Bruhat, J. J. Viennot, S. Datta, A. Cottet, and T. Kontos, Photon mediated interaction between distant quantum dot circuits, Nature Communications 4, 1400 (2013).

\bibitem{Deng}
G.-W. Deng {\it et al.}, Coupling two distant double quantum dots to a microwave resonator, arXiv:1409.4980. 

\bibitem{DiCarlo}
L. DiCarlo {\it et al.}, Demonstration of Two-Qubit Algorithms with a Superconducting Quantum Processor, Nature 460, 240-244 (2009). 

\bibitem{Michel}
S. Shankar {\it et al.}, Stabilizing entanglement autonomously between two superconducting qubits, Nature, 504 419-422 (2013).

\bibitem{Aspect}
A. Aspect, P. Grangier and G. Roger, Experimental Realization of Einstein-Podolsky-Rosen-Bohm Gedankenexperiment: A New Violation of Bell's Inequalities, Phys. Rev. Lett. 49 (2) 91-4 (1982); A. Aspect, J. Dalibard, G. Roger, Experimental Test of Bell's Inequalities Using Time-Varying Analyzers, Phys. Rev. Lett. 49 (25) 1804-7 (1982).

\bibitem{CooperBCS}
L. N. Cooper, Bound electron pairs in a degenerate Fermi gas, Physical Review, vol. 104, 1189-1190 (1956).

\bibitem{PascaleLoic}
Christophe Arnold, Justin Demory, Vivien Loo, Aristide Lema\^itre, Isabelle Sagnes, Mikha\" il Glazov, Olivier Krebs, Paul Voisin, Pascale Senellart, Lo\" ic Lanco, Macroscopic Polarization Rotation Induced by a Single Spin, Nature Communications (2015).

\bibitem{Takis2}
J. J. Viennot, M. C. Dartiailh, A. Cottet, T. Kontos, Coherent coupling of a single spin to microwave cavity photons, Science 349, 408 (2015).

\bibitem{LPA}
E. Bocquillon,  V. Freulon, F.D. Parmentier, J.-M Berroir, B. Pla\c{c}ais, C. Wahl, J. Rech, T. Jonckheere, T. Martin, C. Grenier, D. Ferraro, P. Degiovanni, G. F\`eve, Electron quantum optics in ballistic chiral conductors,  Annalen der Physik, 526, 1 (2014).

\bibitem{Anderson}
P. W. Anderson, Model for the Electronic Structure of Amorphous Semiconductors, Phys. Rev. Lett. 34, 953 (1975).

\bibitem{Holstein}
T. Holstein, Annals of Physics 8, 325 (1959), ISSN 0003-4916.

\bibitem{Mitra}
A. Mitra, I. Aleiner and A. J. Millis, Phonon effects in molecular transistors: Quantum and classical treatment, Phys. Rev. B 69, 245302 (2004).

\bibitem{Grempel}
P. S. Cornaglia, H. Ness and D. R. Grempel, Many Body Effects on the Transport Properties of Single-Molecule Devices, Phys. Rev. Lett. 93, 147201 (2004).

\bibitem{Vinkler}
Y. Vinkler, A. Schiller and N. Andrei, Quantum quenches and driven dynamics in a single-molecule device, Phys. Rev. B 85, 035411 (2012).

\bibitem{Deng2}
G.-W. Deng, L. Henriet, D. Wei, S.-X. Li, H.-O. Li, G. Cao, M. Xiao, G.-C. Guo, M. Schiro, K. Le Hur, G.-P. Guo, A Quantum Electrodynamics Kondo Circuit with Orbital and Spin Entanglement, arXiv:1509.06141.

\bibitem{Audrey}
A. Cottet, T. Kontos and B. Dou\c{c}ot, Electron-photon coupling in Mesoscopic Quantum Electrodynamics,  Phys. Rev. B 91, 205417 (2015).

\bibitem{Olesia}
O. Dmytruk, M. Trif, C. Mora and P. Simon, Cavity quantum electrodynamics with an out-of-equilibrium quantum dot,  arXiv:1510.03748.

\bibitem{Hennessy}
K. Hennessy, A. Badolato, M. Winger, D. Gerace, M. Atat\" ure, S Gulde, S F\" alt, E. L Hu, A Imamoglu, Quantum nature of a strongly coupled single quantum dot-cavity system, 
Nature 445, 896-899 (22 February 2007).

\bibitem{Kozinsky}
I. Kozinsky, H. W. Ch. Postma, O. Kogan, A. Husain, and M. L. Roukes, Basins of Attraction of a Nonlinear Nanomechanical Resonator, Phys. Rev. Lett. 99, 207201 (2007).

\bibitem{Samuelsson}
Christian Bergenfeldt, Peter Samuelsson, Bj\" orn Sothmann, Christian Flindt, Markus B\" uttiker, Hybrid Microwave-Cavity Heat Engine, Phys. Rev. Lett. 112, 076803 (2014).

\bibitem{Jordan}
Bj\" orn Sothmann, Rafael Sanchez, Andrew N. Jordan, Thermoelectric energy harvesting with quantum dots, Nanotechnology 26, 032001 (2015).

\bibitem{Nanoengine}
L. Henriet, A. N. Jordan and K. Le Hur, Electrical Current from Quantum Vacuum Fluctuations in Nano-Engines,  Phys. Rev. B 92, 125306 (2015).

\bibitem{H1}
M. Kulkarni, O. Cotlet, H. E. Tureci, Cavity-coupled double-quantum dot at finite bias: analogy with lasers and beyond, arXiv:1403.3075

\bibitem{H2}
B. Sbierski, M. Hanl, A. Weichselbaum, H. E. T\" ureci, M. Goldstein, L. I. Glazman, J. von Delft, and A. Imamoglu, Proposed Rabi-Kondo Correlated State in a Laser-Driven Semiconductor Quantum Dot, Phys. Rev. Lett. 111, 157402 (2013)

\bibitem{Berkeleypolaron}
F. Mei, V. M. Stojanovic, I. Siddiqi and L. Tian, Analog Superconducting Quantum Simulator for Holstein Polarons, Phys. Rev. B 88, 224502 (2013).

\bibitem{Alicea}
For recent reviews: Jason Alicea, New directions in the pursuit of Majorana fermions in solid state systems, Rep. Prog. Phys. 75, 076501 (2012). See also C. W. J. Beenakker, Search for Majorana fermions in superconductors, Annu. Rev. Con. Mat. Phys. 4, 113 (2013).

\bibitem{Kitaev}
A. Kitaev, Unpaired Majorana fermions in quantum wires, Phys.-Usp. 44 131 (2001).

\bibitem{Read}
N. Read and D. Green, Paired states of fermions in two dimensions with breaking of parity and time-reversal symmetries, and the fractional quantum Hall effect, Phys. Rev. B 61 10267 (2000).

\bibitem{Schmidt}
T. L. Schmidt, A. Nunnenkamp and C. Bruder, Majorana qubit rotations in microwave cavities, Phys. Rev. Lett. 110, 107006 (2013).

\bibitem{Nunnenkamp}
T. L. Schmidt, A. Nunnenkamp and C. Bruder, Microwave-controlled coupling of Majorana bound states, New J. Phys. 15, 025043 (2013).

\bibitem{Cottet}
A. Cottet, T. Kontos and B. Dou\c{c}ot, Squeezing light with Majorana fermions, Phys. Rev. B 88, 195415 (2013).

\bibitem{Trif}
M. Trif and Y. Tserkovnyak, Resonantly Tunable Majorana Polariton in a Microwave Cavity, Phys. Rev. Lett. 109, 257002 (2012).

\bibitem{Simon}
O. Dmytruk, M. Trif and P. Simon, Cavity quantum electrodynamics with mesoscopic topological superconductors,  arXiv:1502.03082. 

\bibitem{GinossarGrosfeld}
E. Ginossar and E. Grosfeld, Tunability of microwave transitions as a signature of coherent parity mixing effects in the Majorana-Transmon qubit, Nat. Commun. 5, 4772 (2014); K. Yavilberg, E. Ginossar and E. Grosfeld, Fermion parity measurement and control in Majorana circuit quantum electrodynamics, arXiv:1411.5699.


\bibitem{ManuelJulia}
D. M. Badiane, L. I. Glazman, M. Houzet and J. S. Meyer, Ac Josephson Effect in Topological Josephson Junctions, C.R. Physique 14, 840 (2013).

\bibitem{Beri}
B. B\' eri and N. R. Cooper,  Topological Kondo effect with Majorana fermions,  Phys. Rev. Lett. 109, 156803 (2012).

\bibitem{Altland}
A. Altland,  B. Beri, R. Egger, A. M. Tsvelik, Bethe ansatz solution of the topological Kondo model, J. Phys. A. 47, 265001 (2014).

\bibitem{Erik}
Erik Eriksson, Christophe Mora, Alex Zazunov, Reinhold Egger, Non-Fermi liquid manifold in a Majorana device, Phys. Rev. Lett. 113, 076404 (2014).

\bibitem{Tsvelikmajo}
A. Altland, B. Beri, R. Egger, A.M. Tsvelik, Multi-channel Kondo impurity dynamics in a Majorana device, Phys. Rev. Lett. 113, 076401 (2014).

\bibitem{CarusottoF}
I. Carusotto, D. Gerace, H. E. Tureci, S. De Liberato, C. Ciuti, and A. Imamoglu, Fermionized Photons in an Array of Driven Dissipative Nonlinear Cavities, Phys. Rev. Lett. 103, 033601 (2009).

\bibitem{Atac}
C.-E. Bardyn and A. Imamoglu, Majorana-Like Modes of Light in a One-Dimensional Array of Nonlinear Cavities, Phys. Rev. Lett. 109, 253606 (2012).

\bibitem{RochLPA}
N. Roch, E. Flurin, F. Nguyen, P. Morfin, P. Campagne-Ibarcq, M. H. Devoret and B. Huard, Widely tunable, non-degenerate three-wave mixing microwave device operating near the quantum limit,  Phys. Rev. Lett. 108, 147701 (2012).

\bibitem{Benjamin}
E. Flurin, N. Roch, F. Mallet, M. H. Devoret and B. Huard, Generating Entangled Microwave Radiation Over Two Transmission Lines, Phys. Rev. Lett. 109, 183901 (2012).

\bibitem{Aron}
C. Aron, M. Kulkarni and H. T\" ureci, Steady-state entanglement of spatially separated qubits via quantum bath engineering, Phys. Rev. A 90, 062305 (2014). 

\bibitem{Tomadin}
A. Tomadin, V. Giovannetti, R. Fazio, D. Gerace, I. Carusotto, H.E. Tureci, A. Imamoglu, Signatures of the super fluid-insulator phase transition in laser driven dissipative nonlinear cavity arrays, Phys. Rev. A 81, 061801(R) (2010).

\bibitem{DMFT}
 A. Georges, G. Kotliar, W. Krauth and M. Rozenberg et al., Dynamical mean-field theory of strongly correlated fermion systems and the limit of infinite dimensions, Rev. Mod. Phys. 68, 13 (1996).

\bibitem{Greiner}
Markus Greiner, Olaf Mandel, Tilman Esslinger, Theodor W. H\" ansch and Immanuel Bloch, Quantum phase transition from a superfluid to a Mott insulator in a gas of ultracold atoms, Nature 415, 39-44 (3 January 2002).

\bibitem{Fisher}
M. P. A. Fisher, P. B. Weichman, G. Grinstein and D. S. Fisher, Boson localization and the superfluid-insulator transition, Phys. Rev. B 40, 546 (1989).

\bibitem{GiamarchiSchulz}
T. Giamarchi and H. J. Schulz, Localization and Interaction in One-Dimensional Quantum Fluids, Europhys. Lett. 3 1287 (1987).

\bibitem{Pollet}
 M. Hohenadler, M. Aichhorn, S. Schmidt and L. Pollet, Dynamical critical exponent of the Jaynes-Cummings-Hubbard model, Phys. Rev. A 84, 041608(R) (2011).
 
\bibitem{Fazio2}
D. Rossini and R. Fazio, Mott-insulating and glassy phases of polaritons in 1D arrays of coupled cavities, Phys. Rev. Lett. 99, 186401 (2007).

 \bibitem{pasek}
 M. Pasek and Y. D. Chong, Network Models of Photonic Floquet Topological Insulators, Phys. Rev. B 89, 075113 (2014).

\bibitem{Camille}
C. Aron, M. Kulkarni and H. Tureci, Photon-mediated interactions: a scalable tool to create and sustain entangled many-body states, arXiv:1412.8477.

\bibitem{Baust}
A. Baust {\it et al.}, Ultrastrong coupling in two-resonator circuit QED,  arXiv:1412.7372.

\bibitem{Hanggi}
G. M. Reuther {\it et al.}, Two-resonator circuit QED: Dissipative Theory, Phys. Rev. B 81, 144510 (2010).

\bibitem{Nissen}
F. Nissen, S. Schmidt, M. Biondi, G. Blatter, H. E. T\" ureci and J. Keeling, Non-equilibrium dynamics of coupled qubit-cavity arrays, Phys. Rev. Lett. 108, 233603 (2012).

\bibitem{Leboite}
A. Le Boit\' e, G. Orso and C. Ciuti, Bose-Hubbard Model: Relation Between Driven-Dissipative Steady-States and Equilibrium Quantum Phases,  arXiv:1408.1330.

\bibitem{Leboite2}
S. Finazzi, A. Leboit\' e, F. Storme, A. Baksic and C. Ciuti, Corner space renormalization method for driven-dissipative 2D correlated systems, arXiv:1502.05651.

\bibitem{Joshi}
Chaitanya Joshi, Felix Nissen, Jonathan Keeling, Quantum correlations in the 1-D driven dissipative transverse field XY model,  Phys. Rev. A 88 063835 (2013).

\bibitem{Biella}
A. Biella, L. Mazza, I. Carusotto, D. Rossini and R. Fazio, Photon transport in a dissipative chain of nonlinear cavities, arXiv:1412.2509.

\bibitem{Smitha}
P. Nalbach, S. Vishveshwara, A. A. Clerk, Quantum Kibble-Zurek physics in the presence of spatially-correlated dissipation,  arXiv:1503.06398.

\bibitem{CamilleChamon}
G. Goldstein, C. Aron, C. Chamon, Driven-dissipative ising model: mean field solution,  arXiv:1502.03046.

\bibitem{A}
Anders S. Sorensen, Eugene Demler, and Mikhail D. Lukin. Fractional Quantum Hall States of Atoms in Optical Lattices. Phys. Rev. Lett. 94 086803 (2005).

\bibitem{B}
M. Hafezi, A. S. Sorensen, E. Demler, and M. D. Lukin. Fractional quantum Hall effect in optical lattices, Phys. Rev. A 76 023613 (2007).

\bibitem{C}
L. Hormozi, G. M\" oller, and S. H. Simon. Fractional Quantum Hall Effect of Lattice Bosons Near Commensurate Flux, Phys. Rev. Lett. 108 256809 (2012).

\bibitem{D}
 R. N. Palmer and D. Jaksch, High-Field Fractional Quantum Hall Effect in Optical Lattices, Phys. Rev. Lett. 96 180407  (2003).
  
  \bibitem{E}
  Nigel R. Cooper and Jean Dalibard, Reaching Fractional Quantum Hall States with Optical Flux Lattices, Phys. Rev. Lett., 110 185301 (2013).
 
 \bibitem{F} 
 N. Y. Yao, A. V. Gorshkov, C. R. Laumann, A. M. La\" uchli, J. Ye, and M. D. Lukin, Realizing Fractional Chern Insulators in Dipolar Spin Systems, Phys. Rev. Lett., 110 185302 (2013).
 
 \bibitem{G}
A. Sterdyniak, B. A. Bernevig, N. R. Cooper, and N. Regnault, Interacting bosons in topological optical flux lattices, Phys. Rev. B, 91 035115 (2015).

 \bibitem{Sougato}
 J. Cho, D. G. Angelakis and S. Bose, Fractional Quantum Hall State in Coupled Cavities, Phys. Rev. Lett. 101, 246809 (2008).
 
 \bibitem{Greentree2}
 A. L.C. Hayward, A. M. Martin and A. D. Greentree, Fractional Quantum Hall Physics in Jaynes-Cummings-Hubbard Lattices, Phys. Rev. Lett. 108, 223602 (2012).
 
 \bibitem{CarusottoHall}
 R. O. Umucalilar and I. Carusotto, Fractional quantum Hall states of photons in an array of dissipative coupled cavities, Phys. Rev. Lett. 108, 206809 (2012).
 
 \bibitem{Lukin}
M. Hafezi, M. D. Lukin and J. M. Taylor, Non-equilibrium Fractional Quantum Hall state of light,  New J. Phys. 15 063001 (2013).

 \bibitem{Lyon}
 David Carpentier, Pierre Delplace, Michel Fruchart, Krzysztof Gawedzki, Cl\' ement Tauber, Construction and properties of a topological index for periodically driven time-reversal invariant 2D crystals,  arXiv:1503.04157.
 
 \bibitem{netanel}
 Mark S. Rudner, Netanel H. Lindner, Erez Berg, Michael Levin, Anomalous edge states and the bulk-edge correspondence for periodically-driven two dimensional systems, Phys. Rev. X 3, 031005 (2013).
  
 \bibitem{simon2}
 N. Jia, A. Sommer, D. Schuster and J. Simon, Time Reversal Invariant Topologically Insulating Circuits,  arXiv:1309.0878.
 
 \bibitem{Macdonald}
 A. B. Khanikaev, S. H. Mousavi, W.-K. Tse, M. Kargarian, A. H. MacDonald and G. Shvets, Photonic Analogue of Two-dimensional Topological Insulators and Helical One-Way Edge Transport in Bi-Anisotropic Metamaterials, Nature Materials 12 233 (2013).
 
 \bibitem{Liang}
 V. V. Albert, L. I. Glazman and L. Jiang, Topological properties of linear circuit lattices. arXiv:1410.1243.
 
 \bibitem{Plotnik}
 Y. Plotnik {\it et al.} Observation of unconventional edge states in photonic graphene'  Nature Materials 13, 57 (2014).
 
 \bibitem{Segev}
 Mikael C. Rechtsman, Yonatan Plotnik, Julia M. Zeuner, Daohong Song, Zhigang Chen, Alexander Szameit, and Mordechai Segev, Topological Creation and Destruction of Edge States in Photonic Graphene, Phys. Rev. Lett. 111, 103901 (2013). 
 
 \bibitem{LPN}
 M Milicevic, T Ozawa, P Andreakou, I Carusotto, T Jacqmin, E Galopin, A Lema\^itre, L Le Gratiet, I Sagnes, and J Bloch, Edge states in polariton honeycomb lattices, 2015 2D Mater. 2 034012.
 
 \bibitem{Kapit}
 Eliot Kapit, Mohammad Hafezi, and Steven H. Simon, Induced Self-Stabilization in Fractional Quantum Hall States of Light, Phys. Rev. X 4, 031039 (2014).
 
 \bibitem{Lebreuilly}
 J. Lebreuilly, M. Wooters, I. Carusotto, Strongly interacting photons in arrays of dissipative nonlinear cavities under a frequency-dependent incoherent pumping, arXiv:1502.04016.
   
\bibitem{Bellec}
M. Bellec,  U. Kuhl, G. Montambaux and F. Mortessagne, Topological transition of Dirac points in a microwave experiment, Phys. Rev. Lett. 110, 033902 (2013).

\bibitem{Pellegrini}
M. Polini, F. Guinea, M. Lewenstein, Hari C. Manoharan and V. Pellegrini, Artificial graphene as a tunable Dirac material, Nature Nanotech. 8, 625 (2013).

\bibitem{Haldane1988}
F. D. M. Haldane, Model for a Quantum Hall Effect without Landau Levels: Condensed-Matter Realization of the "Parity Anomaly", Phys. Rev. Lett. 61, 2015 (1988).

\bibitem{Semenoff}
G. W. Semenoff, Condensed-Matter Simulation of a Three-Dimensional Anomaly, Phys. Rev. Lett. 53, 2449 (1984).

\bibitem{Annica}
T. O. Wehling, A. M. Black-Schaffer, and A. V. Balatsky, Dirac materials, Adv. Phys. 76, 1 (2014).

\bibitem{Wallace}
P. R. Wallace, The Band Theory of Graphite, Phys. Rev. 71, 622 (1947).

\bibitem{Cayssol}
J. Cayssol,  Introduction to Dirac materials and topological insulators, Comptes Rendus Physique, Volume 14, p. 760-778 (2013).

\bibitem{Solano2}
J. S. Pedernales, R. Di Candia, D. Ballester and E. Solano, Quantum Simulations of Relativistic Quantum Physics in Circuit QED, New J. Phys. 15, 055008 (2013).

\bibitem{RaghuSC}
S. Raghu, Xiao-Liang Qi, C. Honerkamp, and Shou-Cheng Zhang, Topological Mott Insulators, Phys. Rev. Lett. 100, 156401 (2008).

\bibitem{Tianhan}
T. Liu, B. Dou\c{c}ot and K. Le Hur, Realizing Topological Mott Insulators from the RKKY Interaction,  arXiv:1409.6237. 

\bibitem{Esslinger}
G. Jotzu, M. Messer, R. Desbuquois, M. Lebrat, T. Uehlinger, D. Greif and T. Esslinger, Experimental realisation of the topological Haldane model Nature 515, 237-240 (2014).

\bibitem{Sengstock}
J. Struck, C. \"Olschl\" ager, M. Weinberg, P. Hauke, J. Simonet, A. Eckardt, M. Lewenstein, K. Sengstock and P. Windpassinger, Tunable gauge potential for neutral and spinless particles in driven lattices, Phys. Rev. Lett. 108, 225304 (2012).

\bibitem{Tarruell}
L. Tarruell, D. Greif, T. Uehlinger, G. Jotzu and T. Esslinger, Creating, moving and merging Dirac points with a Fermi gas in a tunable honeycomb lattice, Nature 483, 302-305 (2012).

\bibitem{Montambaux}
G. Montambaux, F. Piechon, J.-N. Fuchs and M.O. Goerbig, Merging of Dirac points in a two-dimensional crystal, Phys. Rev. B 80, 153412 (2009).

\bibitem{Bloch1}
M. Aidelsburger, M. Lohse, C. Schweizer, M. Atala, J. T. Barreiro, S. Nascimb\`ene, N. R. Cooper, I. Bloch and N. Goldman, Measuring the Chern number of Hofstadter bands with ultracold bosonic atoms, Nature Physics 11, 162-166 (2015). 

\bibitem{Monica}
M. Atala, M. Aidelsburger, J. T. Barreiro, D. Abanin, T. Kitagawa, E. Demler and I. Bloch, Direct Measurement of the Zak phase in Topological Bloch Bands, Nature Physics 9, 795-800 (2013).

\bibitem{Delplace}
P. Delplace, D. Ullmo and G. Montambaux, The Zak phase and the existence of edge states in graphene,  Phys. Rev. B 84, 195452 (2011). 

\bibitem{QAH1}
C.-Z. Chang et al., Experimental Observation of the Quantum Anomalous Hall Effect in a Magnetic Topological Insulator, Science 340, 167 (2013).

\bibitem{Chamon}
D. Green, L. Santos and C. Chamon, Isolated Flat Bands and Spin-1 Conical Bands in Two-Dimensional Lattices, Phys. Rev. B 82, 075104 (2010).

\bibitem{Lieb}
H. H. Lieb, Two theorems on the Hubbard model, Phys. Rev. Lett. 62, 1201 (1989).

\bibitem{Mielke}
A. Mielke, Ferromagnetic ground states for the Hubbard model on line graphs, J. Phys. A: Math Gen. 24, L73 (1991).

\bibitem{Kagomedot}
Hiroyuki Tamura, Kenji Shiraishi, Takashi Kimura, Hideaki Takayanagi, Flat-band ferromagnetism in quantum dot superlattices,  Phys. Rev. B 65, 085324 (2002).

\bibitem{Baboux}
F. Baboux, L. Ge, T. Jacqmin, M. Biondi, A. Lema\^itre, L. Le Gratiet, I. Sagnes, S. Schmidt, H. E. T\" ureci, A. Amo, J. Bloch, Bosonic condensation in a flat energy band, arXiv:1505.05652.

\bibitem{Huber}
Matteo Biondi, Evert P. L. van Nieuwenburg, Gianni Blatter, Sebastian D. Huber, Sebastian Schmidt, Incompressible polaritons in a flat band, Phys. Rev. Lett. 115, 143601 (2015).

\bibitem{Pannetier}
B. Pannetier, J. Chaussy R. Rammal, and J. C. Villegier, Experimental Fine Tuning of Frustration: Two-Dimensional Superconducting Network in a Magnetic Field, Phys. Rev. Lett. 53, 1845 (1984).

\bibitem{HuseChaikin}
Y. Xiao, D. A. Huse, P. M. Chaikin, M. J. Higgins, S. Bhattacharya and D. Spencer, Comparison of Phase Boundaries between Kagome and Honeycomb Superconducting Wire Networks, Phys. Rev. B 65, 214503 (2002).

\bibitem{VidalMosseriBenoit}
J. Vidal, R. Mosseri and B. Dou\c{c}ot, Aharonov-Bohm cages in two-dimensional structures, Phys. Rev. Lett. 81, 5888 (1998).

\bibitem{Berkeley}
G.-Boong Jo, J. Guzman, C. K. Thomas, P. Hosur, A. Vishwanath and D. M. Stamper-Kurn, Ultracold Atoms in a Tunable Optical Kagome Lattice, Phys. Rev. Lett. 108, 045305 (2012).

\bibitem{Fak}
B. Fak {\it et al.}, Kapellasite: A Kagome Quantum Spin Liquid with Competing Interactions, Phys. Rev. Lett. 109, 037208 (2012).

\bibitem{Lecheminant}
P. Lecheminant, B. Bernu, C. Lhuillier, L. Pierre, P. Sindzingre, Order versus Disorder in the Quantum Heisenberg Antiferromagnet on the Kagome lattice: an approach through exact spectra analysis, Phys. Rev. B 56, 2521 (1997).

\bibitem{Lecheminant2}
P. Azaria, C. Hooley, P. Lecheminant, C. Lhuillier, A. M. Tsvelik, Kagome Lattice Antiferromagnet Stripped to Its Basics, Phys. Rev. Lett. 81, 1694 (1998).

\bibitem{Balents}
Shou-Shu Gong, Wei Zhu, Leon Balents, D. N. Sheng, Global Phase Diagram of Competing Ordered and Quantum Spin Liquid Phases on the Kagome Lattice, Physical Review B 91, 075112 (2015).

\bibitem{White}
Simeng Yan, David A. Huse, Steven R. White, Spin Liquid Ground State of the $S=1/2$ Kagome Heisenberg Model,  Science 332, 1173-1176 (2011).

\bibitem{Schollwock}
Fabian Kolley, Stefan Depenbrock, Ian P. McCulloch, Ulrich Schollw\"ock, Vincenzo Alba, Phase diagram of the J1-J2 Heisenberg model on the kagome lattice, Phys. Rev. B 91, 104418 (2015).

\bibitem{Messio}
Laura Messio, Bernard Bernu, Claire Lhuillier, The Kagome antiferromagnet: a chiral topological spin liquid ?, Phys. Rev. Lett., 108, 207204 (2012).

\bibitem{AntoineS}
A. Wietek, A. Sterdyniak, A. M. L\" auchli, Nature of chiral spin liquids on the kagome lattice, Phys. Rev. B 92, 125122 (2015).

\bibitem{Cecile}
C. Repellin, B. Andrei Bernevig, N. Regnault, $Z_2$ fractional topological insulators in two dimensions, Phys. Rev. B 90, 245401 (2014).

\bibitem{Martinisdisorder}
Yu Chen {\it et al.}, Simulating weak localization using superconducting quantum circuits, Nature Communications 5, 5184 (2014).

\bibitem{Stanfordlight}
K. Fang, Z. Yu and S. Fan, Realizing Effective Magnetic field for photons by controlling the phase of dynamic modulation, Nature 782, 6 2012.

\bibitem{Harper}
P. G. Harper, Single Band Motion of Conduction Electrons in a Uniform Magnetic Field, Proceedings of the Physical Society, Section A 68, 874 (1955).

\bibitem{Hofstadter}
D. R. Hofstadter, Energy levels and wave functions of Bloch electrons in rational and irrational magnetic fields, Phys. Rev. B 14, 2239 (1976).

\bibitem{Fang2}
K. Fang and S. Fan, Photonic de Haas-van Alphen effect, Optics Express 21, 18216 (2013).

\bibitem{Kitagawa}
Takuya Kitagawa, Erez Berg, Mark Rudner, Eugene Demler, Topological characterization of periodically-driven quantum systems, Phys. Rev. B 82, 235114 (2010).

\bibitem{Lindner}
Netanel  H.  Lindner,  Gil  Refael,  Victor  Galitski, Floquet Topological Insulator in Semiconductor Quantum Wells, Nature Physics  7, 490-495 (2011).

\bibitem{KaneMele1}
C. L. Kane and E. J. Mele, Quantum Spin Hall Effect in Graphene, Phys. Rev. Lett. 95, 226801 (2005).

\bibitem{KaneMele2}
C.L. Kane and E.J. Mele, $Z_2$ Topological Order and the Quantum Spin Hall Effect, Phys. Rev. Lett. 95, 146802 (2005).

\bibitem{JoelMoore}
J. E. Moore and L. Balents, Topological invariants of time-reversal-invariant band structures, Phys. Rev. B 75, 121306(R) (2007).

\bibitem{Donna}
D. N. Sheng, Z. Y. Weng, L. Sheng, F. D. M. Haldane, Quantum Spin Hall Effect and Topologically Invariant Chern Numbers, Phys. Rev. Lett. 97, 036808 (2006).

\bibitem{FuKane}
Liang Fu and C.L. Kane, Time Reversal Polarization and a $Z_2$ Adiabatic Spin Pump, Phys. Rev. B 74, 195312 (2006). 

\bibitem{Molenkamp}
M. K\" onig {\it et al.}, Quantum Spin Hall Insulator State in HgTe Quantum Wells, Science 318, 766 (2007).

\bibitem{ZhangHughesBernevig}
B. A. Bernevig, T. L. Hughes and S.-C. Zhang, Quantum Spin Hall Effect and Topological Phase Transition in HgTe Quantum Wells, Science 314, 1757 (2006).

\bibitem{Stephan}
S. Rachel and K. Le Hur, Topological insulators and Mott physics from the Hubbard interaction, Phys. Rev. B 82, 075106 (2010).

\bibitem{Wei}
W. Wu, S. Rachel, W.-M. Liu and K. Le Hur, Quantum Spin Hall Insulators with Interactions and Lattice Anisotropy, Phys. Rev. B 85, 205102 (2012).

\bibitem{TianhanSpiral}
T. Liu, B. Dou\c{c}ot, K. Le Hur, Anisotropic Quantum Spin Hall Effect, Spin-Orbital Textures and Mott Transition, Phys. Rev. B 88, 245119 (2013).

\bibitem{Assaad}
M. Hohenadler and F. F. Assaad, Correlation effects in two-dimensional topological insulators,  J. Phys.: Condens. Matter 25, 143201 (2013).

\bibitem{BalentsKim}
W. Witczak-Krempa, G. Chen, Y. Baek Kim and L. Balents, Correlated quantum phenomena in the strong spin-orbit regime, Annual Review of Condensed Matter Physics, Vol. 5: 57-82 (2014).

\bibitem{RMP}
 G. Kotliar, S. Y. Savrasov, K. Haule, V. S. Oudovenko, O. Parcollet, and C. A. Marianetti, Electronic structure calculations with dynamical mean-field theory, Rev. Mod. Phys. 78, 865 (2006).
 
 \bibitem{SergeAntoine}
 Serge Florens, Antoine Georges, Slave-rotor mean field theories of strongly correlated systems and the Mott transition in finite dimensions, Phys. Rev. B 70, 035114 (2004).
 
 \bibitem{Bardyn}
 Charles-Edouard Bardyn, Torsten Karzig, Gil Refael, Timothy C. H. Liew, Chiral Bogoliubons in Nonlinear Bosonic Systems,  arXiv:1503.08824.
 
 \bibitem{Nalitov}
 A. V. Nalitov, G. Malpuech, H. Ter\c{c}as, and D. D. Solnyshkov, Spin-Orbit Coupling and the Optical Spin Hall Effect in Photonic Graphene, Phys. Rev. Lett. 114, 026803 (2015). 
 
 \bibitem{Karzig}
 Torsten Karzig, Charles-Edouard Bardyn, Netanel Lindner, Gil Refael, Topological Polaritons, Phys. Rev. X 5, 031001 (2015).
 
\bibitem{Joel}
C. Xu and J. E. Moore, Stability of the quantum spin Hall effect: effects of interactions, disorder, and $Z_2$ topology, Phys. Rev. B 73, 045322 (2006).

\bibitem{Gurarie}
V. Gurarie, Single particle Green's functions and interacting topological insulators, Phys. Rev. B 83, 085426 (2011).

\bibitem{Levin}
M. Levin and A. Stern, Fractional topological insulators, Phys. Rev. Lett. 103, 196803 (2009).

\bibitem{Beenakker}
C. W. Groth, M. Wimmer, A. R. Akhmerov, J. Tworzyd?o, C. W. J. Beenakker, Theory of the topological Anderson insulator, Phys. Rev. Lett. 103, 196805 (2009).

\bibitem{Prodan}
E. Prodan, T. L. Hughes, and B. A. Bernevig, Entanglement Spectrum of a Disordered Topological Chern Insulator, Phys. Rev. Lett 105, 115501 (2010).

\bibitem{ZhangQi}
Z. Wang, X.-L. Qi and S.-C. Zhang, Topological invariants for interacting topological insulators with inversion symmetry, Phys. Rev. B 85, 165126 (2012).

\bibitem{Kim}
A. Go, W. Witczak-Krempa, G. Sang Jeon, K. Park and Y. Baek Kim, Correlation effects on 3D topological phases: from bulk to boundary, Phys. Rev. Lett. 109, 066401 (2012).

\bibitem{Budich}
Jan Carl B., R. Thomale, G. Li, M. Laubach and S.-C. Zhang, Fluctuation-induced Topological Quantum Phase Transitions in Quantum Spin Hall and Quantum Anomalous Hall Insulators, Phys. Rev. B 86, 201407(R) (2012).

\bibitem{Gurarie2}
T. C. Lang, A. M. Essin, V. Gurarie and S. Wessel, Z2 topological invariants in two dimensions from quantum Monte Carlo, Phys. Rev. B 87, 205101 (2013).

\bibitem{Bloch2}
M. Atala, M. Aidelsburger, M. Lohse, J. T. Barreiro, B. Paredes and I. Bloch, Observation of the Meissner effect with ultracold atoms in bosonic ladders, Nature Physics 10, 588-593 (2014).

\bibitem{Hemmerich}
T. Kock, M. Olschl\" ager, A. Ewerbeck, W.-M. Huang, L. Mathey, A. Hemmerich, Observing Chiral Superfluid Order by Matter-Wave Interference, Phys. Rev. Lett. 114, 115301 (2015).

\bibitem{Polkovnikov2}
M. Bukov and A. Polkovnikov, Stroboscopic versus non-stroboscopic dynamics in the Floquet realization of the Harper-Hofstadter Hamiltonian, Phys. Rev. A 90, 043613 (2014).

\bibitem{OrignacGiamarchi}
E. Orignac and T. Giamarchi, Meissner effect in a bosonic ladder, Phys. Rev. B 64 p. 144515 (2001).

\bibitem{Crepin}
Fran\c{c}ois Cr\' epin, Nicolas Laflorencie, Guillaume Roux, Pascal Simon, Phase diagram of hard-core bosons on clean and disordered 2-leg ladders: Mott insulator - Luttinger liquid - Bose glass,  Phys. Rev. B 84, 054517 (2011).

\bibitem{Giamarchi}
T. Giamarchi, Quantum Physics in One Dimension, Oxford University Press (2003).

\bibitem{AlexKarynMeissner1}
A. Petrescu and K. Le Hur, Bosonic Mott Insulator with Meissner Currents, Phys. Rev. Lett. 111, 150601 (2013).

\bibitem{AlexKarynMeissner2}
A. Petrescu and K. Le Hur, Chiral Mott Insulators, Meissner Effect, and Laughlin States in Quantum Ladders, Phys. Rev. B 91, 054520 (2015).

\bibitem{Maryland}
B. K. Stuhl, H.-I Lu, L. M. Aycock, D. Genkina, I. B. Spielman, Visualizing edge states with an atomic Bose gas in the quantum Hall regime, Science 349, 1514-1518 (2015).

\bibitem{Florence}
M. Mancini, G. Pagano, G. Cappellini, L. Livi, M. Rider, J. Catani, C. Sias, P. Zoller, M. Inguscio, M. Dalmonte, L. Fallani, Observation of chiral edge states with neutral fermions in synthetic Hall ribbons, Science 25 September 2015:Vol. 349 no. 6255 pp. 1510-1513.

\bibitem{Marie}
M. Piraud, F. Heidrich-Meisner, I. P. McCulloch, S. Greschner, T. Vekua and U. Schollw\"{o}ck, Vortex and Meissner phases of strongly-interacting bosons on a two-leg ladder, 
Phys. Rev. B 91, 140406(R), 2015 .

\bibitem{Fabian}
S. Greschner, M. Piraud, F. Heidrich-Meisner, I. P. McCulloch, U. Schollw\" ock, T. Vekua, Spontaneous increase of magnetic flux and chiral-current reversal in bosonic ladders: Swimming against the tide,  arXiv:1504.06564.

\bibitem{Dhar}
A. Dhar, M. Maji, T. Mishra, R. V. Pai, S. Mukerjee and A. Paramekanti, Bose Hubbard Model in a Strong Effective Magnetic Field: Emergence of a Chiral Mott Insulator Ground State, Phys. Rev. A 85, 041602 (R) (2012).

\bibitem{Mueller}
Ran Wei, Erich J. Mueller, Theory of Bosons in two-leg ladders with large magnetic fields, Phys. Rev. A 89, 063617 (2014).

\bibitem{AkiyukiAntoine}
A. Tokuno and A. Georges, Ground States of a Bose-Hubbard Ladder in an Artificial Magnetic Field: Field-Theoretical Approach, New J. Phys. 16, 073005 (2014).

\bibitem{TeoKane}
J. C.Y. Teo and C. L. Kane, From Luttinger liquid to non-Abelian quantum Hall states, Phys. Rev. B 89, 085101 (2014).

\bibitem{KaneLubensky}
C. L. Kane, R. Mukhopadhyay and T. C. Lubensky, The Fractional Quantum Hall effect in an array of quantum wires, Phys. Rev. Lett. 88, 036401(2002).

\bibitem{BertStern}
Eran Sagi, Yuval Oreg, Ady Stern, Bertrand I. Halperin, Imprint of topological degeneracy in quasi-one-dimensional fractional quantum Hall states,  arXiv:1502.01665.

\bibitem{IvanaAlex}
I. Vasic, A. Petrescu, K. Le Hur and W. Hofstetter, Chiral Bosonic Phases on the Haldane Honeycomb Lattice, Phys. Rev. B 91, 094502 (2015).

\bibitem{Cristiane}
Lih-King Lim, C. Morais Smith, Andreas Hemmerich, Staggered-Vortex Superfluid of Ultracold Bosons in an Optical Lattice, Phys. Rev. Lett. 100, 130402 (2008).

\bibitem{AffleckMarston}
I. Affleck and J. B. Marston, The Large-N Limit of the Hubbard Model: Implications For High-T Superconductors, Phys. Rev. B 37, 3774 (1988).

\bibitem{Chakravarty}
S. Chakravarty, R. B. Laughlin, D. K. Morr, and C. Nayak, Hidden Order in the Cuprates, Phys. Rev. B. 63, 094503 (2001).

\bibitem{Bourges}
B. Fauque, Y. Sidis, V. Hinkov, S. Pailhes, C.T. Lin, X. Chaud, Ph. Bourges, Magnetic order in the pseudogap phase of high-TC superconductors, Phys. Rev. Lett. 96, 197001 (2006).

\bibitem{ladder1}
J. O. Fjaerestad, J. B. Marston and U. Schollwoeck, Orbital currents and charge density waves in a generalized Hubbard ladder, Ann. Phys. (N.Y.) 321, 894 (2006).

\bibitem{ladder2}
G. Roux, E. Orignac, S. R. White and D. Poilblanc, Diamagnetism of doped two-leg ladders and probing the nature of their commensurate phases, Phys. Rev. B 76, 195105 (2007).

\bibitem{ladder3}
S. T. Carr, B. N. Narozhny and A. A. Nersesyan, Spinless Fermionic ladders in a magnetic field: Phase Diagram, Phys. Rev. B 73, 195114 (2006).

\bibitem{FisherZoller}
H.P. B\" uchler, M. Hermele, S.D. Huber, Matthew P.A. Fisher, P. Zoller, Atomic quantum simulator for lattice gauge theories and ring exchange models,  Phys. Rev. Lett. 95 (2005) 040402.

\bibitem{DoucotIoffe}
B. Dou\c{c}ot and L. B. Ioffe, Physical Implementation of Protected Qubits, Rep. Prog. Phys. 75 072001 (2012).

\bibitem{Barbarareview}
B. M. Terhal, Quantum Error Correction for Quantum Memories, Rev. Mod. Phys. 87, 307 (2015).

\bibitem{JohnMartinis}
R. Barends {\it et al.}, Superconducting quantum circuits at the surface code threshold for fault tolerance, Nature 508, 500-503 (2014).

\bibitem{Fu}
S. Vijay and L. Fu, Physical Implementation of a Majorana Fermion Surface Code for Fault-Tolerant Quantum Computation,  arXiv:1509.08134.

\bibitem{AltlandEgger}
L. A. Landau, S. Plugge, E. Sela, A. Altland, S. M. Albrecht and R. Egger,  arXiv:1509.05345.

\bibitem{ZollerMarcos}
D. Marcos, P. Widmer, E. Rico, M. Hafezi, P. Rabl, U.-J. Wiese, P. Zoller, Two-dimensional Lattice Gauge Theories with Superconducting Quantum Circuits, Annals of Physics 351, 634 (2014).

\bibitem{RK}
D. S. Rokhsar and S. A. Kivelson, Superconductivity and the Quantum Hard-Core Dimer Gas, Phys. Rev. Lett. 61, 2376 (1988).

\bibitem{MoessnerSondhi}
Moessner, S. L. Sondhi and E. Fradkin, Short-ranged RVB physics, quantum dimer models and Ising gauge theories , Phys. Rev. B 65, 024504 (2002).

\bibitem{Wiese}
S. Chandrasekharan, U.-J. Wiese, Quantum Link Models: A Discrete Approach to Gauge Theories, Nucl.Phys. B492 455-474 (1997).

\bibitem{Kitaev2}
A. Kitaev, Anyons in an exactly solved model and beyond, Ann. Phys. 321, 2 (2006).

\bibitem{Azbel}
M. Y. Azbel, energy spectrum of a conduction electron in a magnetic field, JETP 19, 634 (1964).

\bibitem{GoldmanSpielman}
N. Goldman, I. Satija, P. Nikolic, A. Bermudez, M.A. Martin-Delgado, M. Lewenstein, I. B. Spielman, Engineering Time-Reversal Invariant Topological Insulators With Ultra-Cold Atoms,  Phys. Rev. Lett. 105 255302 (2010). 

\bibitem{Cocks}
Daniel Cocks, Peter P. Orth, Stephan Rachel, Michael Buchhold, Karyn Le Hur, Walter Hofstetter, Time-Reversal-Invariant Hofstadter-Hubbard Model with Ultracold Fermions, Phys. Rev. Lett. 109, 205303 (2012).

\bibitem{Peter}
Peter P. Orth, Daniel Cocks, Stephan Rachel, Michael Buchhold, Karyn Le Hur, Walter Hofstetter, Correlated Topological Phases and Exotic Magnetism with Ultracold Fermions,  J. Phys. B: At. Mol. Opt. Phys. 46 (2013) 134004.

\bibitem{StephanNature}
Mathias S. Scheurer, Stephan Rachel, Peter P. Orth, Dimensional crossover and cold-atom realization of topological Mott insulators, Sci. Rep. 5, 8386 (2015).

\bibitem{Piraud}
Marie Piraud, Zi Cai, Ian P. McCulloch, Ulrich Schollw\"ock, Quantum magnetism of bosons with synthetic gauge fields in one-dimensional optical lattices: a Density Matrix Renormalization Group study, Phys. Rev. A 89, 063618 (2014).

\bibitem{Edmond1}
M. Di Dio, S. De Palo, E. Orignac, R. Citro, M. Luisa Chiofalo, Persisting Meissner state and incommensurate phases of hard-core boson ladders in a flux, arXiv:1506.03986.

\bibitem{pomeau}
Y. Pomeau and S. Rica, Diffraction Non-Lin\' eaire, C. R. Acad. Sci. Paris 317, S\' erie II, 1287 (1993).

\bibitem{huberpho}
Roman S\" usstrunk, Sebastian D. Huber, Observation of phononic helical edge states in a mechanical topological insulator,  Science, 349, 47 (2015).

\bibitem{diehl}
C.-E. Bardyn, M. A. Baranov, C. V. Kraus, E. Rico, A. Imamoglu, P. Zoller, S. Diehl, Topology by dissipation,  arXiv:1302.5135.

\bibitem{JSTATAlex}
Alexandru Petrescu, H. Francis Song, Stephan Rachel, Zoran Ristivojevic, Christian Flindt, Nicolas Laflorencie, Israel Klich, Nicolas Regnault, Karyn Le Hur, Fluctuations and Entanglement spectrum in quantum Hall states,  J. Stat. Mech. (2014) P10005.

\bibitem{Glattli}
L. Saminadayar, D. C. Glattli, Y. Jin, and B. Etienne, Observation of the e/3 Fractionally Charged Laughlin Quasiparticle, Phys. Rev. Lett. 79, 2526 (1997).

\bibitem{Heiblum}
R. de-Picciotto, M. Reznikov, M. Heiblum, V. Umansky, G. Bunin and D. Mahalu, Direct observation of a fractional charge, Nature 389, 162-164 (1997).


\end{thebibliography}
